\RequirePackage{luatex85}
\documentclass[reprint]{revtex4-2}

\usepackage{graphicx}
\usepackage[normalem]{ulem}
\usepackage{amssymb,amsthm,amsfonts,mathrsfs,dsfont,ctable,commath,sansmath,mathtools}
\usepackage{physics}
\usepackage{color}
\usepackage{paralist}
\usepackage{verbatim}

\usepackage[caption=false]{subfig}
\usepackage[unicode=true,pdfusetitle,
bookmarks=true,bookmarksnumbered=false,bookmarksopen=false,
breaklinks=false,pdfborder={0 0 0},pdfborderstyle={},backref=false,colorlinks=true]
{hyperref}

\usepackage{xcolor}

\newtheorem{theorem}{Theorem}
\newtheorem{corollary}{Corollary}
\newtheorem{definition}{Definition}
\newtheorem{proposition}{Proposition}

\newtheorem{lemma}{Lemma}

\newtheorem{assumption}{Assumption}

\newcommand{\rA}{\textnormal{A}} \newcommand{\rB}{\textnormal{B}} \newcommand{\rC}{\textnormal{C}}  \newcommand{\rE}{\textnormal{E}} \newcommand{\rI}{\textnormal{I}}      

  \newcommand{\tC}{\mathcal{C}} \newcommand{\tD}{\mathcal{D}}     \newcommand{\tI}{\mathcal{I}}  \newcommand{\tM}{\mathcal{M}}  \newcommand{\tO}{\mathcal{O}} \newcommand{\tS}{\mathcal{S}} \newcommand{\tT}{\mathcal{T}} \newcommand{\tU}{\mathcal{U}}

\newcommand{\N}{\mathbb{N}}

\newcommand{\R}{\mathbb{R}}

\newcommand{\grp}[1]{\mathsf{#1}}
\newcommand{\spc}[1]{\mathcal{#1}}

\def\d{{\rm d}}

\newcommand{\Span}{{\mathsf{Span}}}

\def\>{\rangle}
\def\<{\langle}

\newcommand{\Vac}[1]{\mathrm{Vac}}

\newcommand{\map}[1]{\mathcal{#1}}

\newcommand{\St}{{\mathsf{St}}}
\newcommand{\Eff}{{\mathsf{Eff}}}
\newcommand{\Pur}{{\mathsf{Pur}}}
\newcommand{\Transf}{{\mathsf{Transf}}}
\newcommand{\Tests}{{\mathsf{Tests}}}

\newcommand{\Proof}{{\bf Proof. \,}}
\newcommand{\myhat}[1]{\mathchoice
	{\widehat{\textstyle#1}} 
	{\widehat{\textstyle#1}} 
	{\widehat{\scriptscriptstyle#1}} 
	{\widehat{\scriptscriptstyle#1}} 
}

\input{qcircuit}

\begin{document}
	
	\title{
		Information-theoretic derivation of energy, speed bounds, and quantum theory}
	
	\author{Lorenzo Giannelli}
	\email{giannell@connect.hku.hk}
	\affiliation{\footnotesize QICI Quantum Information and Computation Initiative, Department of Computer Science, The University of Hong Kong, Pok Fu Lam Road, Hong Kong}
	\affiliation{\footnotesize HKU-Oxford Joint Laboratory for Quantum Information and Computation}
		
	\author{Giulio Chiribella}
	\email{giulio@cs.hku.hk}
	\affiliation{\footnotesize QICI Quantum Information and Computation Initiative, Department of Computer Science, The University of Hong Kong, Pok Fu Lam Road, Hong Kong}
	\affiliation{\footnotesize HKU-Oxford Joint Laboratory for Quantum Information and Computation}
	\affiliation{\footnotesize Quantum Group, Department of Computer Science, University of Oxford, Wolfson Building, Parks Road, Oxford, United Kingdom}
	\affiliation{\footnotesize Perimeter Institute for Theoretical Physics, 31 Caroline Street North, Waterloo, Ontario, Canada}
	
	\begin{abstract}
	
	We provide a derivation of quantum theory in which the existence of an energy observable that generates the reversible  dynamics  follows directly from information-theoretic principles. 
		 Our first principle is that every reversible dynamics is implementable  through a collision model, \textit{i.e.} a sequence of fast collisions with an array of identically prepared systems. Combined with four additional  principles, known as causality, classical decomposability, purity preservation, and strong symmetry, the collision model pins down the quantum framework, sets up a one-to-one correspondence between energy observables and generators of the dynamics, and provides an information-theoretic derivation of the Mandelstam-Tamm bound on the speed of quantum evolutions.     
	\end{abstract}

	\maketitle
	
\paragraph{Introduction.}
	Information-theoretic notions  play an important role in modern physics, from thermodynamics to quantum field theory and gravity.  With the advent of quantum information and computation, new links between  information-theoretic primitives and physical laws have been uncovered, suggesting that information could be the key to understand the counterintuitive laws of quantum mechanics \cite{fuchs2002quantum, brassard2005information}.   Over the past two decades, this research program blossomed  into  a series of reconstructions of the quantum framework from information-theoretic principles \cite{hardy2001quantum,chiribella2011informational,dakic2011quantum,masanes2011derivation, hardy2011reformulating,barnum2014higher,hohn2017quantum,selby2021reconstructing}.  
		
	A limitation of most quantum reconstructions, however, is that they do not provide  direct  insights into  the dynamics of quantum systems. Crucially, they do not provide an information-theoretic characterization of the notion of energy,  and an explanation of its dual role of  observable quantity and  generator of the dynamics.    For long time,     this observable-generator duality   has been known to be a key  feature from which much of the  algebraic structure of quantum theory can be derived \cite{grgin1974duality}. More recently, the observable-generator duality was assumed as an axiom in a  quantum reconstruction~\cite{barnum2014higher}, or  as a requirement for post-quantum dynamics~\cite{branford2018defining, plavala2022operational, jiang2024unification}.   And yet, little is known about the origin of the duality.  How is ``energy'' emerging from ``information,'' and why does it drive the evolution of quantum systems?

	\begin{figure}[t!]
		\hspace*{-.65cm}
		\includegraphics[width=0.55\textwidth]{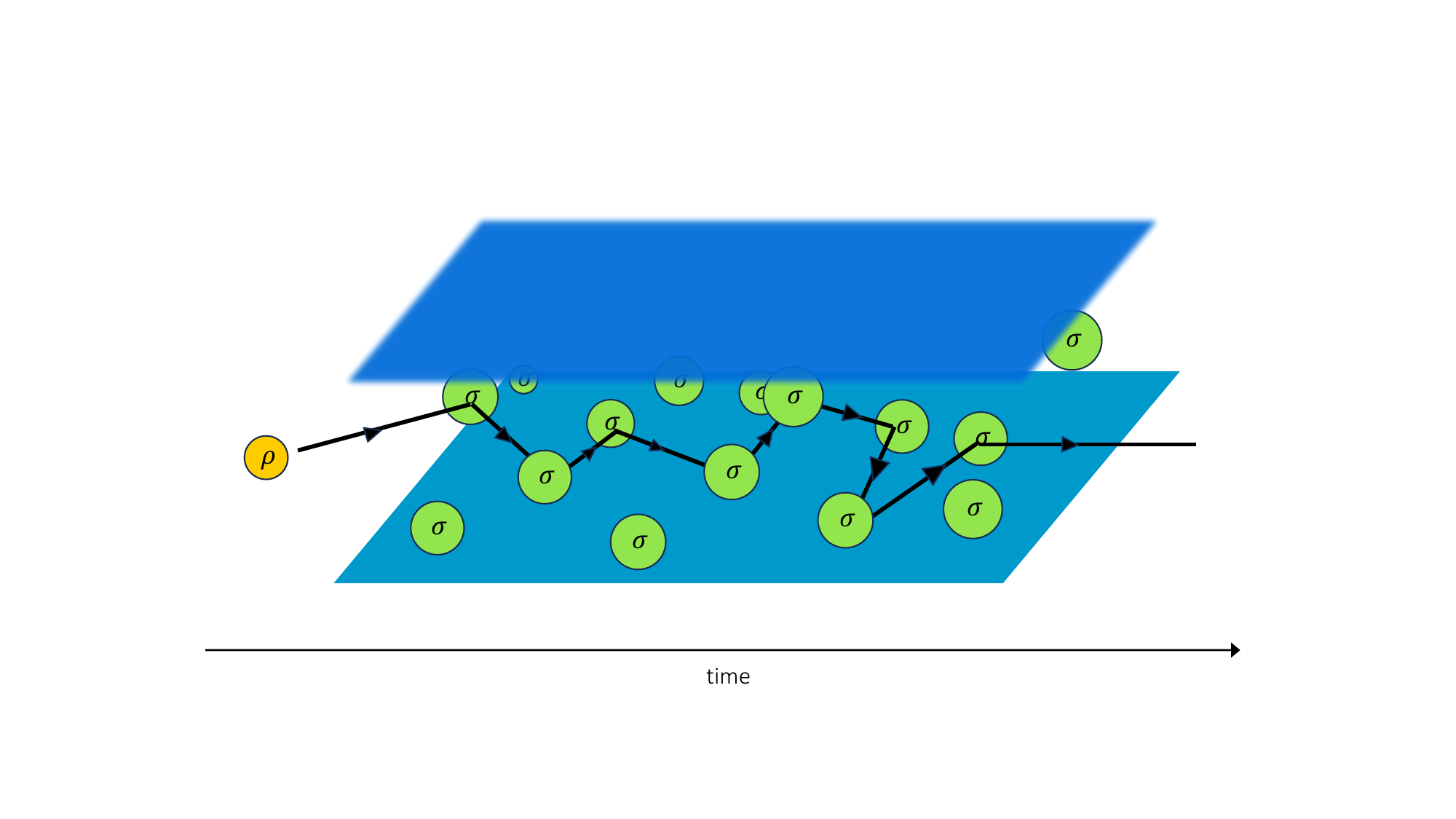} 
		\caption{{\bf Collision model for reversible dynamics.}  A target system in an initial state $\rho$ undergoes a sequence of interactions   with a set of systems, independently and identically prepared  in the reference state $\sigma$. 	Here we require that every reversible dynamics of the target system can be obtained  by varying the state $ \sigma$, in the limit of instantaneous and infinitely frequent interactions. 
		 }
		\label{fig:collision_model}
	\end{figure}

	Here,  we provide a reconstruction of quantum theory in which the key features of  quantum dynamics are reconstructed directly from the principles, without invoking the machinery of   Hilbert spaces. Our first principle is that every reversible dynamics should be implementable by  a sequence of  fast collisions with an array of identically prepared systems, as   illustrated in Figure \ref{fig:collision_model}.  In this model,  we explore the idea that  dynamics  arises from \emph{informational nonequilibrium}, the condition that the system's state deviates  from  the state of the systems it collides with.    We then combine this idea with four   principles that have featured in previous quantum reconstructions: causality \cite{chiribella2010probabilistic, chiribella2011informational} classical decomposability \cite{barnum2014higher}, strong symmetry \cite{barnum2014higher}, and purity preservation \cite{chiribella2015conservation, chiribella2015operational, chiribella2016entanglement}.
	We show that, together,   the above principles pin down the framework of finite-dimensional quantum theory and, at the same time,  provide direct access to  the key features of quantum dynamics,   including in particular the existence of an energy observable in one-to-one correspondence with the generator of the dynamics. We then show that the variance of the energy observable yields a lower bound on the speed of state evolution, thereby providing  a purely information-theoretic derivation of the Mandelstam-Tamm bound in quantum theory  \cite{mandelstam1945uncertainty}. Our results establish  a direct link between information-theoretic principles  and  dynamical notions such as  energy and speed.

\paragraph{Collisional models in general probabilistic theories.}
	A general class of physical  theories, including quantum theory and potential candidates of post-quantum theories, can be formulated as general probabilistic theories \cite{hardy2001quantum, barrett2007information, barnum2007generalized, hardy2011foliable, hardy2013formalism, hardy2016reconstructing}. More specifically, here we adopt the framework of operational probabilistic theories (OPTs)~\cite{chiribella2010probabilistic, chiribella2011informational, hardy2013formalism, dariano2017quantum}, describing  a set of physical systems, closed under composition, a set of transformations thereof, closed under parallel and sequential composition.

	Let us start with a brief overview of the OPT framework (a more in-depth presentation can be found in the Supplemental Material \cite{supp} and in the original papers). Mathematically, the  structure of an OPT is underpinned by the graphical language of symmetric monoidal categories   \cite{abramsky2004categorical, abramsky2008categorical, coecke2010quantum, selinger2010survey, coecke2016generalised, coecke2017picturing}. For a pair of systems $\rA$  and $\rB$,  $\rA\otimes \rB$ denotes the composite system consisting of subsystems $\rA$ and $\rB$.  The trivial system $\rI$, representing degrees of freedom that are not relevant in the OPT, satisfies the relation $\rA \otimes  \rI  =  \rI \otimes \rA =  \rA$ for every system $\rA$.  
	
	The set of  physical transformations with input system $\rA$ and output system $\rA'$ is denoted by $\Transf  (\rA  \to  \rA')$.   For general systems $\rA,\rA',\rB,\rB'$,  two transformations $\map T  \in  \Transf (\rA \to  \rA')$ and  $\map S  \in  \Transf (\rB \to  \rB')$ can be performed in parallel, giving rise to the transformation $\map T\otimes \map S  \in \Transf (\rA  \otimes \rB \to  \rA'\otimes \rB')$.    If $\rA'  =  \rB$,  the transformations $\map T$ and $\map S$ can also be performed in a sequence,  giving rise to the transformation $\map S  \circ \map T  \in  \Transf (\rA  \to  \rB')$.    
 
	For every system $\rA$,  there exists an identity transformation, denoted by   $\map I_{\rA} \in \Transf (\rA\to \rA)$, which leaves the system unchanged.  A transformation $\map T  \in  \Transf (\rA\to \rB)$ is called reversible if there exist another  transformation     $\map S\in \Transf (\rB\to \rA)$ such that $ \map S  \circ \map T  =  \map I_{\rA}$ and $\map T\circ \map S  = \map I_{\rB}$.   The reversible transformations in $\Transf (\rA\to \rA) $ form a group, hereafter denoted by $\grp G_{\rA}$.  In quantum theory, reversible transformations are associated to unitary operators $U:  \spc H_\rA \to \spc H_\rA$, where $\spc H_\rA$ is the Hilbert space associated to system $\rA$, while general transformations  in $\Transf (\rA \to \rB)$ are described by quantum operations \cite{hellwig1969pure, hellwig1970operations, kraus1983states}, that is, completely positive, trace non-increasing linear maps transforming linear operators on $\spc H_\rA$ into linear operators on $\spc H_\rB$.  
 
In quantum theory,  the possible experiments are described by {\em quantum instruments} \cite{davies1976quantum}, that is, collections of quantum operations that sum up to trace-preserving maps.  	In a general OPT, an $N$-outcome experiment transforming an input system $\rA$ into an output system $\rB$ is described by a collection of transformations  $(\map T_i)_{i=1}^N$, with the transformation $\map T_i  \in     \Transf (\rA \to \rB)$ corresponding to the $i$-th outcome.    The OPT specifies the set of all possible experiments, with the  constraint that such a set  must be closed under parallel composition, sequential composition, and coarse-graining (see the Supplemental Material for more details). 
 
	Experiments with a single outcome ($N=1$) are associated to a single transformation,  called {\em deterministic}.  The set of deterministic transformations with input $\rA$ and output $\rB$ is denoted by $\Transf_1  (\rA \to \rB)$.  Transformations from the trivial system to system $\rA$ are called {\em states}, while transformations from system $\rA$ to the trivial system are called {\em effects}.  We will  use the notation  $\St  (\rA) :  =  \Transf  (\rI \to \rA)$ and $\Eff  (\rA):  =  \Transf (\rA  \to \rI)$ for the sets of states and effects, and  $\St_1  (\rA) :  =  \Transf_1  (\rI \to \rA)$ and $\Eff_1  (\rA):  =  \Transf_1 (\rA  \to \rI)$ for the subsets of deterministic states and deterministic effects, respectively.   

	OPTs associate  transformations with trivial input and trivial output ($\rA=  \rB=  \rI$) to probabilities.  In particular, the combination of a state $\rho \in \St (\rA)$ with an effect $e \in  \Eff (\rA)$ yields a probability $e\circ \rho$.    As a consequence, states can be represented as functions on the set of effects, and {\em vice-versa}.  In turn, this fact implies that states and effects can be embedded into vector spaces over the real field, hereafter denoted by $\St_\R  (\rA) $ and $\Eff_\R  (\rA)$, respectively. More generally, each set of  physical transformations $\Transf(\rA\to \rB)$ can be embedded in a real vector space, hereafter denoted by $\Transf_\R (\rA\to \rB)$.  In the following we will make the standard assumption that the vector spaces  $\Transf_\R (\rA\to \rB)$ are finite-dimensional  for all systems $\rA$ and $\rB$, and that the sets  $\Transf(\rA\to \rB)$ are compact convex sets. Under this assumption,  the group of reversible transformations $\grp{G}_\rA$ is a compact Lie group and admits a faithful matrix representation for every system $\rA$ (see Supplemental Material~\cite{supp} for a proof). 

	We now formulate a definition of collision model in general probabilistic theories. Collision models have been extensively studied in the theory of quantum open systems \cite{rau1963relaxation, ziman2002diluting, scarani2002thermalizing, rybar2012simulation, ciccarello2013collision, cattaneo2021collision, ciccarello2022quantum}, with applications to non-equilibrium quantum thermodynamics \cite{scully2003extracting, strasberg2017quantum} and quantum machine learning \cite{lloyd2014quantum}. In a collision model, a target system  evolves through a sequence of pairwise  interactions with an array of independent and identically prepared  systems, as depicted in Fig.~\ref{fig:collision_model}. We assume that the interaction is the same at all steps of the sequence,  and gives rise to a reversible joint evolution, specified by a one-parameter subgroup $(\map S_t)_{t \in \R}$  of the group $\grp G_{\rA \otimes \rA'}$, where $\rA$ is the target system and $\rA'$ is the system it collides with.
 As a result of the  interaction, the target system undergoes  the transformation
	\begin{equation}
	\label{eq:GPT-unit-block-main}
		\map C_{\tau, \sigma} :  =   (\map I_A  \otimes u_{\rA'})  \circ   \map S_\tau  \circ (  \map I_A \otimes \sigma) \,, 
	\end{equation}
	where  $\tau $ is the interaction time,   $\sigma \in \St_1 (\rA')$  is the initial state of system $\rA'$, and  $u_{\rA'}  \in  \Eff_1 (\rA')$ is a deterministic effect, representing the process of discarding system $\rA'$ after the interaction.   Hereafter, we will call $\sigma$  the {\em reference state} of the collisional model.

	We now  consider the effective transformation resulting from an asymptotically long sequence of collisions, taking place in a finite time $t$.  We will assume that {\em {(i)}} the number of collisions grows linearly with $t$, as  $n=  \lfloor t/\tau\rfloor$, 
	and  {\em (ii)} the collisions are instantaneous ($\tau \to 0$). In this setting, the effective evolution  of system $\rA$ over time $t$ is given by $\tU_{t ,  \sigma} :  =  \lim_{\tau \to 0}      \left( \tC_{ \tau  ,  \sigma }\right)^{  \left \lfloor \frac{    t   }{\tau} \right \rfloor}$. We name $(\map U_{t,\sigma})_{t\in \R}$ the {\em collisional dynamics} generated by $\sigma$. In the Supplemental Material  we prove the following: 
	\begin{theorem}
	\label{thm:GPT-programmed-ev}
		For every reference  state $\sigma$, the collisional dynamics $(\map U_{t,\sigma})_{t\in \R}$ is a one-parameter  Lie group, of the form  $\tU_{t,\sigma} = e^{ G_\sigma t}$, with  $G_\sigma  := (  \tI_{\rA} \otimes u_{\rA'}   )  \frac{ \d  \tS_\tau }{\d\tau} \big|_{\tau  =  0}   ( \map I_A \otimes \sigma)  $. 
	\end{theorem}   
	Here, the derivative $ \d  \tS_\tau /\d\tau$ is well-defined because $(\tS_\tau)_{\tau \in \R}$ is  a one-parameter subgroup of a compact Lie group  with  a faithful matrix representation,   and therefore it is also a Lie group~\cite{baker2003matrix, hall2015lie}.   

An important  consequence of Theorem \ref{thm:GPT-programmed-ev} is that the transformation  $\map U_{t, \sigma}$ is physically reversible: for example, it  can be reversed by replacing $\map S_\tau$ with its inverse $\map S_{-\tau}$ in Eq.~\eqref{eq:GPT-unit-block-main}, while keeping the state $\sigma$ fixed.   Note that the map $G^{(\rA)}:  \sigma  \mapsto  G_\sigma$ transforms states of system $\rA$ into elements of the Lie algebra $\mathfrak{g}_\rA$ associated to the Lie group $\grp G_\rA$.     Physically, the elements of $\mathfrak{g}_\rA$ represent the  generators of the possibly dynamics of system $\rA$.

\paragraph{Informational equilibrium.}
	We now consider {\em symmetric} collision models, for which  $\rA  = \rA'$.  In these models, we say that the system and its environment are  at {\em informational equilibrium} if they  are in the same state ($\rho  =  \sigma$). The model  is {\em stationary at   equilibrium} if the condition
	\begin{align}
	\label{infoeq}
		\tS_\tau  (\rho \otimes \rho)  =  \rho \otimes \rho  \qquad \forall \rho \in  \St_1 (\rA),  \,   \forall \tau  \in  \R  
	\end{align}  
	holds. This condition expresses the idea that  dynamics only takes place when the system and its environment are  in different states. A straightforward consequence of stationarity is the following:
	\begin{lemma}
		If  a symmetric collision model is stationary at equilibrium, then the reference state  $\sigma$ is invariant under the collisional dynamics $(  \map U_{t, \sigma})_{t\in \R}$. 
	\end{lemma}

In the rest of this paper, we explore the idea that every dynamics can be generated by a symmetric collisional model that is stationary at equilibrium.  Specifically, we say that a collision model is {\em universal} if every reversible transformation $\tU  \in  \grp G_\rA$  is of the form  $ \map U  =    \map U_{t,\sigma}$ for some suitable time $t$ and some suitable  state $\sigma \in \St_1  (\rA)$.  Our first principle, called {\em Dynamics from Informational Nonequilibrium (DIN)}, is  that  there exists a symmetric collision model that is both  stationary and universal.  
	
	In the Supplemental Material, we show that  quantum theory on complex Hilbert spaces satisfies DIN,  and we explicitly provide a symmetric, stationary, and universal collision model. In contrast, we show that for quantum theory on real Hilbert spaces, every stationary collision model is trivial ({\em i.e.} it only implements the identity transformation), and therefore fails to be universal.

	An important consequence of DIN is the following theorem, proven  in the Supplemental Material~\cite{supp}:
	\begin{theorem}
	\label{thm:1to1-state-generator}
		In a theory satisfying DIN,  
		the map $G^{(\rA)}  :  
		\St_1 (\rA) \to  \mathfrak{g}_\rA,  \,,  
		 \sigma \mapsto G_\sigma$   is injective if and only if system $\rA$ has a unique invariant state.
	\end{theorem}
 Theorem \ref{thm:1to1-state-generator} shows that a duality between  states and generators of the reversible dynamics is equivalent to  a simple property, the 	Uniqueness of the Invariant State (UIS).  UIS is a well-studied property, satisfied by quantum theory \cite{chiribella2010probabilistic, chiribella2011informational, chiribella2016entanglement}  and playing a key role in the foundations of thermodynamics \cite{chiribella2014microcanonical}.  
 
 	We conclude this section with an interesting aside. For systems with a unique invariant state $\chi_\rA$,   the evolution $\map U_{t, \sigma}$ can be inverted by replacing the state $\sigma$ and the time $t$ with  a new state $\sigma_{\rm inv}:  = ( \chi_\rA  -  p_\sigma\,  \sigma )/(1-p_\sigma)$ and  a new time $t_{\rm inv} :  =  t \,(  1-p_\sigma)  /p_\sigma$, where $p_\sigma$ is the maximum probability of the state $\sigma$ in a convex decomposition of $\chi_\rA$ (see the Supplemental Material for details).   It is interesting to observe that  $t_{\rm inv}$ is  generally different from $t$, thereby implying that, if our collision model is taken to be fundamental, dynamical evolutions, while in principle reversible, have a privileged direction that is physically easier to implement, in the sense that it takes a shorter evolution time.  For systems with high-dimensional state space,  the inversion time, thereby implying a form of physical  irreversibility.  For example, quantum dynamics generated by pure states $\sigma$ have an inversion time $t_{\rm inv}   =  t\,  (d-1)$, where $d$ is the system's dimension.

\paragraph{Duality between observables and generators.}	In the Supplemental Material, we show that  the state-generator duality set up by Theorem~\ref{thm:1to1-state-generator} can be turned into an observable-generator duality whenever the theory satisfies an additional property known as  strong self-duality \cite{koecher1958die, vinberg1960homogeneous}, {\em i.e.} whenever  the cones spanned by the sets $\St (\rA)$ and $\Eff (\rA)$ are isomorphic and the pairing defined by $\<  e,  \rho  \>  :  =  e\circ \rho \, ,  \forall \rho  \in\St (\rA) , \, \forall e\in \Eff (\rA)$  can be expressed as a Euclidean inner product.
  	\begin{theorem}[Observable-generator duality]
		\label{theo:energy-obs}
		In a theory satisfying DIN,    UIS,  and strong self-duality, the elements of the Lie algebra $\mathfrak{g}_\rA$ are in one-to-one linear correspondence with the elements of the subspace  $\map O_{\rA}  :  =  \{      H  \in  \Eff_\R  (\rA) ~|~  H  \circ \chi_\rA =  0\}$. Moreover, every element $H  \in  \map O_\rA$ is invariant under the dynamics generated by the corresponding element $G_H \in  \mathfrak{g}_\rA$.   
		\end{theorem}
		Theorem~\ref{theo:energy-obs} implies that every generator of the reversible dynamics is canonically associated to an {\em energy observable} $H  \in \map O_\rA$,  whose expectation value can be estimated by performing suitable measurements on the system (see the Supplemental Material for more details).  In summary,  Theorem~\ref{theo:energy-obs} provides a derivation of the notion of energy from the idea that dynamics arises from collisions   out of informational equilibrium, combined  with the properties of  strong self-duality and uniqueness of the invariant state. In the following, we  will replace these two properties with more basic  requirements. 

\paragraph{Derivation of quantum theory.}
	Finite-dimensional quantum theory can be uniquely characterized by DIN, plus four additional assumptions that have appeared in  other quantum reconstructions. Here we review these assumptions briefly,  leaving a detailed discussion to the Supplemental Material \cite{supp}. 
		
	The first assumption is  {\em Causality (C)}, expressing the fact that the outcomes of present experiments should be independent of the setting of future experiments. The second  assumption is {\em Classical Decomposability (CD)}  \cite{barnum2014higher}, namely the requirement that every state can be prepared as a random mixture of perfectly distinguishable pure states. This assumption provides a starting point for the quantification of information in terms of  entropies~\cite{short2010entropy, barnum2012entropy, chiribella2016entanglement}.   	The third  assumption   is   {\em Purity Preservation (PP)} \cite{chiribella2016entanglement}, namely the requirement that  composing two pure transformations in parallel or in sequence gives rise to another pure transformation (informally, a pure transformation is one that cannot be viewed as the coarse-graining of a set of transformations; see the Supplemental Material  for more details.) The fourth assumption is {\em Strong Symmetry (SS)} \cite{barnum2014higher}, the condition that every two sets of perfectly distinguishable pure states can be reversibly converted into one another if they have the same cardinality.  SS  guarantees that two sets of states that can perfectly encode the same amount of classical information are equivalent.

	In the Supplemental Material~\cite{supp},  we prove that the combination of C, CD,  PP, and SS  implies that the states of the theory form an irreducible Euclidean Jordan algebra \cite{jordan1934algebraic, barnum2020composites}. This result  implies in particular that every theory satisfying  C, CD, PP, and SS   also satisfies strong self-duality and UIS.  We then combine this fact with Theorem~\ref{theo:energy-obs}, showing that DIN, UIS, and strong self-duality imply the observable-generator duality.    Finally, we leverage a result of Barnum,   M\"uller, and Ududec  \cite{barnum2014higher}, who showed that irreducible Jordan algebras plus observable-generator duality  imply quantum theory on complex Hilbert spaces.  In summary,  DIN, C, CD, PP, and SS imply the standard Hilbert space framework. 

\paragraph{Speed bound.}
	The above derivation of the quantum framework can be used to reconstruct properties of quantum dynamics directly from the principles, thereby establishing a closer connection between information-theoretic and dynamical features. 
	For example, in the Supplemental Material \cite{supp} we use the principles to show that every dynamics has a maximal set of perfectly distinguishable stationary states  $\{  \psi_i\}_{i=1}^d$,  corresponding to energy eigenstates in the Hilbert space framework. In addition, the energy observable  associated to the dynamics can be decomposed as $H   =  \sum_{i=1}^d    E_i  \,   e_{\psi_i}$, where $(e_{\psi_i})_{i=1}^d$ are the effects associated to the states $(\psi_i)_{i=1}^d$ by strong self-duality, and $(E_i)_{i=1}^d$ are real numbers.  The effects  $(e_{\psi_i})_{i=1}^d$   form a measurement, and the expectation value of the energy observable  on a given state $\rho$ can be computed as $\<  H \>_\rho   :  =  \sum_i  E_i \,  \<e_{\psi_i} , \rho\>$.  This expression  suggests to interpret  $(e_{\psi_i})_{i=1}^d$   as the {\em canonical energy measurement} and  the numbers $(E_i)_{i=1}^d$ as the possible values of the energy, distributed with probability $p_\rho  (E_i)  =  \<e_{\psi_i}, \rho\>$,  $\forall  i\in \{1,\dots, d\}$.   
	  
	  We now show that the principles  yield an upper bound on the speed of state changes in terms of the variance of the energy observable $H$, providing an information-theoretic  derivation of the celebrated Mandelstam-Tamm bound~\cite{mandelstam1945uncertainty}.    
	 To this purpose, we  introduce  a notion of speed in  general probabilistic theories: 
	\begin{definition}[Speed of state change]
		Let $(\tU_t)_{t \in \R}$ be a one-parameter subgroup of $\grp G_{\rA}$, representing  a possible dynamics of system $\rA$, 
		let $\{\rho_t  =   \tU_t  (\rho)~|~  t\in  \R\}$ be the trajectory of an initial state $\rho$,  and let $t_0$ and $t_1 \ge t_0$ be two moments of time.   The average speed of state change from time $t_0$ to time $t_1$ is  defined as 
		\begin{equation}
			v_\rho  (t_0, t_1)\coloneqq \frac{\| \rho_{t_1}  - \rho_{t_0} \|}{t_1-t_0} \, , 
		\end{equation}
		where $\| \cdot \|$ is an arbitrary  norm on  $\St_\R  (\rA)$.    
		Similarly, the instantaneous speed at time $t$  is defined as   $v_\rho  (t)  := \lim_{  \delta t \to 0}    v_\rho  ( t ,  t+  \delta t )$. 		
	\end{definition}
	In the following we will take the norm $\|  \cdot \|$ to be the Euclidean norm induced by self-duality, namely $  \|  \sum_{j}   \,   c_j  \,  \rho_j\|  : =  \sqrt{  \sum_{j,k}   \,  c_j   \,  c_k  \, e_{\rho_j}    (  \rho_k)}$ for every linear combination of  states $(\rho_j)$ with real coefficients $(c_j)$.  With this choice of norm,  we show that  the instantaneous speed at time $t$ is $v_\rho (t)    =   \| G ( \rho) \| $  for every reversible dynamics with time-independent generator  $ G$ \cite{supp}.   Using this fact, we obtain the following:
	\begin{theorem}[Speed limit]\label{thm:speed-bound}
		DIN,  C, CD, PP, and SS imply that the time $\Delta t$ taken by a system to transition from an initial state $\rho_{t_0}$ to a final state $\rho_{t_1}$ through a reversible dynamics  satisfies the bound 
		\begin{equation}
			\label{eq:generalized-speed-bound}
			\Delta t  \ge    \frac{D  (\rho_{t_0},\rho_{t_1}) }{ \Delta  H} \, ,  
		\end{equation}
		where  $D(\rho,    \rho')  :  =   \|   \rho  -  \rho'\| /\sqrt 2 $ is the normalized Euclidean distance between two  states $\rho$ and $\rho'$, while $\Delta H$
		 is the standard deviation  of the canonical energy measurement when performed on the state $\rho$. 
	\end{theorem}
	The proof is provided in the Supplemental Material \cite{supp}. 
	For two perfectly distinguishable pure states,  the bound becomes   $  \Delta  t   \ge   1  /\Delta E$, which in the case of quantum theory  coincides with the Mandelstam-Tamm bound up to a dimensional factor $h/4$, where $h$ is Planck's constant  \cite{mandelstam1945uncertainty}.   Hence, Eq. (\ref{eq:generalized-speed-bound}) can be regarded as an alternative, purely information-theoretic derivation of the Mandelstam-Tamm bound.

	\paragraph{Conclusions.}   In this work we explored the idea that reversible dynamics arises from informational nonequilibrium in a collisional model. 
	  In this setting,  we provided an information-theoretic derivation of the notion of energy, and of its dual role of observable and generator of the dynamics.   Our derivation can be extended to a full reconstruction of quantum theory, in which the principles give direct access to the key aspects of quantum dynamics, such as the  Mandelstam-Tamm speed limit.  

	\medskip{}
	\begin{acknowledgments}
		{\bf Acknowledgments.} We thank Tamal Guha, Alexander Wilce, and Marco Erba for helpful comments during early presentations of this work. LG thanks Antoine de Saint Germain for valuable discussions.
		This work was supported by the Chinese Ministry of Education (MOST) through grant 2023ZD0300600,   by the Hong Kong Research Grant Council through  
		the Senior Research Fellowship Scheme SRFS2021-7S02 and the Research Impact Fund R7035-21F,  by the John Templeton Foundation  through the ID\# 62312 grant, as part of the ‘The Quantum Information Structure of Spacetime’ Project (QISS), and  by the State Key Laboratory of Quantum Information Technologies and Materials, Chinese University of Hong Kong.  The opinions expressed in this publication are those of the authors and do not necessarily reflect the views of the John Templeton Foundation. Research at the Perimeter Institute is supported by the Government of Canada through the Department of Innovation, Science and Economic Development Canada and by the Province of Ontario through the Ministry of Research, Innovation and Science. 
	\end{acknowledgments}

\bibliographystyle{apsrev4-2}
\bibliography{references}	
	
\onecolumngrid
	
\clearpage
	
\section*{SUPPLEMENTAL MATERIAL}

\section{Overview of the framework of Operational Probabilistic Theories (OPTs)}

A general class of physical theories, including quantum theory and potential post-quantum candidates, can be  in the framework of general probabilistic theories (GPTs). A GPT describes physical systems in terms of convex geometry and linear algebra \cite{hardy2001quantum, barrett2007information, barnum2007generalized, chiribella2010probabilistic, hardy2011foliable, hardy2013formalism,dariano2017quantum}.   Operational Probabilistic Theories (OPTs)  \cite{chiribella2010probabilistic, chiribella2011informational, hardy2013formalism, dariano2017quantum} are GPTS in which the composition of physical systems is regarded as a primitive notion:  every OPT comes with a full specification of all the composite systems that arise in a given physical theory.  This compositional approach has a rigorous foundation  in the framework of category theory \cite{mac2013categories,coecke2017picturing}, although, in fact, no knowledge of category theory is necessary to understand or use the OPT framework. In the following, we provide an intuitive (and yet sufficiently rigorous) overview of the main facts and techniques used to derive the results of this paper. 

\subsection{Systems}

An OPT describes the possible processes undergone by  a set of physical systems. The systems are represented diagrammatically by wires, such as
\begin{equation*}
	\begin{aligned}
		\Qcircuit @C=1.2em @R=.8em @!R {  & \qw & \poloFantasmaCn{\rA}\qw & \qw & \qw & }
	\end{aligned} \; .
\end{equation*}
For a pair of systems $\rA$ and $\rB$, $\rA \otimes \rB$ denotes the composite system consisting of subsystems $\rA$ and $\rB$. The composition is associative, namely $\rA \otimes (\rB \otimes \rC) = (\rA \otimes \rB) \otimes \rC$ for every triple of systems $\rA,\rB,\rC$. In the following we will often  omit the symbol of tensor and use the shorthand notation  $\rA\rB: =  \rA\otimes \rB$.  Every OPT includes a special system $\rI$, called the trivial system and satisfying the relation $\rA \otimes \rI = \rI \otimes \rA = \rA$ for every system $\rA$.  The trivial system is typically omitted in the diagrams.

\subsection{Transformations}

For systems $\rA$ and $\rA'$, we denote by $\Transf(\rA\to\rA')$ the set of transformations with input system $\rA$ and output system $\rA'$.   Graphically, a transformation $\tT \in \Transf(\rA \to \rA')$ is represented by the following diagram
\begin{equation*}
	\begin{aligned}
		\Qcircuit @C=1.2em @R=.8em @!R { & \poloFantasmaCn{\rA}\qw & \gate{\tT} & \poloFantasmaCn{\rA'}\qw & \qw & }
	\end{aligned} \, .
\end{equation*}
For general systems $\rA,\rA',\rB,\rB'$,  two transformations $\map T  \in  \Transf (\rA \to  \rA')$ and  $\map S  \in  \Transf (\rB \to  \rB')$ can be performed in parallel, giving rise to the transformation $\map T\otimes \map S  \in \Transf (\rA  \otimes \rB \to  \rA'\otimes \rB')$, graphically represented as
\begin{equation*}
	\begin{aligned}
		\Qcircuit @C=1.2em @R=.8em @!R {
			& \poloFantasmaCn{\rA}\qw & \gate{\tT} & \poloFantasmaCn{\rA'}\qw & \qw & \\
			& \poloFantasmaCn{\rB}\qw & \gate{\tS} & \poloFantasmaCn{\rB'}\qw & \qw & }
	\end{aligned} \, .
\end{equation*}
If $\rA' = \rB$, the transformations $\map T$ and $\map S$ can also be performed in a sequence,  giving rise to the transformation $\map S  \circ \map T  \in  \Transf (\rA  \to  \rB')$, graphically represented as
\begin{equation*}
	\begin{aligned}
		\Qcircuit @C=1.2em @R=.8em @!R {
			& \poloFantasmaCn{\rA}\qw & \gate{\tT} & \poloFantasmaCn{\rA'}\qw & \gate{\tS} & \poloFantasmaCn{\rB'}\qw & \qw & }
	\end{aligned} \, .
\end{equation*}
By definition, the parallel and sequential composition are taken to be associative.	

For every system $\rA$,  there exists an identity transformation, denoted by   $\map I_{\rA} \in \Transf (\rA\to \rA)$, which leaves the system unchanged. The identity transformation satisfies the property  $  \map T  \circ \map I_{\rA}   =   \map I_{\rB} \circ \map T  =  \map T$, for every transformation $\map T \in  \Transf (\rA\to \rB)$, and for every pair of  systems  $\rA$ and $\rB$.

Mathematically, the above definitions amount of the fact that  systems are objects in a symmetric monoidal category, and physical transformations are morphisms in that  category.
\subsection{Tests}

A \textit{test} represents an experiment transforming an input system $\rA$ into an output system $\rB$.  Here, we will focus on experiments with a finite number of outcomes, denoted by $N$. Mathematically, a test is a collection of transformations  $(\map T_i)_{i=1}^N$, where the transformation $\map T_i  \in \Transf (\rA \to \rB)$ represents the physical process taking place when  the $i$-th outcome of the experiment occurs. The set of all tests from system $\rA$ to system $\rB$, denoted by $\Tests(\rA \to \rB)$, is part of the specification of the OPT.

Tests can be composed in parallel: the parallel composition of a test $(\map T_i)_{i \in I} \in \Tests(\rA \to \rA')$ with a test $(\map S_j)_{j \in J} \in \Tests(\rB \to \rB')$ yields a test in $\Tests(\rA\otimes \rB \to \rA'\otimes \rB')$ defined by $(\map T_i)_{i \in I} \otimes (\map S_j)_{j \in J} = (\map T_i \otimes \map S_j)_{(i,j) \in I\times J} $, where $I\times J$ denotes the Cartesian product of the sets $I$ and $J$. 

Similarly, tests can be composed sequentially:  the sequential composition of a test $(\map T_i)_{i \in I} \in \Tests(\rA \to \rB)$ with a test $(\map S_j)_{j \in J} \in \Tests(\rB \to \rC)$ is the test in $\Tests(\rA \to \rC)$ defined by $(\map T_i)_{i \in I} \circ (\map S_j)_{j \in J} = (\map T_i \circ \map S_j)_{(i,j) \in I\times J} $.    		

Experiments with a single outcome ($N=1$) are associated to a single transformation. A transformation occurring in a single-outcome test is  called {\em deterministic}.  The set of deterministic transformations with input $\rA$ and output $\rB$ is denoted by $\Transf_1  (\rA \to \rB)$. The identity transformation $\map I_\rA$ is assumed to be deterministic for every system $\rA$ (in fact, this assumption can be omitted in the definition of the OPT framework,  because it follows automatically once the probabilistic structure is introduced.)

The tests in $ \Tests(\rA \to \rI)$ are usually called {\em measurements}, and represent experiments  in which the system is discarded after the measurement. 	

\subsection{States, effects, and scalars}

The transformations from the trivial system to system $\rA$ are called {\em states} of system $\rA$, and represent different ways in which the system can be initialized.  We will  use the notation  $\St  (\rA) :  =  \Transf  (\rI \to \rA)$ for the set of states of system $\rA$, and  $\St_1  (\rA) :  =  \Transf_1  (\rI \to \rA)$ for the subset of deterministic states. Graphically, a state $\rho \in \St(\rA)$ is represented by a  diagram of the form  
\begin{equation*}
	\begin{aligned}
		\Qcircuit @C=1.2em @R=.8em @!R {
			& \prepareC{\rho} & \poloFantasmaCn{\rA}\qw & \qw & }
	\end{aligned} \, .
\end{equation*}

Transformations from system $\rA$ to the trivial system are called {\em effects} and form the set   $\Eff  (\rA):  =  \Transf (\rA  \to \rI)$. Analogously to the case of states, we will  use the notation $\Eff_1  (\rA) :  =  \Transf_1  (\rA \to \rI)$ for the subset of deterministic effect of system $\rA$. Graphically, an effect $e \in \Eff(\rA)$ is represented by the following diagram
\begin{equation*}
	\begin{aligned}
		\Qcircuit @C=1.2em @R=.8em @!R { 
			& \qw & \poloFantasmaCn{\rA}\qw & \measureD{e} & }
	\end{aligned} \, .
\end{equation*}
A common assumption is the {\em Causality Principle} \cite{chiribella2010probabilistic}, which is equivalent to stating that for every system $\rA$ the set $\Eff_1  (\rA)$ contains one and only one effect $u_\rA$.  Later in this paper we will make this assumption, but it is important to stress that Causality is not included in the definition of the OPT framework,  and that some of our results even hold when Causality is not assumed.


Transformations from the trivial system $\rI$ to itself are called {\em scalars}. 
For scalars,  the sequential and parallel composition commute and coincide:  given a scalar $s\in\Transf (\rI\to \rI)$ and an arbitrary transformation $\tT \in \Transf (\rA \to \rB)$, with arbitrary input system $\rA$ and arbitrary output system $\rB$, one has  the equalities
\begin{equation}\label{eq:commuting-scalar}
	s  \otimes  \tT   =  \tT\otimes s  =  (s\otimes \tI_{\rB})    \circ (\tI_{\rI}\otimes \tT )=  (\tT  \otimes \tI_{\rI}) \circ (\tI_{\rA}\otimes s)   \, .
\end{equation}

\subsection{Probabilistic structure}

The probabilistic structure of an OPT associates scalars  to probabilities, by specifying  a function $\mathsf{Prob}:\Transf(\rI \to \rI) \to \left[0,1\right] \subset \R$ satisfying the conditions
\begin{equation}\label{eq:normalized}
	\sum_{x\in X} \mathsf{Prob} (s_x) = 1  \qquad \forall (s_x)_{x \in X} \in \Tests(\rI \to \rI) \, ,
\end{equation} 
and  
\begin{align}\label{eq:prodprob} 
	\mathsf{Prob}(s \otimes t) = \mathsf{Prob}(s) \cdot \mathsf{Prob}(t) \qquad \forall s ,  t  \in \Transf (\rI, \rI) \,.
\end{align}
Eqs.~\eqref{eq:normalized} and~\eqref{eq:prodprob}  are sometimes referred to as \textit{consistency} and  \textit{independence}, respectively \cite{hardy2011reformulating,hardy2013formalism,chiribella2014dilation}.   

Note that, since the identity transformation  on system $\rI$ is assumed to be deterministic, Eq. (\ref{eq:normalized}) implies $\mathsf{Prob} (\map I_{\rI}) = 1$.   

Transformations that give rise to the same probabilities in all possible circuits are indistinguishable in all possible experiments allowed by the OPT.  As such, they are usually identified. The identification is achieved as follows: two transformations  $\tT$ and $\tT'$ in $\Transf(\rA\to\rB)$  are called {\em probabilistically equivalent} if
\begin{equation*}
	\mathsf{Prob}\left(
	\begin{aligned}
		\Qcircuit @C=1.2em @R=.8em @!R {       
			&
			\multiprepareC{1}{\rho}&
			\poloFantasmaCn{\rA}\qw&	
			\gate{\tT}&
			\poloFantasmaCn{\rB}\qw&
			\multimeasureD{1}{e}& \\
			&
			\pureghost{\rho}&
			\qw&
			\poloFantasmaCn{\rE}\qw&	
			\qw&
			\ghost{e}&
			\\
		}
	\end{aligned} \right) = \mathsf{Prob} \left(
	\begin{aligned}
		\Qcircuit @C=1.2em @R=.8em @!R {       
			&
			\multiprepareC{1}{\rho}&
			\poloFantasmaCn{\rA}\qw&	
			\gate{\tT'}&
			\poloFantasmaCn{\rB}\qw&
			\multimeasureD{1}{e}& \\
			&
			\pureghost{\rho}&
			\qw&
			\poloFantasmaCn{\rE}\qw&	
			\qw&
			\ghost{e}&
			\\
		}
	\end{aligned} \right)
\end{equation*}
for every system $\rE$, every state  $\rho\in\St(\rA\rE)$, and every effect $e \in \Eff(\rB\rE)$.  By taking the quotient with respect to this equivalence relation, one obtains  a new OPT called the {\em quotient OPT} \cite{chiribella2014dilation}.  

In the following, we will always consider quotient OPTs, which can equivalently characterized as the OPTs in which two transformations  $\tT$ and $\tT'$ in $\Transf(\rA\to\rB)$ are distinct ($\tT \not =  \tT'$) if and only if they give rise to different probabilities in at least one experiment, that is, if and only if 
\begin{equation}
	\label{eq:equivalence}
	\mathsf{Prob}\left(
	\begin{aligned}
		\Qcircuit @C=1.2em @R=.8em @!R {       
			&
			\multiprepareC{1}{\rho}&
			\poloFantasmaCn{\rA}\qw&	
			\gate{\tT}&
			\poloFantasmaCn{\rB}\qw&
			\multimeasureD{1}{e}& \\
			&
			\pureghost{\rho}&
			\qw&
			\poloFantasmaCn{\rE}\qw&	
			\qw&
			\ghost{e}&
			\\
		}
	\end{aligned} \right) \not = \mathsf{Prob} \left(
	\begin{aligned}
		\Qcircuit @C=1.2em @R=.8em @!R {       
			&
			\multiprepareC{1}{\rho}&
			\poloFantasmaCn{\rA}\qw&	
			\gate{\tT'}&
			\poloFantasmaCn{\rB}\qw&
			\multimeasureD{1}{e}& \\
			&
			\pureghost{\rho}&
			\qw&
			\poloFantasmaCn{\rE}\qw&	
			\qw&
			\ghost{e}&
			\\
		}
	\end{aligned} \right)
\end{equation}
for some system $\rE$, some state  $\rho\in\St(\rA\rE)$, and some effect $e \in \Eff(\rB\rE)$.

In the particular case of states, the additional system $\rE$ is not needed:
\begin{proposition}
	In a quotient OPT, two states $\rho$ and $\rho'$ in  $ \St(\rA)$ are distinct ($\rho \ne \rho'$) if and only if it exist an effect $e \in \Eff(\rA)$ such that
	\begin{equation}\label{bambino}
		\mathsf{Prob}\left(
		\begin{aligned}
			\Qcircuit @C=1.2em @R=.8em @!R {       
				&
				\prepareC{\rho}&
				\poloFantasmaCn{\rA}\qw&	
				\measureD{e}& \\
			}
		\end{aligned} \right) \ne \mathsf{Prob} \left(
		\begin{aligned}
			\Qcircuit @C=1.2em @R=.8em @!R {       
				&
				\prepareC{\rho'}&
				\poloFantasmaCn{\rA}\qw&	
				\measureD{e}& \\
			}
		\end{aligned} \right) \, .
	\end{equation}
\end{proposition}
\Proof   In the case of states,  Eq.~\eqref{eq:equivalence} gives
\begin{equation}\label{mamma}
	\mathsf{Prob}\left(
	\begin{aligned}
		\Qcircuit @C=1.2em @R=.8em @!R {       
			&
			\prepareC{\rho}&
			\poloFantasmaCn{\rA}\qw&
			\multimeasureD{1}{f}& \\
			&
			\prepareC{\eta}&
			\poloFantasmaCn{\rE}\qw&	
			\ghost{f}&
			\\
		}
	\end{aligned} \right) \ne \mathsf{Prob} \left(
	\begin{aligned}
		\Qcircuit @C=1.2em @R=.8em @!R {       
			&
			\prepareC{\rho'}&
			\poloFantasmaCn{\rA}\qw&
			\multimeasureD{1}{f}& \\
			&
			\prepareC{\eta}&
			\poloFantasmaCn{\rE}\qw&
			\ghost{f}&
			\\
		}
	\end{aligned} \right)
\end{equation}
and we can define
\begin{equation*}
	\begin{aligned}
		\Qcircuit @C=1.2em @R=.8em @!R {       
			&
			\poloFantasmaCn{\rA}\qw&
			\measureD{e}&
			\\
		}
	\end{aligned} :=
	\begin{aligned}
		\Qcircuit @C=1.2em @R=.8em @!R {       
			&
			&
			\poloFantasmaCn{\rA}\qw&
			\multimeasureD{1}{f}& \\
			&
			\prepareC{\eta}&
			\poloFantasmaCn{\rE}\qw&
			\ghost{f}&
			\\
		}
	\end{aligned} \, .
\end{equation*}
With this definition, Eq.~\eqref{mamma} becomes Eq.~\eqref{bambino}. \qed

\medskip  
A similar result holds for effects.  Furthermore, scalars are identified with probabilities:  
\begin{proposition}
	In a quotient OPT, two scalars scalars $s$ and $s'$ are equal if and only if the corresponding probabilities are equal $\mathsf{Prob}  (s)  = \mathsf{Prob}  (s')$.
\end{proposition} 	

\Proof   Eq.~\eqref{bambino}, used for $\rA  =  \rI$,  implies that the scalars $s$ and $s'$ are distinct if and only if there exists another scalar $t$ such that $\mathsf{Prob}  (t\circ s)  \not= \mathsf{Prob}  (t \circ s')$.    On the other hand, Eqs.~\eqref{eq:commuting-scalar} and~\eqref{eq:prodprob} imply   $\mathsf{Prob}  (t\circ s)    = \mathsf{Prob}  (t  \otimes s) = \mathsf{Prob}  (t) \,  \mathsf{Prob}  (s) $ and $\mathsf{Prob}  (t\circ s')    = \mathsf{Prob}  (t  \otimes s') = \mathsf{Prob}  (t) \,  \mathsf{Prob}  (s')$.   Hence,   $s$ and $s'$ are distinct if and only if  there exists a scalar $t$ such that  $\mathsf{Prob}  (t) \,  \mathsf{Prob}  (s')  \not  =   \mathsf{Prob}  (t) \,  \mathsf{Prob}  (s')$, that is, if and only if  $\mathsf{Prob}  (s)  \not  =   \mathsf{Prob}  (s')$.  (Note that, if $\mathsf{Prob}  (s)  \not  =   \mathsf{Prob}  (s')$ one automatically has  $\mathsf{Prob}  (t) \,  \mathsf{Prob}  (s')  \not  =   \mathsf{Prob}  (t) \,  \mathsf{Prob}  (s')$ for $t=  \map I_{\rI}$.)  \qed

\medskip  

Based on  the above Proposition, from now on we will identify scalars with probabilities and we will omit the function $\sf Prob$.

\subsection{Embedding into real vector spaces}

Since scalars are probabilities, states can be regarded as functions on the set of effects, and {\em vice-versa}. As such, the set of states and the set of effects  of a given system $\rA$ can be embedded into vector spaces over the real field, hereafter denoted by $\St_\R  (\rA) $ and $\Eff_\R  (\rA)$, respectively.    These vector spaces are generated by the convex  cones  $\St_+(\rA):  =  \{   \xi~|~   \xi  =  \sum_i   \,  a_i \,  \rho_i ,  \,  \forall i: \,    a_i\ge   0  \, ,    \rho_i\in \St (\rA) \}$       and $\Eff_+(\rA) :  = \{  x~|~    \sum_i  a_i  \,  e_i  ,  \,\forall i:   \,   a_i  \ge 0  \, , e_i \in \Eff (\rA)  \}$, respectively.

For a pair of systems  $\rA$ and $\rB$,  the set of product states $\{\alpha \otimes \beta ~|~  \alpha \in \St (\rA),   \,  \beta \in \St (B)\}$  spans the tensor product vector space $\St_\R(\rA)  \otimes \St_\R  (\rB)$. Since the product states are valid states of the composite system $\rA\otimes \rB$, one has the inclusion
\begin{align}
	\St_\R(\rA)  \otimes \St_\R  (\rB)     \subseteq \St_\R  (  \rA \otimes \rB)
\end{align}
associated to the composite system $\rA \otimes \rB$. If this inclusion is an equality, then the composite system $\rA \otimes \rB$ is said to satisfy Local Tomography \cite{dariano2006how, dariano2007operational, araki1980characterization, wootters1990local, hardy2001quantum, barrett2007information, barnum2007generalized, chiribella2010probabilistic, hardy2011reformulating, chiribella2012quantum, dariano2017quantum, chiribella2021process}.  In turn, an OPT is said to satisfy Local Tomography if, for every pair of systems $\rA$ and $\rB$,  the composite system $\rA\otimes \rB$  satisfies Local Tomography.  

In this work, we will not assume Local Tomography. Hence, we will allow the vector space of the composite system $\rA \otimes \rB$ to be  generally larger than the tensor product of the vector spaces of the component systems $\rA$ and $\rB$.  In this case, some care is needed when dealing with physical transformations, as discussed in the remainder of this subsection.  

For every pair of systems $\rA$ and $\rB$, every transformation $ \map T  \in   \Transf (\rA \to \rB)$ induces a linear map $\widehat{\map T }  :  \St_\R  (\rA)  \to  \St_\R  (\rB)$ between the vector space spanned by the states of system $\rA$ and the vector space spanned by the states of system $\rB$.    When  Local Tomography holds, the linear map $\widehat{\map T}$ completely specifies the transformation $\map T$, and therefore it is not necessary to distinguish between $\map T$ and $\widehat{\map T}$.   In this case, the set of physical transformations $\Transf (\rA\to \rB)$ is identified with a subset of linear maps in $L  \Big(  \St_\R  (\rA)  ,   \St_\R  (\rB)   \Big)$, the vector space of all linear maps from $\St_\R  (\rA)$ to $ \St_\R  (\rB)$.   

In general, however, the identification between physical transformations and  linear maps is less straightforward: when Local Tomography fails to hold,  specifying  the action of a transformation on the vector space $\St_\R (\rA)$ is not sufficient to specify the local action of the transformation on the vector spaces $\St_\R (\rA\otimes \rE)$ associated to composite systems of $\rA$ and $\rE$, for some generic system  $\rE$  \cite{chiribella2010probabilistic}. In this case, specifying the transformation $\map T$ may generally require specifying all the linear maps $\widehat{\map T \otimes \map I_{\rE}}$ associated to all possible  systems $\rE$.    Note that, nevertheless, transformations can always be regarded as elements of a (possibly very large) vector space over the real field:  the  transformation $\map T$ can be identified with the function $f_{\map T}: \rE \mapsto f_{\map T}   (\rE)   : =  \widehat {  \map T\otimes \map I_{\rE}}$ from the set of all possible systems $\rE$ to the set of linear maps from $ \St_\R  (\rA \otimes \rE) $ to $\St_\R (\rB \otimes \rE)$. Equivalently, the transformation $\map T$ can be identified with the direct sum of linear maps \cite{chiribella2014dilation}.
\begin{align}\label{directsum} 
	\map L_{\map T}   :  = \bigoplus_{\rE}   \,  \widehat {  \map T\otimes \map I_{\rE}} \, .
\end{align} 
Either way, transformations can be  viewed as elements of a vector space, hereafter denoted by $\Transf_{\R}  (\rA \to \rB)$. Using the representation of Eq. (\ref{directsum}), we have the identification
\begin{align}\label{directsumspaces}
	\Transf_\R  (\rA \to \rB)   \simeq   \bigoplus_{\rE}     \,  L  \Big(  \St_\R  (\rA \otimes \rE)  ,   \St_\R  (\rB\otimes \rE)   \Big) \, . 
\end{align}

\subsection{Coarse-graining and pure transformations}  
The vector space structure of the set of transformations can be used to define an operation of coarse-graining of tests:  
\begin{definition}
	Let  $(\map T_i)_{i \in I} $   and $(\map S_j)_{j \in J}$ be two tests with input $\rA$ and output $\rB$.  We say that  $(\map T_i)_{i \in I} $ is a {\em coarse-graining}	of $(\map S_j)_{j \in J}$ if there is a partition of $J$ into	disjoint subsets $(J_i)_{i\in I}$ such that 
	\begin{align}\label{coarsegraining}
		\map T_i  =  \sum_{j  \in    J_i}   \,  \map S_j   \qquad\forall i\in  I \, .
	\end{align} 
\end{definition}  
Intuitively, we can regard coarse-graining as the result of an experimenter considering different outcomes together.  It is generally assumed that the set of tests is closed under coarse-graining: for every test $(\map S_j)_{j \in J}  \in  \Tests (\rA \to \rB)$ and every partition  $(J_i)_{i\in  I}$ of $J$ into disjoint subsets, the transformations $(  \map T_i)_{i\in  I}$ defined in Eq. (\ref{coarsegraining}) represent a valid test. 

A transformation that cannot be viewed as a coarse-graining is called pure:
\begin{definition}[Pure transformation]
	A transformation $\tT \in \Transf(\rA \to \rB)$ is \textit{pure} if for every test  $\{\tS_j\}_{j \in J} \in \Tests(\rA \to \rB)$ and every subset $J_*  \subseteq  J$,  one has the implication
	\begin{equation*}
		\tT = \sum_{j \in J_*} \tS_j \quad \Longrightarrow \quad \tS_j = p_j \tT \text{ for every } j \in J_* \, ,
	\end{equation*}
	for some probability distribution $(p_j)_{j \in J_*}$.
\end{definition}
The set of pure transformation of type $\rA\to\rB$ is denoted by $\Pur\Transf(\rA\to\rB)$. We use the notation $\Pur\St(\rA):= \Pur\Transf(\rI \to \rA)$ and $\Pur\Eff(\rA) := \Pur \Transf(\rA \to \rI)$ to denote the set of pure states and pure effects of system $\rA$, respectively.

\subsection{The operational norm}
The vector space spanned by the states of a given system can be equipped with a norm, called the {\em operational norm}~\cite{chiribella2010probabilistic}, which quantifies the distinguishability of states in the same way as the trace norm quantifies the distinguishability of states in quantum theory.  The definition of the operational norm is as follows:
\begin{definition}[Operational norm]
	Let $\xi \in \St_\R(\rA)$ be an arbitrary element of the vector space $\St_\R(\rA)$.   Then, the operational norm of $\rho$ is
	\begin{equation}
		\| \rho \| := \sup_{a \in \Eff (A)} a(\rho) - \inf_{b \in \Eff (A)} b(\rho) \, .
	\end{equation}
\end{definition}
In turn,  the operational norm for in the states' vector space induces an operational norm in the transformations' vector space,  defined as~\cite{chiribella2010probabilistic}:

\begin{definition}[Operational norm for transformations]
	Let $\tT \in \Transf_\R(\rA\to\rB)$, the operational norm of a transformation $\tT$ is defined as
	\begin{equation}\label{opnorm}
		\| \tT \|_{\rm op} := \sup_{\rE} \sup_{\rho \in \St(\rA\rE), \, \| \rho\|  \not =  0} \frac{\| (\tT\otimes \tI_\rE) \rho \|}{\| \rho \|} \, ,
	\end{equation}
	where $\tI_\rE$ is the identity transformation for system $\rE$ and  $\| \cdot \|$ is the operational norm for states.
\end{definition}

\subsection{Basic assumptions}

The OPT framework covers a broad variety of theories. Usually, however, it is convenient to restrict the attention to OPTs that satisfy a few basic assumptions, such as finite-dimensionality, convexity, and closure  of the sets of transformations.   In this work, we will make these assumptions: 

\begin{assumption}[finite-dimensionality, convexity, and closure]
	\label{as:closure}
	For every pair of systems $\rA$ and $\rB$, we assume that 
	\begin{enumerate}
		\item the vector space $\Transf_\R(\rA \to \rB)$ is finite-dimensional,
		\item  the set $\Transf(\rA \to \rB)$ is convex, and
		\item  the set $\Transf(\rA \to \rB)$ is  closed with respect to the operational norm.
	\end{enumerate}
\end{assumption}

Under these assumptions, the set of all transformations   $\Transf(\rA\to\rB)$ is a compact convex set for every pair of systems $\rA$ and $\rB$.


\section{The group  of reversible transformations}

A transformation $\map T  \in  \Transf (\rA\to \rB)$ is called {\em reversible} if there exists another  transformation $\map S\in \Transf (\rB\to \rA)$ such that $ \map S  \circ \map T  =  \map I_{\rA}$ and $\map T\circ \map S  = \map I_{\rB}$.   The reversible transformations from a system $\rA$ to itself  form a group, hereafter denoted by $\grp G_{\rA}$.     

In this section we show that, under 	Assumption~\ref{as:closure}  (finite-dimensionality, convexity, and closure), the group of reversible transformations $\grp G_\rA$   is a compact Lie group for every system $\rA$.  In turn, this will imply that every one-parameter subgroup  $\grp G_\rA$ is a one-parameter Lie group, and  can be expressed as the exponential of a suitable  generator.

The proof proceeds in three steps: we will show that, for every system $\rA$,
\begin{enumerate}
	\item the group of reversible transformations $\grp G_\rA$ is compact  with respect to the operational norm,
	\item $\grp G_\rA$ has a faithful representation $\pi_\rA$ on a finite-dimensional vector space, and
	\item   the image of $\grp G_\rA$ under the representation $\pi_\rA$ is a compact group.
\end{enumerate}

\subsection{Compactness of $\grp G_\rA$}

Let us start from the first step.  
\begin{lemma}[Lemmas 30 and 31 in Ref.~\cite{chiribella2010probabilistic}]
	\label{lemma:G_compact}
	In an OPT satisfying Assumption \ref{as:closure},  the group  $\grp G_\rA$ is closed and compact with respect to the operational norm for every system $\rA$.
\end{lemma} 
\Proof The group $\grp G_\rA$ is a subset of the vector space  $ \Transf_\R(\rA\to\rA)$, which is finite-dimensional  by Assumption \ref{as:closure}.     We now show that  $\grp G_\rA$ is closed with respect to the operational norm.   Let $\{\tU_i\}_{i\in \N}$ be a sequence of reversible transformations converging to some transformation $\tT\in \Transf(\rA\to\rA)$.   Now, consider the sequence $\{\tU^{-1}_j\}_{j \in \N}$.   By the compactness of $\Transf(\rA \to \rA)$, there is a converging subsequence $\{\tU^{-1}_{j_k}\}_{k  \in \N}$ that converges to some transformation $\tS$. Hence, we have
\begin{equation*}
	\tT \circ \tS= \lim_{i \to \infty}   \lim_{k\to \infty}   \tU_{i} \circ \tU_{j_k}^{-1} = \lim_{k\to\infty}\tU_{j_k} \circ \tU_{j_k}^{-1} = \tI_\rA
\end{equation*}
and
\begin{equation*}
	\tS \circ \tT=  \lim_{i \to \infty}   \lim_{k\to \infty}   \tU_{j_k}^{-1} \circ \tU_{i} = \lim_{k\to\infty} \tU_{j_k}^{-1} \circ \tU_{j_k} = \tI_\rA \, .
\end{equation*}
Hence $\tT$ is a reversible transformation and $\tS$ is its inverse.  In conclusion,  $\grp G_\rA$ is closed with respect to the operational norm.  Since  $\grp G_\rA$ is closed and finite-dimensional, it is compact.	\qed

\subsection{Faithful representation of $\grp G_\rA$ on a finite-dimensional vector space}	
We are now ready to move to the second step:  showing that $\grp G_\rA$ has a faithful  representation on a finite-dimensional vector space. For OPTs satisfying Local Tomography, this step is immediate from Assumption~\ref{as:closure}, which guarantees that the state spaces of all systems are finite-dimensional.  Here, however, we do not assume Local Tomography. Nevertheless, we show that Assumption \ref{as:closure} still guarantees the existence of a faithful representation on a finite-dimensional vector space.   Even more strongly, here we show that the set of all (not necessarily reversible) transformations $\Transf(\rA \to \rB)$ admits a faithful representation on a finite-dimensional vector space for every pair of systems $\rA$ and $\rB$.

In general,  we have seen that a transformation $\map T\in \Transf (\rA\to \rB )$ can be faithfully represented by a direct sum of linear maps,  as in Eq. (\ref{directsum}).  In this representation,   the number of terms in the direct sum is infinite.  We now show that the finite-dimensionality condition in Assumption \ref{as:closure} allows us to restrict the direct sum to a finite number of terms. 
The core of the argument is to show that the transformations in  $ \Transf (\rA\to \rB )$ are  uniquely determined by their action on a   finite set of states. 

\begin{definition}[Tomographically faithful sets of states \cite{chiribella2021process}]
	For a positive  integer $K$, let $(\rE_1,\dots,  \rE_K)$ be a $K$-tuple of (not necessarily distinct) systems, and let $(\rho_1 ,    \cdots,  \rho_K)$ be a  $K$-tuple of states, with $\rho_i  \in  \St  (  \rA  \otimes E_i)$  for every $i\in  \{1,\dots, K\}$.    We say that $(\rho_1 ,    \cdots,  \rho_K)$  is {\em tomographically faithful}  for  $\Transf(\rA \to \rB)$ if, for every pair of transformations $\map T , \map T'  \in \Transf (\rA \to \rB)$, the condition  
	\begin{align}
		(\map T'\otimes \map I_{\rE_i})  (  \rho_i)  =    (\map T\otimes \map I_{\rE_i})  (  \rho_i)   \qquad \forall i\in  \{1,\dots, K\} 
	\end{align}  
	implies $\map T' =  \map T$.
\end{definition}
The existence of a finite, tomographically faithful set of states expresses the possibility of doing process tomography with a finite number of experimental settings. This property was identified in Ref. \cite{chiribella2021process} as a requirement for general probabilistic theories, and was shown to follow from several combinations of axioms. In the following, we show that the finite-dimensionality assumption (Assumption \ref{as:closure}) guarantees that this requirement is satisfied.  

Let us start by providing an equivalent condition to tomographic faithfulness:   a $K$-tuple of states $ \boldsymbol \rho = (\rho_1 ,    \cdots,  \rho_K)$ is tomographically faithful if and only if the linear map  $\map M_{\boldsymbol \rho}  :  \Transf_\R  (\rA \to \rB) \to \bigoplus_{i=1}^K \, \St_\R (\rB \otimes \rE_i) $ defined by 
\begin{align}\label{Mrho}
	\map M_{\boldsymbol \rho}  (\map T) :=  \bigoplus_{i=1}^K \, (\map T\otimes \map I_{\rE_i}) ( \rho_i)    \qquad \forall  \map T  \in   \Transf  (\rA\to \rB)
\end{align}
is injective. In turn, injectivity of the map $\map M$ is equivalent to the condition that the domain and the range of $\map M$ have the same dimension, that is, 
\begin{align}
	\dim  {\sf Range}  (\map M_{\boldsymbol  \rho} )  =  \dim   \Transf_{\R}  (\rA \to \rB)  \, .
\end{align}
We will now show that this condition can be satisfied by choosing a sufficiently large  set of states, whenever the transformations' vector space  $\Transf_{\R}  (\rA \to \rB)$ is finite-dimensional.  
The proof uses the following lemma:
\begin{lemma}\label{lem:largerdimension}
	Let  $ \boldsymbol \rho = (\rho_1 ,    \cdots,  \rho_K)$ be a $K$-tuple of states, with $\rho_i  \in \St (\rA\otimes \rE_i)$ for every $i\in  \{1,\dots,K\}$, and let $\map M_{\boldsymbol{\rho}}$ be the associated map defined in Eq. (\ref{Mrho}).   If $\dim  {\sf Range}  (\map M_{\boldsymbol  \rho} )  <  \dim   \Transf_{\R}  (\rA \to \rB)$, then there exists an auxiliary system $E_{K+1}$ and a state $\rho_{K+1}  \in  \St (  \rA  \otimes \rE_{K+1})$ such that the $(K+1)$-tuple $\boldsymbol  {  \rho}' = ( \rho_1,\dots, \rho_K, \rho_{K+1}) $ satisfies the condition $\dim  {\sf Range} (\map M_{\boldsymbol  \rho'} )   \ge \dim {\sf Range} (\map M_{\boldsymbol \rho} )+1$.  
\end{lemma}
\Proof Let $r:  =   \dim  {\sf Range}  (\map M_{\boldsymbol  \rho} )$,  and  let $  (\map T_i)_{i=1}^{r}$ be $r$ elements of $\Transf (\rA  \to \rB)$ such that $      \big(\map M_{\boldsymbol  \rho}    (  \map T_i)  \big)_{i=1}^r$  are linearly independent.       If $\dim  {\sf Range}  (\map M_{\boldsymbol  \rho} )  <  r$, then there must exist a transformation  $\map T_{r+1} \in \Transf (\rA  \to \rB) $ that is    linearly independent from  $    (  \map T_i)_{i=1}^r$.    Since  $      \big(\map M_{\boldsymbol  \rho}    (  \map T_i)  \big)_{i=1}^r$  are $r$  linearly independent vectors in an $r$-dimensional vector space,   $\map M_{\boldsymbol \rho}   (\map T_{r+1})  $ is necessarily linearly dependent on  $\big(\map M_{\boldsymbol  \rho}    (  \map T_i)  \big)_{i=1}^r$, namely
\begin{align}\label{aaaa}
	\map M_{\boldsymbol \rho}    (\map T_{r+1})   =  \sum_{i=1}^r    \,  x_i\,     \map M_{\boldsymbol \rho}    (\map C_{i}) \, ,   
\end{align}  
for suitable coefficients $\{x_i \}_{i=1}^r \subset  \R$.
On the other hand, since  $\map T_{r+1} \in \Transf (\rA \to \rB) $ is linearly independent from  $ ( \map T_i)_{i=1}^r$, one has 
\begin{align}\label{dddd}
	\map T_{r+1}  \not  =  \sum_{i=1}^r    \,  x_i\,     \map T_{i} \, ,    
\end{align}  
for any possible choice of coefficients $\{ x_i\}_{i=1}^r \subset \R$. This condition implies that there exists an auxiliary system $\rE_{K+1}$ and a state $\rho_{K+1}  \in  \St (  \rA  \otimes \rE_{K+1})$ such that
\begin{align}\label{ddddy}
	(\map T_{r+1}  \otimes \map I_{\rE_{K+1}} )(\rho_{K+1}) \not  =     \sum_{i=1}^r    \,  x_i\,     (\map T_{i} \otimes \map I_{\rE_{K+1}})  (\rho_{K+1 })\, .    
\end{align}  
Setting $\boldsymbol  \rho'  :  =  (  \rho_1, \dots,  \rho_{K},  \rho_{K+1})$, we now claim that $  \big( \map M_{\boldsymbol \rho'}   (\map T_i)\big)_{i=1}^{r+1}$ are linearly independent.  To prove this fact, suppose that $(y_i)_{i=1}^{r+1}$ are real coefficients such that 
\begin{align*}
	\sum_{i=1}^{r+1}    \,  y_i\,     \map M_{\boldsymbol \rho'}    (\map T_{i})   =  0 \, .   
\end{align*}  
We now show that $y_{r+1} = 0$. The proof is by contradiction: if $y_{r+1}$ were different from $0$, then we would have
\begin{align*}
	\map M_{\boldsymbol \rho'}    (\map T_{r+1}) =  \sum_{i=1}^r    \,  \frac{y_i}{y_{r+1}}  \,   \map M_{\boldsymbol \rho'}    (\map T_{i})   \, ,  
\end{align*}
which implies, in particular,    	
\begin{align}\label{bbbb}
	\map M_{\boldsymbol \rho}    (\map T_{r+1}) =  \sum_{i=1}^r    \,  \frac{y_i}{y_{r+1}}  \,   \map M_{\boldsymbol \rho}    (\map T_{i})   \, .  
\end{align}
and 
\begin{align}\label{cccc}
	(\map T_{r+1} \otimes  \map I_{\rE_{K+1}})  (\rho_{K+1}) =  \sum_{i=1}^r    \,  \frac{y_i}{y_{r+1}}  \, (\map T_{i} \otimes  \map I_{\rE_{K+1}})  (\rho_{K+1})  \, .  
\end{align}
Comparing  Eq. (\ref{bbbb}) with Eq. (\ref{aaaa}) and using the fact that the vectors $ \big(  \map M_{\boldsymbol \rho}    (\map T_{i})  \big)_{i=1}^r $ are linearly independent, we  would  obtain the equality  
\begin{align}
	\frac{  y_i}{y_{r+1}}  =  x_i   \qquad \forall  i\in  \{ 1,\dots,  r\} \, .   
\end{align}
Substituting it into Eq. (\ref{cccc}), we would obtain  
\begin{align*}
	(\map T_{r+1} \otimes  \map I_{\rE_{K+1}})  (\rho_{K+1}) =  \sum_{i=1}^r    \,  x_i  \, (\map T_{i} \otimes  \map I_{\rE_{K+1}})  (\rho_{K+1})  \, , 
\end{align*}
which however is in contradiction with Eq. (\ref{ddddy}).  Hence, we conclude that $y_{r+1}$ must be  $0$. 

Since $y_{r+1} = 0$, the linear independence of the vectors $ \big(\map M_{\boldsymbol \rho} (  \map T_i) \big)_{i=1}^r$ implies $y_i=0$ for every $i\in \{1,\dots, r\}$. 

In conclusion, we have shown that the vectors $ \big(\map M_{\boldsymbol \rho} (  \map T_i) \big)_{i=1}^{r+1}$ are linearly independent, and therefore $\dim {\sf Range} (\map M_{ \boldsymbol \rho}) \ge r+1$. \qed  

\medskip  

Using the above lemma,  we can prove that every finite-dimensional set of transformations admits a finite tomographically faithful set of states: 
\begin{proposition}[Existence of a finite, tomographically faithful set of states]\label{prop:tomofaith}
	If the vector space $\Transf_\R  (\rA \to \rB)$ is finite-dimensional,  then there exists an integer $N\ge 1$, an $N$-tuple of auxiliary systems $(  \rE_i)_{i=1}^N$, and a $N$-tuple of states $(\rho_i)_{i=1 }^N$, with $\rho_i \in \St  (\rA \otimes \rE_i)$ for every $i  \in \{1,\dots, N\}$, such that $( \rho_1, \dots, \rho_N)$ is tomographically faithful for $\Transf(\rA\to\rB)$. 
\end{proposition}
\Proof  For an arbitrary $K$-tuple of system $(\rE_i)_{i=1}^K$ and an arbitrary $K$-tuple of states  $\boldsymbol \rho:  = (\rho_i)_{i=1 }^K$ with $ \rho_i \in  \St  (\rA\otimes \rE_i) \, , \forall i$, let $  r = \dim {\sf Range} (\map M_{\boldsymbol \rho})$, where $\map M_{\boldsymbol \rho}$ is the linear map defined in Eq.~\ref{Mrho}). If $r <  \dim  \Transf_\R  (\rA \to \rB)$, Lemma~\ref{lem:largerdimension} guarantees that there exists a system $\rE_{K+1}$ and a state $\rho_{K+1}\in\St (\rA \otimes \rE_{K +1 })$ such that  the $(K+1)$-tuple of states  $\boldsymbol \rho': = (\rho_i)_{i=1 }^{K+1}$ satisfies the condition $\dim {\sf Range}  (\map M_{\boldsymbol \rho'})\ge r+1$.
Iterating this construction a finite number of times,  we then obtain a finite integer $N$ and an $N$-tuple of states $\boldsymbol \rho^{*}$ such that $\dim {\sf Range}  (\map M_{\boldsymbol \rho^*}) =  \dim \Transf_\R  (\rA \to \rB)$. Since the dimension of the range  of $\map M_{\boldsymbol \rho^*}$ is equal to the dimension of its domain, we conclude that $\map M_{\boldsymbol \rho^*}$ is injective. Hence, the $N$-tuple of states $\boldsymbol \rho^*$ is tomographically faithful.  \qed  

\medskip	

The above proposition implies that all sets of transformations admit a faithful representation as linear transformations on  finite-dimensional vector spaces:  

\begin{theorem}[Representation of physical transformations]\label{theo:finitedimrep}
	If the vector space $\Transf_\R  (\rA \to \rB)$ is finite-dimensional,  then the set of physical transformations  $\Transf  (\rA \to \rB)$ can be faithfully represented by linear maps between finite-dimensional vector spaces. Specifically, there exist two finite-dimensional vector spaces $V_{\rA}$ and $V_{\rB}$ and a linear map $\pi_{\rA\to \rB}  : \Transf_\R  (\rA \to \rB) \to   L(V_{\rA},  V_{\rB}) $ (where $L(V_{\rA},  V_{\rB})$ is the set of linear transformations from $V_{\rA}$ to $V_\rB$) such that $\pi_{\rA\to \rB}   (\map T')   =  \pi_{\rA\to \rB}   (\map T)$ implies $\map T'  = \map T$ for every pair of transformations $\map T, \map T' \in \Transf (\rA\to \rB)$.  	\end{theorem}	

\Proof By Proposition \ref{prop:tomofaith},   $\Transf_\R  (\rA \to \rB)$ admits a finite, tomographically faithful set of states. Let  $(\rho_i)_{i=1}^N$ be such a set, for some positive integer $N\ge 1$ and  some set of systems $(\rE_i)_{i=1}^N$ such that  $\rho_i \in\St (\rA  \otimes \rE_i)  ~\forall i\in  \{1 ,  \dots,  N\}$.  
Then, define the vector spaces 
\begin{align*}
	V_\rA  :  =  \bigoplus_{i=1}^N  \,  \St_\R  (\rA  \otimes \rE_i) \, ,
\end{align*}
and 
\begin{align*}
	V_\rB  :  =  \bigoplus_{i=1}^N  \,  \St_\R  (\rB  \otimes \rE_i) \, ,
\end{align*}
and note that they are both finite-dimensional, since they are direct sums of finite-dimensional vector spaces. 

Now, consider the set of linear transformations from $V_{\rA}$ to $V_\rB$, denoted by $L(V_{\rA}, V_{\rB})$, and define   the linear map   $\pi_{\rA\to \rB}:  \Transf_\R(\rA\to \rB)  \to  L(V_\rA, V_{\rB})$ via the relation
\begin{align} \label{eq:injective-representation}
	\pi_{\rA\to \rB}  (\map  T)    :  =     \bigoplus_{i=1}^N \,  \widehat { \map T  \otimes \map I_{\rE_i}}    \qquad \forall \map T \in  \Transf (\rA \to \rB)\, ,
\end{align}	
where, for every $i\in  \{1,\dots,  N\}$,   $\widehat { \map T  \otimes \map I_{\rE_i}}$  is the linear map describing the local action of the physical  transformation $\map T$ when applied on system $\rA\otimes \rE_i$.    

The linear map $\pi_{\rA\to \rB}$ is a representation of the set of physical transformation as linear transformations between the vector spaces $V_{\rA}$ and $V_{\rB}$.   Since the $N$-tuple $(\rho_i)_{i=1}^N$ is tomographically faithful, the representation $\pi_{\rA\to \rB}$ is faithful: indeed, for every pair of  transformations $\map T, \map T'  \in \Transf (\rA\to \rB)$, the condition $\pi_{\rA\to \rB} (\map T') =  \pi_{\rA\to \rB} (\map T)$ implies 
\begin{align}
	\pi_{\rA\to \rB} (\map T')  \left(\bigoplus_{i=1}^N \, \rho_i\right)  =  \pi_{\rA\to \rB} (\map T)   \left(\bigoplus_{i=1}^N \, \rho_i\right)  \, ,
\end{align}
or equivalently, 
\begin{align}
	\widehat{\map T' \otimes \map I_{E_i} } (\rho_i) 
	= \widehat{\map T \otimes \map I_{E_i}} (\rho_i )  \qquad \forall i\in  \{1,\dots,  N\} \, ,
\end{align}	
which is the same as 
\begin{align}
	(\map T' \otimes \map I_{E_i} ) (\rho_i) 
	= (  \map T \otimes \map I_{E_i}  ) (\rho_i )  \qquad \forall i\in  \{1,\dots,  N\} \, .
\end{align}	
Since $(\rho_i)_{i=1}^N$ is tomographically faithful, this condition implies $\map T'  =  \map T$.  Hence, the representation $\pi_{\rA\to \rB}$ is faithful.  
\qed  
\medskip		

In  the special case of transformations from a physical system to itself ($\rB  = \rA$),  the set of transformations $\Transf(\rA\to \rA)$ is a monoid:  it contains the identity transformation $\map I_\rA$ and it is closed under composition.    the above theorem yields the following:     
\begin{corollary}\label{cor:finitedimrep}
	If the vector space $\Transf_\R  (\rA \to \rA)$ is finite-dimensional,  then  the monoid $\Transf(\rA\to \rA)$  has a faithful representation  on a finite-dimensional vector space: specifically, there exists a finite-dimensional vector space $V_\rA$ and a linear map $\pi_{\rA}:  \Transf_\R  (\rA \to \rA)  \to  L(V_\rA)$  (where $L (V_\rA)$ is the set of linear transformations from $V_\rA$ to itself) such that 
	\begin{enumerate}
		\item $\pi_{\rA}$ is faithful  ($\pi_\rA (\map T')  =  \pi_\rA(\map T)$ implies $\map T'  =  \map T$ for every pair $\map T,\map T' \in \Transf (\rA\to \rA)$), 
		\item $\pi_\rA$ is a homomorphism of monoids, namely 
		\begin{enumerate}
			\item $\pi_{\rA}  (\map I_\rA)   =  \map I_{V_\rA}$, where $\map I_{V_\rA}$ is the identity map on $V_\rA$, and
			\item $\pi_\rA  (\map S\circ \map T)   = \pi_\rA  (\map S)\circ   \pi_\rA  ( \map T) $ for every pair of transformations $\map S, \map T  \in \Transf (\rA\to \rA)$.
		\end{enumerate} 
		Moreover, $\pi_\rA$ is a faithful representation of the group $\grp G_\rA$.  
	\end{enumerate}
\end{corollary}	
\Proof  By Theorem \ref{theo:finitedimrep}, all the transformations in  $\Transf (\rA \to \rA)$ can be represented as linear maps on a finite-dimensional vector space $V_\rA$ using the representation  $\pi_{\rA\to \rA}:  \Transf_\R  (\rA\to \rA)  \to  L(V_\rA,V_\rA)$. 

Let us use the shorthand notation $\pi_\rA :  = \pi_{\rA\to \rA}$.   We now show that that $\pi_{\rA}$ is a monoid homomorphism.   The condition $\pi_{\rA}  (\map I_\rA)   =  \map I_{V_\rA}$  is guaranteed by Eq. (\ref{eq:injective-representation}), which implies  
\begin{align} 
	\pi_{\rA}  (\map  I_\rA)      =     \bigoplus_{i=1}^N \,  \widehat { \map I_\rA  \otimes \map I_{\rE_i}}     =   \bigoplus_{i=1}^N \,   \map I_{\St_\R   (\rA \rE_i) }     \equiv \map I_{V_\rA}     \, .
\end{align}	  
Eq. (\ref{eq:injective-representation}) implies  	the relation 
\begin{align}
	\nonumber  \pi_{\rA}  (\map  S \circ \map T)      &=     \bigoplus_{i=1}^N \,  \myhat { (\map S \circ \map T)  \otimes \map I_{\rE_i}}  \\
	\nonumber     &=     \bigoplus_{i=1}^N \,  \myhat { (\map S   \otimes \map I_{\rE_i}) \circ  (\map T   \otimes \map I_{\rE_i})  }  \\  
	\nonumber     &=     \bigoplus_{i=1}^N \,  \widehat { \map S   \otimes \map I_{\rE_i}} \circ \widehat{ \map T   \otimes \map I_{\rE_i}}   \\    
	\nonumber     &=   \left[  \bigoplus_{i=1}^N \,  \widehat { \map S   \otimes \map I_{\rE_i}} \right]  \circ \left[  \bigoplus_{i=1}^N \,  \widehat{ \map T   \otimes \map I_{\rE_i}}  \right] \\     
	&=     \pi_{\rA}  (\map  S ) \circ   \pi_{\rA}  (\map  T) \qquad \forall \map S, \map T \in \Transf(\rA\to \rB) \, . 
\end{align}
Finally, the restriction of $\pi_\rA$ to the group of reversible transformations $\grp G_\rA$ is a faithful group representation.   This fact follows automatically from the fact that $\pi_\rA$ is a monoid homomorphism. 
\qed 

\medskip  

To conclude, recall that Assumption \ref{as:closure} guarantees that all transformations' vector spaces are finite dimensional.  Hence, we obtained the following: 
\begin{corollary}\label{cor:finitedimrep}
	In an OPT satisfying Assumption \ref{as:closure}, the following fact hold: 
	\begin{enumerate}
		\item For every pair of systems $\rA$ and $\rB$, there exists two finite-dimensional vector spaces $V_\rA$ and $V_\rB$, and an injective linear   map $\pi_{\rA\to \rB} :  \Transf_\R  (\rA\to \rB)  \to  L(V_\rA,  V_\rB)$, 
		\item For every system $\rA$, the linear map $\pi_\rA  : =  \pi_{\rA\to \rA}$ provides a faithful representation of the monoid of physical transformations $\Transf (\rA\to \rA)$, and a faithful representation of the group of reversible transformations $\grp G_\rA$.
	\end{enumerate}
\end{corollary} 

\subsection{The image of $\grp G_\rA$ under the representation $\pi_\rA$ is a compact group}

\begin{proposition}
	\label{prop:G-repr-closed}
	In an OPT satisfying Assumption \ref{as:closure},   the linear map  $\pi_{\rA} :  \Transf_\R  (\rA\to \rA)  \to  L(V_\rA)$    is an isometry and $\pi_\rA (\grp G_\rA)$ is a compact group.
\end{proposition}
\Proof  We start by constructing a norm on $L(V_\rA)$ such that $\pi_\rA$ is an isometry.  To this purpose, we proceed through the following steps: 
\begin{enumerate}	
	\item Consider  the subspace of $S\subset  L(V_\rA)$ spanned by $\pi_\rA (\grp G_\rA)$, namely  
	\begin{align}
		S:  = \Span_\R\left\{ \pi_\rA  (\grp G_\rA)\right\} \,.
	\end{align}  This subspace can be equipped with a norm  induced by the operational norm on $\grp G_\rA$ by setting 
	\begin{align}  \left\|   \sum_i   \,  x_i  \,  \pi_\rA  (  \map U_i)   \right\|_S  :  =  \left\|  \sum_i  \,x_i \,  \map U_i\right\|_{\rm op} \,,
	\end{align} 
	for every set of real coefficients $\{x_i\}$ and every set of reversible transformations $\{  U_i\}\subset \grp G_\rA$. 
	
	Note that the norm of an element $X  \in  S$ is independent on the way this element is represented as a linear combination:   if  a set of real coefficients $\{y_j\}_j$  and a set of reversible transformations $\{  V_j\}\subset \grp G_\rA$ are such that $X   =    \sum_i   \,  x_i  \,  \pi_\rA  (  \map U_i)  =  \sum_j \,  y_j  \, \pi_\rA  (\map V_j)  $, then  we have 
	\begin{align}
		\pi_\rA  \left(    \sum_i   \,  x_i  \,   \map U_i  -    \sum_j \,  y_j  \, \map V_j\right)      =  \pi_\rA  (X)  -  \pi_\rA  (X)      =  0  \, ,
	\end{align}
	which implies 
	\begin{align}
		\sum_i   \,  x_i  \,   \map U_i  =    \sum_j \,  y_j  \, \map V_j 
	\end{align}
	since $\pi_\rA$ is an injective linear map.  
	Hence, we have the equality 
	\begin{align}
		\nonumber    \|X\|_S    &=   \left\|   \sum_i   \,  x_i  \,  \pi_\rA  (  \map U_i)   \right\|_S\\
		\nonumber   &  =  \left\|  \sum_i  \,x_i \,  \map U_i\right\|_{\rm op} \\
		\nonumber & =     \left\|     \sum_j \,  y_j  \, \map V_j    \right\|_{\rm op}\\
		&  =    \left\|     \sum_j \,  y_j  \, \pi_\rA  (\map V_j )   \right\|_S \, .
	\end{align}
	In summary,   the norm $\|X\|_S$ is independent of the choice of linear combination used to represent $X$.
	
	\item   Extend the norm $\|  \cdot \|_S$ to a norm on $L(\rA)$.  To this purpose, we  decompose $L(V_\rA)$ as 
	\begin{align}
		L(V_\rA) = S\oplus S_\perp\, ,  
	\end{align}
	where 	 $S_\perp$ is the orthogonal complement of the set $S$ with respect to the Hilbert-Schmidt inner product in $L(V_\rA)$. Hence,  every element $\map X \in L(V_\rA)$ can be uniquely decomposed  as $X  =  X_\parallel  \oplus X_\perp$, with $X_\parallel \in  S$ and $X_\perp  \in S_\perp$.       Then,  we pick an arbitrary norm on $S_\perp$, denoted by $\|  \cdot \|_{S_\perp}$,  and for every element $X\in  L(V_\rA)$, we set   
	\begin{equation*}
		\|  X\|   :  =  \|  X_\parallel\|_S  +  \|  X_\perp \|_{S_\perp} \, .  
	\end{equation*} 
	This equation defines a norm on  $L(\rA)$.  
	
	\item Show that $\pi_\rA$ is an isometry. For every   element   $\map M  \in \Transf_\R  (\rA\to \rA)$, the image $\pi_\rA  (\map M)$ is in the subspace $S$.  Hence, we have 
	\begin{align}
		\|   \pi_{\rA}  (\map M)  \|   =  \|   \pi_{\rA}  (\map M)  \|_{S}   =  \|  \map M  \|_{\rm op} \qquad\forall \map M  \in    \Transf_\R  (\rA\to \rA) \,m
	\end{align} 
	meaning that $\pi_\rA$ is an isometry.
\end{enumerate}

We now show that  $\pi_\rA (\grp G_\rA)$ is compact.  By definition, $\pi_\rA (\grp G_\rA)$  is a group of linear transformations of $V_\rA$ into itself.  Since $V_\rA$ is finite-dimensional, the vector space $L(V_\rA)$ of linear transformations on $V_\rA$ is also finite-dimensional. Hence, to prove compactness of $\pi_\rA (\grp G_\rA)$ it is enough to show that $\pi_\rA (\grp G_\rA)$ is closed with respect to some norm $\| \cdot\|$ on   $L(V_\rA)$.   To this purpose, let $\{ \pi_\rA ( \map U_i)\}_{i\in \N} \subseteq \pi_\rA(\grp G_\rA)$ be a Cauchy sequence, that is, a sequence such that 
\begin{align}  
	\lim_{i\to \infty}  \lim_{j\to \infty}   \| \pi_\rA  (\map U_i)   -  \pi_\rA   (\map U_j)  \|   =   0  
\end{align}  
Then, also $\{  \map U_i\}_{i\in \N} \subset \grp G_\rA$ must be a Cauchy sequence: indeed, we have the equality
\begin{align}
	\|   \pi_\rA ( \map U_i)   -   \pi_\rA ( \map U_j)\|    =      \|   \pi_\rA ( \map U_i)   -   \pi_\rA ( \map U_j)\|_S     =  \|    \map U_i  -   \map U_j\|_{\rm op}  \qquad \forall i,j\, ,
\end{align}
which implies 
\begin{align}  
	\lim_{i\to \infty}  \lim_{j\to \infty}   \|  \map U_i   - \map U_j  \|_{\rm op}  =   \lim_{i\to \infty}  \lim_{j\to \infty}   \| \pi_\rA  (\map U_i)   -  \pi_\rA   (\map U_j)  \|   =   0 \, . 
\end{align}  
Now, since $\grp G_\rA$ is closed (Lemma \ref{lemma:G_compact}),   there exists an element $\map U\in \grp G_\rA$ such that 
\begin{align}
	\lim_{i\to \infty}  \map U_i   =  \map U  \, .
\end{align}
It is then obvious that the sequence $\{ \pi_\rA ( \map U_i)\}_{i\in \N}$ converges to $\pi_\rA  (\map U)$: indeed, one has 
\begin{align}
	\nonumber \lim_{i\to \infty}   \left\|  \pi_\rA  (\map U_i)   -   \pi_\rA  (\map U)  \right\|   &    =  \lim_{i\to \infty}   \left\|  \pi_\rA  (\map U_i     -   \map U)  \right\|\\ 
	\nonumber &     =  \lim_{i\to \infty}   \left\|  \pi_\rA  (\map U_i     -   \map U)  \right\|_S  \\
	\nonumber &     =  \lim_{i\to \infty}   \left\|  \map U_i     -   \map U  \right\|_{\rm op}  \\
	\nonumber   &  =  0 \, .
\end{align}  
In summary, every Cauchy sequence in   $\pi_\rA (\grp G_\rA)$ converges to an element of $\pi_\rA (\grp G_\rA)$, {\em i.e.} $\pi_\rA (\grp G_\rA)$ is closed, and therefore compact.  
\qed 

\medskip

\subsection{Characterization of the group of reversible transformations and of its algebra}
We now  prove compactness of the group of reversible transformations.

\begin{proposition}
	\label{prop:lie-group}
	In an OPT satisfying Assumption \ref{as:closure},  $\grp G_\rA$ is a compact Lie group for every system $\rA$.
\end{proposition}
\Proof By Corollary~\ref{cor:finitedimrep}, $\grp G_\rA$ has a faithful representation $\pi_\rA$ on a finite-dimensional vector space $V_\rA$, \textit{i.e.} $\pi_\rA (\grp G_\rA) \subset \grp{GL}(V_\rA)$ where $\grp{GL}(V_\rA)$ is the general linear group of $V_\rA$. By Proposition~\ref{prop:G-repr-closed}, the group $\pi_\rA(\grp G_\rA)$ is compact.  To conclude, we invoke a known result in representation theory  (\textit{viz.} Theorem~5.13 of Ref.~\cite{folland2016course}), stating that a group that has  a faithful, finite-dimensional, and compact representation is  a compact Lie group \qed 

\medskip

Hereafter, we denote by $\mathfrak{g}_\rA$ the Lie algebra associated to the Lie group $\grp{G}_\rA$. We conclude this section with a definition and a basic result of group theory.
\begin{definition}[One-parameter subgroup]
	Given a real finite-dimensional vector space $V$, a function $\tD : \mathbb{R} \rightarrow \grp{GL}(V)$ is called a one-parameter Lie subgroup of $\grp{GL}(V)$ if
	\begin{itemize}
		\item $\tD$ is continuous;
		\item $\tD(0)=I$;
		\item $\tD(t+s) = \tD(t) \tD(s)$ for all $t, \, s \in \mathbb{R}$.
	\end{itemize}
\end{definition}
Often, instead of the continuity of $\tD$, it is even required the differentiability of $\tD$. However, for one-parameter subgroup of matrix Lie groups, differentiability and continuity are equivalent conditions, \textit{viz.} Ref.~\cite{hall2015lie}.

Given a one-parameter Lie subgroup $\tD(t) \subseteq \grp \pi_\rA (G_\rA)$, there exists a unique \textit{generator} $G$ such that $\tD(t) = e^{G t}$ for every $t \in \R$ (see \textit{e.g.} Theorem~8 in Ref.~\cite{lawson2015matrix}). Finally, note that, since $\grp G_\rA$ is a compact Lie group represented by transformations on a real vector space, it can be equivalently represented as a subgroup of the group of orthogonal matrices.  When $\grp G_\rA$ is connected, it is a subgroup of the group of the special orthogonal group. In this case, the corresponding generators form the algebra of real skew-symmetric matrices, or a subalgebra thereof.

\section{Collision Models in General probabilistic theories}

A collision model consists of three main ingredients: the target system that undergoes the evolution, an array of independent and identically prepared systems that compose the field (and play an analogous role to the environment in open system evolution), and the reversible interaction that describes the collision between the target system and the other systems. The sequence of these pairwise collisions results in the overall dynamics generated by the collision model. We start by formalizing these ideas in operational terms.

\begin{definition}[Collision model]
	A collision model is specified by the tuple $(\rA,\rA',\tS_{t}, u)$, where
	\begin{itemize}
		\item $\rA$ and $\rA'$ are systems named the \emph{target} and the \emph{reference systems}, respectively;
		\item $(\tS_{t})_{t \in \R}$ is a one-parameter subgroup of $\grp{G}_{\rA\otimes \rA'}$;
		\item $u$ is a deterministic effect of the reference system.
	\end{itemize}
\end{definition}

Given a collision model $(\rA,\rA',\tS_{t}, u)$ and a state $\sigma \in \St_1(\rA')$, hereafter called the \textit{reference state}, we name \textit{collision} the transformation $\tC_{\tau, \sigma} \in \Transf(\rA \to \rA)$ defined as follows
\begin{equation}
	\label{eq:GPT-unit-block}
	\Qcircuit @C=1.2em @R=.8em @! R {
		&
		\qw\poloFantasmaCn{\rA}&
		\gate{\tC_{\tau, \sigma}}&
		\qw\poloFantasmaCn{\rA}& \qw }  
	\, \coloneqq \,
	\begin{aligned}
		\Qcircuit @C=1.2em @R=.8em @! R {
			&
			\qw\poloFantasmaCn{\rA}&
			\multigate{1}{\tS_{\tau}}&
			\qw\poloFantasmaCn{\rA}&	
			\qw \\
			\prepareC{\sigma}&
			\qw\poloFantasmaCn{\rA'}&
			\ghost{\tS_{\tau}}&
			\qw\poloFantasmaCn{\rA'}&
			\measureD{u}\\
		}
	\end{aligned}\, .
\end{equation}

We use the notation $(\tC_{\tau, \sigma})^n$ to indicate the transformation corresponding to $n$ subsequent collisions. Graphically,
\begin{equation}
	\label{eq:collision-model}
	\left(
	\Qcircuit @C=1.2em @R=.8em @! R {
		&
		\qw\poloFantasmaCn{\rA}&
		\gate{\tC_{\tau ,  \sigma}}&
		\qw\poloFantasmaCn{\rA}&
		\qw
	}
	\right)^n	\, \coloneqq \, 
	\underbrace{
		\begin{aligned}
			\Qcircuit @C=1.2em @R=.8em @! R {
				&
				\qw\poloFantasmaCn{\rA}&
				\multigate{1}{\tS_{\tau}}&
				\qw&
				\qw\poloFantasmaCn{\rA}&
				\qw&
				\qw&
				\multigate{1}{\tS_{\tau}}&
				\qw\poloFantasmaCn{\rA}&
				\qw\\
				\prepareC{\sigma}&
				\qw\poloFantasmaCn{\rA'}&
				\ghost{\tS_{\tau}}&
				\qw\poloFantasmaCn{\rA'}&
				\measureD{u}&
				\prepareC{\sigma}&
				\qw\poloFantasmaCn{\rA'}&
				\ghost{\tS_{\tau}}&
				\qw\poloFantasmaCn{\rA'}&
				\measureD{u}\\
			}
		\end{aligned}
		\,
		\begin{aligned}
			&\dots \\
			&
		\end{aligned}
		\,
		\begin{aligned}
			\Qcircuit @C=1.2em @R=.8em @! R {
				&
				\qw\poloFantasmaCn{\rA}&
				\multigate{1}{\tS_{\tau}}&
				\qw\poloFantasmaCn{\rA}&	
				\qw \\
				\prepareC{\sigma}&
				\qw\poloFantasmaCn{\rA'}&
				\ghost{\tS_{\tau}}&
				\qw\poloFantasmaCn{\rA'}&
				\measureD{u}\\
			}
		\end{aligned}
	}_{n \text{  times}} \; .
\end{equation}

In collision models, the overall evolution time $t$ of the target system is typically large compared to the single-collision time $\tau$, while the total number of collisions $n$ is taken to be large,  with $n  = t/\tau $. 
Since $n$ is  an integer, the  total time $t$ is a multiple of the collision time $\tau$. Equivalently, one can take $t$ to be an arbitrary real number and set  $n  := \lfloor t/\tau \rfloor$. This rounding has no effect in the limit $\tau \to 0$,  as the limit can be computed by  taking $\tau = t/n \to 0$ for integer $n$ tending to infinity. 	

Here, we evaluate the action of a collision model in the limit of infinitesimal interactions, \textit{i.e.} $\tau \rightarrow 0$ or equivalently $n \rightarrow \infty$. In this case, the overall evolution of the target system, denoted by $\tU_{t, \sigma}$, is defined by
\begin{equation}
	\label{eq:N-unit-block}
	\Qcircuit @C=1.2em @R=.8em @! R {
		&
		\qw\poloFantasmaCn{\rA}&
		\gate{\tU_{t, \sigma}}&
		\qw\poloFantasmaCn{\rA}& \qw
	}  
	\, \coloneqq \, \lim_{n \rightarrow \infty}  \, \left(
	\Qcircuit @C=1.2em @R=.8em @! R {
		&
		\qw\poloFantasmaCn{\rA}&
		\gate{\tC_{ t/n, \sigma}}&
		\qw\poloFantasmaCn{\rA}& \qw
	} \right)^{n} \, .
\end{equation}
The dynamics $\tU_{t,\sigma}$ is the overall evolution generated through the collision model. More precisely:
\begin{definition}[Collisional dynamic]
	For a collision model $(\rA, \rA', \tS_{t}, u)$, we define the \textit{collisional dynamics} generated by the (reference) state $\sigma \in \St(\rA')$ the transformation $(\tU_{t, \sigma})_{t \in \R}$, where $\tU_{t, \sigma}$ is defined as in Eq.~\eqref{eq:N-unit-block}.
\end{definition}

We show in Section~\ref{sec:generalized-collision-model} that for any collision model $(\rA, \rA', \tS_{\tau}, u)$, the collisional dynamics $(\tU_{t, \sigma})_{t \in \R}$  generated by any reference state $\sigma$  is a one-parameter subgroup of the group of reversible transformations $\grp{G}_\rA$ and we provide an explicit expression that describes it.

\subsection{Proof of Theorem~1}
\label{sec:generalized-collision-model}

\begin{theorem}[Generator of the collisional dynamics]
	\label{thm:limit-cm}
	For a collision model $(\rA, \rA', \tS_{t}, u)$, for every reference state $\sigma \in \St(\rA')$, the collisional dynamics $(\map U_{t,\sigma})_{t\in \R}$ is a one-parameter subgroup of the Lie group $\grp{G}_\rA$, and has the form $\tU_{t, \sigma}= e^{G_\sigma \, t}$, where
	\begin{equation}
		\label{eq:state-generator-correspondence}
		G_{\sigma} (\rho) \coloneqq (\tI_\rA \otimes u)  \frac{\d \tS_\tau}{\d \tau }\bigg|_{\tau=0} ( \rho \otimes \sigma) \, ,
	\end{equation}
	for every $\rho \in \St(\rA)$.
\end{theorem}

\Proof Let us start by considering a single collision, diagrammatically represented in Eq.~\eqref{eq:GPT-unit-block}. Since $(\tS_t)_{t \in \R}$ is a one-parameter subgroup of $\grp{G}_{\rA\otimes \rA'}$, there exists a generator $G_{\text{TOT}}$ for which  $\tS_t= \exp(G_{\text{TOT}}\, t )$ for any $t \in \R$.

Also, let $\rho\in \St(\rA)$ be the initial state of the target system. In the limit of small interaction time $\tau$, every collision can be Taylor-expanded as follow
\begin{equation}
	\label{eq:taylor-expansion}
	\begin{aligned}
		\tC_{\tau,\sigma} (\rho) =& (\tI_\rA \otimes u) \tS_\tau (\rho \otimes \sigma)  =  (\tI_\rA \otimes u) (\tI_{\rA\rA'} + \tau \, G_\text{TOT}  + O(\tau^2)) (\rho \otimes \sigma)  = \\
		=& \rho + \tau  (\tI_\rA \otimes u)  G_\text{TOT} (\rho \otimes \sigma)  + O(\tau^2) \, .
	\end{aligned}
\end{equation}

Given $G _{\sigma} = (\tI_\rA \otimes u)  G_\text{TOT} ( \tI_\rA \otimes \sigma) $, let $e^{G_\sigma t}$ be the exponential of $G _{\sigma}$ multiplied for a parameter $t\in\mathbb{R}$. For small values of $t$, namely $t \simeq \tau$, we can Taylor expand $e^{G_\sigma \tau}$ in a neighborhood of $\tau=0$: $e^{G_\sigma \tau} = \tI_\rA + G_\sigma \tau + O(\tau^2)$. Comparing the above expansion with Eq.~\eqref{eq:taylor-expansion}, we obtain
\begin{equation}
	\label{eq:taylor-bound}
	\tC_{\tau,\sigma} - e^{G_\sigma \tau} = O(\tau^2) \, .
\end{equation}

Then, we can write
\begin{align*}
	e^{G_\sigma t} =& (e^{G_\sigma t/n})^n = (e^{G_\sigma t/n})^{n-1} \circ e^{G_\sigma t/n} = \\
	=& (e^{G_\sigma t/n})^{n-1} \circ \tC_{t/n,\sigma} + O(t^2/n^2) = \ldots = ( \tC_{t/n,\sigma})^n + n \cdot O( t^2/n^2) = \\
	=& (\tC_{t/n, \sigma})^n + O(t^2/n)\, .
\end{align*}
Finally, by taking the limit for $n\rightarrow \infty$, we have that $O(t^2/n) \rightarrow 0$, and we are left with
\begin{equation}
	\label{eq:collisional-dynamics}
	\lim_{n\rightarrow \infty} \tC_{t/n,\sigma}^n = e^{ G_{\sigma} t} \, .
\end{equation}

Since $\tS_\tau$ is a physical transformation, and so is $C_{\tau,\sigma}$ for every $\tau \in \R$, $\sigma \in \St(\rA')$, Eq.~\eqref{eq:collisional-dynamics} combined with the closure of the set of transformations, \textit{viz.} Assumption~\ref{as:closure}, implies that also the limit of a sequence of physical transformation is physical, \textit{i.e.} $e^{ G_{\sigma} t} \in \Transf (\rA\to\rA)$,  $\forall \, t \, \in \mathbb{R}$.

Furthermore, since $\tS_\tau$ is a one-parameter subgroup of the reversible transformations, its inverse, $\tS_{-\tau}$, must also be a physical (reversible) transformation. Then, by substituting $\tS_\tau$ with $\tS_{-\tau}$ in Eq.~\eqref{eq:GPT-unit-block}, the minus sign is transmitted at the second term on the right hand side of Eq.~\eqref{eq:taylor-expansion} as follows
\begin{equation*}
	\begin{aligned}
		\tC_{-\tau,\sigma} (\rho) =& (\tI_\rA \otimes u) \tS_{-\tau} (\rho \otimes \sigma)  =  (\tI_\rA \otimes u) (\tI_{\rA\rA'} - \tau \, G_\text{TOT}  + O(\tau^2)) (\rho \otimes \sigma)  = \\
		=& \rho - \tau  (\tI_\rA \otimes u)  G_\text{TOT} (\rho \otimes \sigma)  + O(\tau^2) \, .
	\end{aligned}
\end{equation*}
The bound in Eq.~\eqref{eq:taylor-bound} assumes the following form:
\begin{equation*}
	\tC_{-\tau,\sigma} - e^{G_\sigma (-\tau)} = O(\tau^2) \, .
\end{equation*}
In conclusion, by the same procedure as before, Eq.~\eqref{eq:collisional-dynamics} now reads
\begin{equation*}
	\lim_{n\rightarrow \infty} \tC_{-t/n,\sigma}^n = e^{ - G_{\sigma} t} \, .
\end{equation*}
Ergo, not only $e^{ G_{\sigma} t}$ is a physical transformation, but also its inverse is, namely $e^{ - G_{\sigma} t} \in \Transf(\rA\to\rA)$. In conclusion, $(e^{ G_{\sigma} t})_{t \in \R}$ is a one-parameter subgroup of $\grp{G}_\rA$.	\qed

\medskip

Finally, we would like to remark two points. First, since Eq.~\eqref{eq:GPT-unit-block} is not in general a reversible transformation, it might look odd that the limit of non-reversible transformations is reversible. Intuitively speaking, the reason is that Eq.~\eqref{eq:GPT-unit-block} is actually reversible at the first order. Since in the limit we are considering an infinite sequence of infinitesimal evolutions, only the first order survives and $e^{ G_{\sigma} t}$ turns out to be reversible.

Second, we want to point out that there is no contradiction between the reversibility of the collisional dynamics and the quantum No-programming theorem \cite{nielsen1997programmable}, namely the impossibility of an universal programmable quantum processor in finite-dimensional quantum theory. Indeed, in order to achieve any unitary deterministically, a collision model would require an infinite number of reference systems.

\subsection{Inverse evolution for generalized collision models}

Clearly, the transformation $\map U_{t, \sigma}$ is reversible for every $\sigma$ and for every $t$. One way is to keep the state $\sigma$ fixed and switch to the collision model $(\rA,\rA',\map S_t^{-1},u)$, since it holds that $S_t^{-1}=\map S_{-t}$. 

Interestingly, there is also an alternative way to invert the collisional dynamics $\map U_{t,\sigma}$, by fixing the joint evolution of the collision model $\map S_\tau$ and replacing the state $\sigma$ and the time $t$ with a new state $\sigma_{\text{inv}}$ and a new time $t_{\text{inv}}$.

For this result we need two additional properties: our collision model needs to be stationary at equilibrium (\textit{viz.} Definition~\ref{def:stationarity-equilibrium}), and the invariant state needs to be unique.

\begin{definition}[Invariant state]
	A state $\chi \in \St_1(\rA)$ is \textit{invariant} if
	\begin{equation*}
		\map U \circ \rho = \rho \qquad \text{ for all } \map U \in \grp G_\rA \, .
	\end{equation*}
\end{definition}

If $\chi$ is the unique invariant state of a system $\rA$, then $\chi$ is also internal.

\begin{definition}[Internal state]
	\label{def:internal-state}
	A state $\rho \in \St_1(\rA)$ is internal, if for every state $\omega \in \St_1(\rA)$, there exist a probability $p \in \left(0,1\right]$ and a state $\omega_\rC \in \St_1(\rA)$ such that $\rho = p \, \omega + (1-p) \, \omega_\rC$.
\end{definition}

\begin{lemma}
	\label{lemma:unique-invariant-implies-internal}
	If $\chi$ is the unique invariant state of system $\rA$, then $\chi$ is also an internal state of system $\rA$.
\end{lemma}
\Proof Let $\sigma \in \St_1(\rA)$, we are going to prove that there exists a $p \in \left[0,1\right]$ and a state $\sigma_C \in \St_1(\rA)$ such that $\chi = p \sigma + (1-p) \sigma_C$. Since $\grp{G}_\rA$ is compact, let $\d \tU$ denote the Haar measure on $\grp{G}_\rA$, then
\begin{equation*}
	\omega := \int_{\grp{G}_\rA} \tU \sigma \, \d \tU \, .
\end{equation*}
For any reversible transformation $\tS \in \grp{G}_\rA$, $\tS \omega = \tS \int_{\grp{G}_\rA} \tU \sigma \, \d \tU =  \int_{\grp{G}_\rA} \tS \tU \sigma \, \d \tU = \int_{\grp{G}_\rA} \tU' \sigma \, \d \tU' = \omega $. By the uniqueness of the invariant state, $\chi = \omega$. By Assumption~\ref{as:closure}, $\St_1(\rA)$ is convex, hence Carathéodory's theorem implies that there exist a finite number of points $M$ and a probability distribution $\{p_i\}_{i=1}^M$ such that $\chi = \sum_{ i=1 }^M p_i \, \tU_i \sigma$, for some $\tU_i \in \grp{G}_\rA$. Finally, $\chi = \tU^{-1}_1 \chi = p_1 \, \sigma + \sum_{ i=2 }^M p_i \, \tU^{-1}_1 \circ \tU_i \, \sigma$. \qed

\begin{proposition}
	\label{prop:inverse-programming-GPT}
	For a stationary collision model $(\rA, \tS_t, u)$, in a theory with uniqueness of the invariant state, for every reference state $\sigma \in \St_1(\rA)$ generating the dynamics $\tU_{t, \sigma} = e^{G_\sigma t}$, where $G_\sigma = (\tI \otimes u) \frac{\d \tS_\tau}{\d \tau }\big|_{\tau=0} ( \cdot \otimes \sigma) $, the reverse dynamics is generated by the reference state
	\begin{equation*}
		\sigma_\text{inv} = \frac{ \chi - p_\sigma \sigma }{  1-p_\sigma} \in\St_1(\rA) \, ,
	\end{equation*}
	where $p_\sigma \in \left(0,1\right]$ such that $\chi = p_\sigma \sigma + (1-p_\sigma) \sigma_\rC  $ for $\sigma_\rC \in \St_1(\rA)$.
\end{proposition}
\Proof Exploiting the linearity of the circuit in Eq.~\eqref{eq:GPT-unit-block}, replacing $\sigma$ with $\sigma_\text{inv}$, we get
\begin{equation*}
	\tC_{t,\sigma_\text{inv}} = \frac{1}{1-p_\sigma} \tC_{t,\chi} - \frac{p_\sigma}{1-p_\sigma} \tC_{t,\sigma} \, .
\end{equation*}

Since $\chi$ is invariant, by Lemma~\ref{lemma:invariant-state-generates-trivial-dynamics}, $\tC_{t,\chi} = I$ for any $t \in \R$. Then, by Taylor expanding $\tC_{t,\sigma}$ we obtain
\begin{align*}
	\tC_{t,\sigma_\text{inv}}  &= \frac{1}{1-p_\sigma} I - \frac{p_\sigma}{1-p_\sigma} \left( I + t G_\sigma + O(t^2) \right)  = \\
	&= I -  \frac{ t \, p_\sigma}{1-p_\sigma} G_\sigma  + O(t^2) = \\
	&= \tC_{t', -\sigma}  + O({t'}^2)\, ,
\end{align*}
where in the last line we defined $t'  \coloneqq t \cdot p_\sigma/ (1-p_\sigma)$.

By taking the limit as in Eq.~\eqref{eq:collisional-dynamics}, we get
\begin{align*}
	\map U_{t, \sigma_\text{inv}} &= \lim_{n \rightarrow \infty} \tC_{t/n,\sigma_\text{inv}}^n =  \lim_{n \rightarrow \infty} \left( \tC_{t/n,\sigma_\text{inv}}^{n-1} \circ \tC_{t'/n, -\sigma} + O\left(\frac{{t'}^2}{n^2}\right) \right) =  \\
	&=  \lim_{n \rightarrow \infty} \left( \tC_{t'/n, -\sigma}^n + O\left(\frac{{t'}^2}{n}\right) \right) = \lim_{n \rightarrow \infty} \tC_{t'/n, -\sigma}^n = e^{ (- G_{\sigma}) t'} =\\
	&= {\map U_{t', \sigma}}^{-1} \, . 
\end{align*} \qed

\medskip

In conclusion, to generate the inverse evolution of the collisional dynamics $U_{t, \sigma}$, we can use the collisional dynamics generated by the reference state $\sigma_\text{inv} = \frac{\chi - p_\sigma \sigma}{1-p_\sigma} $ for a time $t_\text{inv} = t \cdot  (1-p_\sigma) /p_\sigma$.

\section{Informational equilibrium}
\label{sec:informational-equilibrium}

In the following, we consider \textit{symmetric} collision models, \textit{i.e.}, collision models where $\rA \equiv \rA'$, then modifying the notation from $(\rA,\rA',\tS_t,u)$ to $(\rA, \tS_t,u)$ for convenience.

\begin{definition}[Stationarity at the equilibrium]
	\label{def:stationarity-equilibrium}
	A symmetric collision model is \textnormal{stationary at equilibrium} if every state generates the trivial dynamics on itself. More precisely, if $S_t(\rho \otimes \rho) = \rho \otimes \rho$ for every $\rho \in \St(\rA)$ and for every $t \in \R$.
\end{definition}

The condition of stationarity at equilibrium can be equivalently expressed in terms of the collisional dynamics.

\begin{lemma}[Lemma~1 of the main text]
	\label{lemma:trivial-dynamics-on-itself}
	If a symmetric collision model is stationary at equilibrium, then for every state $\sigma \in  \St_1  (\rA)$, $\sigma$ is invariant under the collisional dynamics $(  \map U_{t, \sigma})_{t\in \R}$. 
\end{lemma}
\Proof By differentiating the left and right hand side of the condition of stationarity at equilibirum we get
\begin{equation}\label{eq:diff-stationarity}
	\frac{\d S_t}{\d t} \bigg|_{t = 0} (\rho \otimes \rho) = 0 \quad \text{ for every }\rho \in \St_1(\rA) \text{ and for every }t\in \R \, .
\end{equation}
Also, by Theorem~\ref{thm:limit-cm}
\begin{equation*}
	\map U_{t, \sigma} = e^{G_\sigma t} = \sum_{n=1}^{\infty}\frac{t^n \, G_\sigma^n}{n!} \, ,
\end{equation*}
where $G_\sigma = (\map I \otimes u) \frac{\d \map S_t}{\d t} \big|_{t = 0} (\cdot \otimes \sigma)$. Hence by Eq.~\eqref{eq:diff-stationarity}, $G_\sigma (\sigma) = (\map I \otimes u) \frac{\d \map S_t}{\d t} \big|_{t = 0} (\sigma \otimes \sigma) = 0$. In conclusion,
\begin{equation*}
	\map U_{t, \sigma} (\sigma) = \sum_{n=0}^{\infty}\frac{t^n \, G_\sigma^n}{n!} \sigma  = \sigma + \sum_{n=1}^{\infty}\frac{t^n \, G_\sigma^{n-1}}{n!} G_\sigma (\sigma) = \sigma \, .
\end{equation*} \qed 

\medskip

\begin{definition}[Universal collision model]
	A symmetric collision model is \textnormal{universal} if every reversible transformation $\tU  \in  \grp G_\rA$  is of the form  $ \map U  = \map U_{t, \sigma}$ for some suitable time $t$ and some suitable state $\sigma \in \St_1 (\rA)$.
\end{definition}

Note that universality implies that for every reversible transformation $\tU \in \grp G_\rA$, there exist a state $\sigma \in \St_1(\rA)$ and a time $t_* \in \R$ such that $\tU = e^{G_\sigma t_*}$. In other words, the linear span of the states of system $\rA$ is surjectively mapped to an algebra of the group of reversible transformation $\grp G_\rA$. Since for a given Lie group, the corresponding Lie algebra is unique up to isomorphism, the linear span of the set of states is mapped to the algebra $\mathfrak{g}_\rA$.

\begin{definition}[DIN]
	We say that a theory satisfies {\emph Dynamics from Informational Nonequilibrium (DIN)} if it admits a symmetric collision model that is stationary at equilibrium and universal.
\end{definition}

In the following, specifically in Sections~\ref{sec:quantum-collision-models} and~\ref{sec:real_collision_model}, we analyze the DIN assumption in the framework or quantum theory on complex and on real Hilbert spaces, respectively. We show that quantum theory on complex Hilbert spaces satisfies DIN by explicitly providing a universal collision model that is stationary at equilibrium. This fact supports our intuition that collision models are at the foundations of any quantum-like dynamics.

In contrast, we show that quantum theory on real
Hilbert spaces admits symmetric collision models, but they
cannot be universal.

\subsection{Quantum theory on complex Hilbert spaces satisfies DIN}
\label{sec:quantum-collision-models}

In this section we investigate quantum collision models, \textit{i.e.} collision models in finite-dimensional quantum theory. The definition of collision model we provided in Section~\ref{sec:generalized-collision-model} can be adapted to the quantum case as it is. In detail, each collision between a target quantum system A and a reference system $\rA'$ is implemented through the quantum channel
\begin{equation}
	\label{eq:quantum-unit-block}
	\tC_{t, \sigma} 
	\, \coloneqq \,
	\begin{aligned}
		\Qcircuit @C=1.2em @R=.8em @! R {
			&
			\qw\poloFantasmaCn{\rA}&
			\multigate{1}{\tS_{\tau}}&
			\qw\poloFantasmaCn{\rA}&	
			\qw \\
			\prepareC{\sigma}&
			\qw\poloFantasmaCn{\rA'}&
			\ghost{\tS_{\tau}}&
			\qw\poloFantasmaCn{\rA'}&
			\measureD{\Tr}\\
		}
	\end{aligned}\, ,
\end{equation}
where $\sigma $ is a valid density matrix, $\tS_\tau$ is a one-parameter subgroup of the group of unitary operators, and $\Tr$ is the trace operation on the ancillary system $\rA'$. As in the general case, the overall evolution that system $\rA$ undergoes is then the sequential application of the above circuit:
\begin{equation}
	\label{eq:quantum-collision-model}
	\left( \tC_{t, \sigma} 	\right)^n	\, \coloneqq \, 
	\underbrace{
		\begin{aligned}
			\Qcircuit @C=1.2em @R=.8em @! R {
				&
				\qw\poloFantasmaCn{\rA}&
				\multigate{1}{\tS_{\tau}}&
				\qw&
				\qw\poloFantasmaCn{\rA}&
				\qw&
				\qw&
				\multigate{1}{\tS_{\tau}}&
				\qw\poloFantasmaCn{\rA}&
				\qw\\
				\prepareC{\sigma}&
				\qw\poloFantasmaCn{\rA'}&
				\ghost{\tS_{\tau}}&
				\qw\poloFantasmaCn{\rA'}&
				\measureD{u}&
				\prepareC{\sigma}&
				\qw\poloFantasmaCn{\rA'}&
				\ghost{\tS_{\tau}}&
				\qw\poloFantasmaCn{\rA'}&
				\measureD{\Tr}\\
			}
		\end{aligned}
		\,
		\begin{aligned}
			&\dots \\
			&
		\end{aligned}
		\,
		\begin{aligned}
			\Qcircuit @C=1.2em @R=.8em @! R {
				&
				\qw\poloFantasmaCn{\rA}&
				\multigate{1}{\tS_{\tau}}&
				\qw\poloFantasmaCn{\rA}&	
				\qw \\
				\prepareC{\sigma}&
				\qw\poloFantasmaCn{\rA'}&
				\ghost{\tS_{\tau}}&
				\qw\poloFantasmaCn{\rA'}&
				\measureD{\Tr}\\
			}
		\end{aligned}
	}_{n \text{  times}} \; .
\end{equation}

In Ref.~\cite{lloyd2014quantum} it was shown that taking $\map S_\tau$ to be the partial $\texttt{SWAP}$, namely $\map S_\tau = e^{-i \, \texttt{SWAP}\, \tau } \cdot e^{i \, \texttt{SWAP} \, \tau }$, $\left( \tC_{t, \sigma} 	\right)^n$ well approximates the unitary evolution $\rho \mapsto e^{-i\sigma t} \, \rho \, e^{i\sigma t}$ for a sufficiently big $n$. Here, we extend the result of Ref.~\cite{lloyd2014quantum} to show that \textit{i)} the approximation becomes exact in the continuous-time limit and that \textit{ii)} any unitary evolution generated by a bounded Hamiltonian can be perfectly achieved in the continuous-time limit.

\subsubsection{Continuous-time limit of the partial \texttt{SWAP} quantum collision model}
\label{sec:quantum-CM}

We consider the quantum collision model of Ref.~\cite{lloyd2014quantum}, where the one-parameter group governing Eq.~\eqref{eq:quantum-unit-block} is $\tS_{\tau} (\, \cdot \otimes \sigma)= e^{-i \, \texttt{SWAP} \, \tau} \, (\, \cdot \otimes \sigma ) \, e^{i \, \texttt{SWAP} \, \tau}$,  where we remind that the  \texttt{SWAP} operator is defined by the relation $ \texttt{SWAP}  (\ket{\psi} \otimes \ket{\phi}) = \ket{\phi} \otimes \ket{\psi} $ for all pure states $\ket{\psi}, \, \ket{\phi}$.

\begin{proposition}
	Given a quantum collision model $(\rA,\rA',\tS_\tau, \Tr_{\rA'})$, where $(\tS_\tau)_{\tau \in \R}$ is the one-parameter group generated by the quantum \texttt{SWAP},
	\begin{equation}
		\label{eq:quantum-limit-collision-model}
		\map U_{t, \sigma} := \lim_{n\rightarrow \infty} \tC_{t/n,\sigma}^n = e^{-i \sigma t}  \, \cdot  \, e^{i \sigma t} \, .
	\end{equation}
\end{proposition}

\Proof We first evaluate the action of Eq.~\eqref{eq:quantum-unit-block} on an arbitrary quantum state $\rho$ of the target system, and then calculate it in the limit of infinite repetitions acting for an infinitesimal time.

For any density matrix $\rho$, 
\begin{equation*}
	\tS_{\tau} (\rho \otimes \sigma) = e^{-i \, \texttt{SWAP} \, \tau} \, ( \rho \otimes \sigma ) \, e^{i \, \texttt{SWAP} \, \tau} = \cos^2(\tau) \, \rho \otimes \sigma + \sin^2(\tau) \, \sigma \otimes \rho -  i \cos (\tau)\, \sin (\tau) \left[\texttt{SWAP}, \rho \otimes \sigma \right],
\end{equation*}
where $ \left[ A,B \right] = AB-BA $ is usual commutator operation, and in the second equality we used the property that $e^{-i \, \texttt{SWAP} \, \tau} = \cos \tau\, I -i \sin \tau\,  \texttt{SWAP} $, with $I$ being the identity operator. Partial tracing on the second system $\rA'$ yields
\begin{equation*}
	\Tr_{\rA'} \left[ \tS_\tau (\rho \otimes \sigma) \right]= \cos^2(\tau) \, \rho  + \sin^2(\tau) \, \sigma  -  i \cos (\tau)\, \sin(\tau) \Tr_{\rA'} \left[ \texttt{SWAP} \,(\rho \otimes \sigma) -  (\rho \otimes \sigma) \, \texttt{SWAP}  \right] \, .
\end{equation*}
Now, let us focus on the term $\Tr_{\rA'} \left[  \texttt{SWAP} \,(\rho \otimes \sigma)\right]$. Using bra-ket notation we can write $\rho = \sum_{i,j} \rho_{ij} \ket{i}\bra{j}$ and analogously for $\sigma= \sum_{i,j} \sigma_{ij} \ket{i}\bra{j}$. The partial trace then takes the form
\begin{align*}
	\Tr_{\rA'} \left[ \texttt{SWAP} \,(\rho \otimes \sigma)\right] =& \sum_{k,i,j,i',j'}\rho_{i,j}\sigma_{i',j'} I_{\rA} \otimes \bra{k}_{\rA'} ( \texttt{SWAP} \, \ket{i}_{\rA} \ket{i'}_{\rA'} \bra{j}_{\rA} \bra{j'}_{\rA'}) I_{\rA} \otimes \ket{k}_{\rA'} =  \\
	=&  \sum_{k,i,j,i',j'}\rho_{i,j}\sigma_{i',j'} I_{\rA} \otimes \bra{k}_{\rA'} ( \ket{i'}_{\rA} \ket{i}_{\rA'} \bra{j}_{\rA} \bra{j'}_{\rA'}) I_{\rA} \otimes \ket{k}_{\rA'}\\
	=&  \sum_{k,j,i'}\rho_{k,j}\sigma_{i',k}  \ket{i'}_{\rA} \bra{j}_{\rA} = \sigma \cdot \rho \, ,
\end{align*}
where $\sigma \cdot \rho$ denotes the matrix product between $\sigma$ and $\rho$. Analogously $\Tr_{\rA'} \left[ (\rho \otimes \sigma) \, \texttt{SWAP}  \right] = \rho \cdot \sigma $, so in conclusion
\begin{equation*}
	\tC_{\tau, \sigma} (\rho) = \cos^2(\tau) \, \rho  + \sin^2(\tau) \, \sigma  - i \cos (\tau)\, \sin(\tau)  \left[ \sigma, \rho \right] \,.
\end{equation*}
When $\tau$ is small, we can Taylor expand the previous expression in a neighborhood of 0 as follows
\begin{equation*}
	\tC_{\tau, \sigma} (\rho) =  \rho -  i \tau  \left[ \sigma, \rho \right] + O(\tau^2) \,.
\end{equation*}
By comparing the previous equation with the Taylor expansion of the reversible transformation $\map D_\tau (\rho) := e^{-i \sigma \tau } \, \rho \,e^{i \sigma \tau } = \rho - i \tau \left[\sigma, \rho\right] + O(\tau^2) $, we notice that they are equal up to the first order:
\begin{equation*}
	\tC_{\tau, \sigma} - \tD_{\tau} = O(\tau^2) \, .
\end{equation*}
Therefore, proceeding as in the proof of Theorem~\ref{thm:limit-cm}, we have that 
\begin{align*}
	\map D_{t, \sigma} =& (\map D_{t/n})^n = (\map D_{t/n})^{n-1} \circ \map D_{t/n} = \\
	=& (\map D_{t/n})^{n-1} \circ \map D_{t/n} + O(t^2/n^2) = \ldots = (\map D_{t/n})^n + n \cdot O(t^2/n^2) = \\
	=& (\map D_{t/n})^n + O(t^2/n)\, .
\end{align*}
Finally, by taking the limit for $n\rightarrow \infty$, $O(t^2/n) \rightarrow 0$, so proving the thesis. \qed

\subsubsection{Partial \texttt{SWAP} quantum collision model for arbitrary Hamiltonians.}
\label{sec:quantum-CM-inverse}

We now show that the \texttt{SWAP} quantum collision model can achieve, in the continuous time limit, any possible quantum unitary evolution for a suitable state of the reference system.

\begin{proposition}
	\label{prop:inverse-programming-quantum}
	For any hermitian operator $H$ with finite spectral radius,  \textit{i.e.} $\max\{|\lambda_1|,\dots,|\lambda_n|\} < +\infty $ where $\lambda_1,\dots,\lambda_n$ are the eigenvalues of $H$, the time evolution generated by $H$ can be achieved with the partial \texttt{SWAP} collision model using as reference state the density matrix 
	\begin{equation}
		\label{eq:generator}
		H^\prime= \frac{\lambda_* I + H}{\lambda_* \cdot d + \Tr \left[ H \right] } \, ,
	\end{equation}
	where I is the identity matrix, $d$ the dimension of the state space, and $\lambda_*$ the absolute value of the smallest eigenvalue of $H$.
\end{proposition}

\Proof It is easy to check that $H^\prime$ so defined is a valid density matrix: it is hermitian (because $H$ is), semi-positive definite, and it has unit trace. By linearity of the circuit in Eq.~\eqref{eq:quantum-unit-block}, $$\tC_{\tau,H^\prime} = \frac{\lambda_*}{\lambda_*\cdot d +\Tr\left[H\right]} \tC_{\tau, I } + \frac{1}{\lambda_*\cdot d + \Tr\left[H\right] } \tC_{\tau,H}.$$
Now, for any density matrix $\rho$, $\tC_{\tau, I }(\rho) = d \, \rho  + O(\tau^2) $ while $\tC_{\tau, H }(\rho) = \Tr\left[H\right] \rho  -  i \tau  \left[ H, \rho \right] + O(\tau^2)$. Reassembling the pieces together we have
\begin{equation*}
	\tC_{\tau,H^\prime} (\rho)= \rho -  \frac{i \tau}{\lambda_*\cdot d + \Tr\left[H\right]}  \left[ H, \rho \right] + O(\tau^2) \, .
\end{equation*}
On the other hand, let $\map D_{t, H} (\rho) := e^{-i H t } \, \rho \,e^{i H t } $ for any state $\rho$, then
\begin{equation*}
	\map D_{t, H} (\rho) = e^{-i H t } \, \rho \,e^{i H t } = \rho - i \, t \left[H, \rho\right] + O(t^2) \, .
\end{equation*}
Therefore, by considering $t'= t \cdot (\lambda_*\cdot d + \Tr\left[H\right] )$, we have that 
\begin{align*}
	\map D_{t, H} =& (\map D_{t/n, H})^n = (\map D_{t/n, H})^{n-1} \circ \tU_{t/n, H} = \\
	=& (\map D_{t/n, H})^{n-1} \circ \tC_{t'/n, H'} + O(t^2/n^2) = \ldots = (\tC_{t'/n, H'})^n + n \cdot O(t^2/n^2) = \\
	=& (\tC_{t'/n, H'})^n + O(t^2/n)\, .
\end{align*}
In conclusion,
\begin{equation}
	\label{eq:general-ev}
	\map U_{t', H'}:=\lim_{n \rightarrow \infty} \tC_{t'/n, H^\prime}^n = e^{-i H t } \, \cdot \, e^{i H t } \, \quad \text{for } t'= t \cdot (\lambda_*\cdot d + \Tr\left[H\right] ) \, .
\end{equation}
\qed

\medskip

From Eq.~\eqref{eq:general-ev} it is immediate to notice that such collision model needs to run for a time $t'$ in general different from $t$ in order to simulate an evolution for a time $t$. Let us analyze in more detail the case where, given a time evolution generated by a density matrix $\sigma$, we want to return to the initial state by performing the inverse evolution on such state. The density matrix that yields the inverse evolution is given by Eq.~\eqref{eq:generator}, in this case simplified to
\begin{equation*}
	\sigma'= \frac{\lambda_{max} I -\sigma}{\lambda_{max} \cdot d -1}\, ,
\end{equation*}
where $\lambda_{max}$ is the spectral radius of $\sigma$. Then, the time our strategy needs to run to simulate $\tU_{t, -\sigma}$ will be $t' = t \cdot (\lambda_{max} \cdot d -1)$. Since $\lambda_{max} \ge 1/d$ for a density matrix $\sigma$, we can distinguish the following three cases
\begin{equation}
	\label{eq:time-inverse-quantum}
	\begin{cases}
		&t' > t \quad \text{ if } \lambda_{max} > 2/d \\
		&t' = t \quad \text{ if } \lambda_{max} = 2/d \\
		&t' < t \quad \text{ if } 1/d \le \lambda_{max} < 2/d
	\end{cases} \, .
\end{equation}

This expression implies that, when $\sigma$ is a pure state ($p_{\max}  =  1$),   the time required by the inverse evolution  is $d-1$ times the time required for the direct evolution.  It is interesting to observe that  $t_{\rm inv}$ can be exponentially large for physical systems consisting of a large number of components (for which $d$ grows exponentially), thereby implying that certain evolutions, while in principle reversible, are  physically hard to invert once the joint evolution $\map S_\tau$ has been fixed.

\subsection{Quantum theory on real Hilbert spaces violates DIN}
\label{sec:real_collision_model}

Real quantum theory (RQT) is formulated in the same way as  standard quantum theory with the only difference that systems correspond to Hilbert spaces on the  real (instead of the complex) field.  Given a system and its corresponding real Hilbert space $\R^d$ for some positive integer $d$, pure states are represented as unit vectors in $\R^d$, mixed states as positive semi-definite, symmetric, trace-one, matrices, and reversible transformations as orthogonal matrices. In RQT, all states can be decomposed as convex combinations of perfectly distinguishable pure states and set of perfectly distinguishable pure states of the same cardinality can be mapped to each other by means of orthogonal transformation. We show the under the stationarity at equilibrium assumption, see Definition~\ref{def:stationarity-equilibrium}, the only collisional dynamics allowed in real quantum theory is the trivial dynamics.

\subsubsection{Informational equilibrium implies trivial collision model in real quantum theory}
In real quantum theory,  the interaction between the system and its environment is described by a  one-parameter group  $(\map S_t)_{t\in\R}$, consisting of linear maps of the form $ \map S_t  (\cdot)=  S_t (\cdot) S_t^\top$, where each $S_t  :  \R^d\otimes \R_d \to \R^d \otimes \R^d$ is an  orthogonal transformation. 
The stationarity at equilibrium assumption  (Definition~\ref{def:stationarity-equilibrium}) amounts to the condition 
\begin{align}\label{s@e}
	\map S_t ( \rho \otimes \rho ) = \rho \otimes \rho 
\end{align}
for every $t\in \R$ and for every $\rho \in \tD(\R^d)$, where $\tD(\R^d)$ is the set of real $d\times d$ density matrices.

Let $\{ |i\> \}_{i=1}^d$ be the canonical  basis for $\R^d$. Applying the condition (\ref{s@e}) to the pure states $\{ |i\>\}_{i=1}^d$, we obtain the relation  
\begin{align}
	S_t \,( |i\>\<i| \otimes |i\>\<i| )\, S_t^{\top} = |i\>\<i| \otimes |i\>\<i| \,,
\end{align} 
which implies
\begin{align}
	\label{doha1}		S_t \, ( |i\> \otimes |i\>) &= \lambda_i | i\> \otimes |i\>  \qquad \forall i\in \{1,\dots,d\} 
\end{align}
for suitable eigenvalues $(\lambda_i)_{i=1}^d$.

Similarly,  applying the condition (\ref{s@e}) to the pure states $\left\{ |\phi_{ij}^\pm\>  :  =  \frac{|i\>  \pm  |j\>}{\sqrt 2} \right\}_{1\le i<j\le d}$, we obtain the conditions
\begin{align}
	\label{doha2}		S_t  \, (|\phi_{ij}^\pm\> \otimes |\phi_{ij}^\pm\>)   &= \lambda^\pm_{ij} \,  |\phi_{ij}^\pm\> \otimes |\phi_{ij}^\pm\>    \qquad \forall 1\le i<j\le d \, ,
\end{align}
for suitable eigenvalues $(\lambda_{ij}^\pm)_{1\le i<j\le d}$.

Inserting Eq. (\ref{doha1}) into Eq. (\ref{doha2}),  we then obtain  
\begin{align}\label{doha3}
	(\lambda_{ij}^\pm   -  \lambda_i)\, (|i\>\otimes |i\> )   +   ( \lambda_{ij}^\pm  \,-\lambda_j)  \,  |j\>\otimes |j\>    ) =  S_{t}   \, ( |i\>\otimes |j\> +   |j\>\otimes |i\>  )  -  \lambda_{ij}^\pm   \,  ( |i\>\otimes |j\> +   |j\>\otimes |i\>  )  \qquad \forall 1\le i  <  j\le d  \,.
\end{align}

Now, since $S_t$ maps orthogonal  vectors into orthogonal vectors,  the vector $S_t  (  |i\>\otimes |j\>  +  |j\>\otimes |i\>)$ is orthogonal to the vector $S_t  ( |i\>\otimes |i\>)  =  \lambda_i \,  |i\>  \otimes |i\>$.    Hence, multiplying Eq. (\ref{doha3}) one the left by $\<i|\otimes \<i|$ we obtain  
\begin{align}
	\lambda_{ij}^\pm  - \lambda_i       = 0  \qquad \forall i,j  \in  \{1,\dots,  d\} \, . 
\end{align} 
Similarly, the vector $S_t  (  |i\>\otimes |j\>  +  |j\>\otimes |i\>)$ is orthogonal to the vector $S_t  ( |j\>\otimes |j\>)  \equiv  |j\> \otimes |j\>$.  Multiplying Eq. (\ref{doha3}) on the left by $\<j|\otimes\<j|$ we obtain  
\begin{align}
	\lambda_{ij}^\pm  - \lambda_j       = 0 \qquad \forall i,j  \in  \{1,\dots,  d\}   \, . 
\end{align} 
In summary, we have $\lambda_i=\lambda_j  = \lambda^\pm_{ij}    =: \lambda $, and 
\begin{align}
	\nonumber S_t  \,  (  |i\> \otimes |i\>)   &=  \lambda \,  (  |i\> \otimes |i\>)     \qquad \forall i\in  \{1,\dots, d\}\\
	\nonumber 
	S_t\,  |\phi_{ij}^+\>  &  \lambda \,  |\phi_{ij}^+\>  \qquad \forall 1\le i< j \le d \\
	\label{basta} S_t\,  |\phi_{ij}^-\>  &  \lambda \,  |\phi_{ij}^-\>  \qquad \forall 1\le i< j \le d \,. 
\end{align}  
Since the vectors $\{|i\> \otimes |i\>\}_{i =1}^d \cup  \{  |\phi_{ij}^+\>\}_{1\le i  <  j  \le d}  \cup  \{  |\phi_{ij}^-\>\}_{1\le i  <  j  \le d}$, Eq. (\ref{basta}) implies $S_t  =  \lambda \,  I$, where $I$ is the identity operator on $\R^d\otimes \R^d$.   

Note that the eigenvalue  $\lambda$  must satisfy the condition $\lambda^2  =1$: indeed, taking the transpose  on both sides of the equation  $S_t  \,  (|i\>\otimes |i\> )  =  \lambda\, |i\>\otimes |i\>$, yields the equation  $(\<i| \otimes \<i|) \,  S_t^\top  =  \lambda \, \<i|\otimes\<i|$, which implies  
\begin{align}
	\lambda^2    =  ( \lambda\, \<i| \otimes \<i|) \,  (  \lambda \,  |i\>\otimes |i\>)   =  (\<i| \otimes \<i|) \,  S_t^\top    S_t  \,  (|i\>\otimes |i\> )   =  1 \, .
\end{align}  
Hence, the linear map  $\map S_t:     \rho \mapsto    S_t  \, \rho  \, S_t^\top$ satisfies the condition 
\begin{align}
	\map S_t        =    \lambda^2  \,  \map I  = \map I \qquad \forall t\in \R\, ,   
\end{align}
that is, the dynamics $(\map S_t)_{t\in \R}$ is trivial. 

\subsubsection{A universal collision model for real qubits}

We now show a collision model for two-dimensional systems in real quantum theory, with the feature that the model is universal (it can generate arbitrary dynamics) but not stationary at equilibrium.   

For a two-dimensional system $\rA$ in RQT  (a \emph{rebit}), the pure normalized state of the system can be represented by vectors of the form $|\psi\> = p |0\> + (1-p) |1\> $, for $p \in \left[0,1\right]$ (loosely speaking the state space is simply the convex hull of a circle). The mixed states are  density matrices of the form $\rho = \begin{pmatrix} p & q \\ q & 1-p \end{pmatrix}$, for $ p ,\, q \in \left[0,1\right]$ such that $\rho$ is positive semidefinite. Regarding the dynamics, the only non-trivial one-parameter subgroup of the group of reversible transformation is $D_t = e^{iY t}$, where $Y$ is the Pauli-$y$ matrix. Indeed, the group of reversible transformation is the group of orthogonal matrices, the corresponding Lie algebra is the algebra of skew-symmetric matrices and, in dimension two, $\{iY\}$ generates such algebra.

Now, let us introduce the symmetric collision model $(\rA, \tS_{\tau}, \Tr_\rA)$, for $\tS_{\tau} = e^{\tau \, G_{\rm TOT} }$, with $G_{\rm TOT} := iY \otimes |0\>\<0|$. For any $\sigma \in \rA$, the corresponding collisional dynamics will be governed by the generator $G_\sigma = (\tI \otimes u) \, G_{\rm TOT} \, (\, \cdot \otimes \sigma) = i p_\sigma \, Y$, where $p_\sigma = \< 0 | \sigma |0\>$. Therefore, the collision model $(\rA, \tS_{\tau}, \Tr_\rA)$ allows to recover all the reversible dynamics of system $\rA$, including the trivial dynamics for $\sigma = |1\>\<1|$. In particular, rebit states generate dynamics describing the same trajectory with, in general, a different rate. For example, the states $\sigma = |0\>\<0|$ and the state $\sigma' = |+\>\<+|$ generate the dynamics $D_t = e^{iY t}$ and  $D'_t = e^{iY t/2}$, respectively.

\subsection{Implications of stationarity at equilibrium and universality}

\begin{lemma}
	\label{lemma:swap-of-generator-and-target}
	Let $(\rA, \tS_t, u)$ be a symmetric collision model that is stationary at equilibrium, and let $G_\text{TOT}=\frac{\d \tS_{\tau}}{\d \tau}\big|_{\tau = 0}$. Then for every pair of states $\rho,\rho' \in \St_1(\rA)$, $ G_\text{TOT} (\rho \otimes \rho') = - \, G_\text{TOT} (\rho' \otimes \rho) $.
\end{lemma}
\Proof We start by noticing that $\tS_\tau (\rho \otimes \rho) = \rho \otimes \rho$ for every $\rho \in \St_1(\rA)$ is equivalent to require that $G_\text{TOT} (\rho \otimes \rho) = 0$ for every $\rho \in \St_1(\rA)$. Then, for every pair of states $\rho,\rho' \in \St(\rA)$, $0 = G_\text{TOT} \big((\rho + \rho') \otimes (\rho + \rho') \big) = G_\text{TOT} (\rho \otimes \rho) + G_\text{TOT} (\rho' \otimes \rho) + G_\text{TOT} (\rho \otimes \rho') + G_\text{TOT} (\rho' \otimes \rho') = G_\text{TOT} (\rho' \otimes \rho) + G_\text{TOT} (\rho \otimes \rho')   $. Therefore, $G_\text{TOT} (\rho' \otimes \rho) = - \, G_\text{TOT} (\rho \otimes \rho')$. \qed

\medskip

\begin{lemma}
	\label{lemma:invariant-state-generates-trivial-dynamics}
	Let $(\rA, \tS_t, u)$ be a symmetric collision model that is stationary at equilibrium, then $(\tU_{t, \sigma})_{t\in\R} = \tI$ for every $t\in \R$ if $\sigma$ is an invariant state of system $\rA$. If  $(\rA, \tS_t,u)$ is also universal, then $(\tU_{t, \sigma})_{t\in\R} = \tI$ for every $t\in \R$ if and only if $\sigma$ is an invariant state of system $\rA$.
\end{lemma}
\Proof Let $\sigma$ be an invariant state of system $\rA$, by Lemma~\ref{lemma:swap-of-generator-and-target},
\begin{equation*}
	G_\sigma (\rho) = (\map I \otimes u) \frac{\d \map S_t}{\d t} \bigg|_{t = 0} (\rho \otimes \sigma) = - (\map I \otimes u) \frac{\d \map S_t}{\d t} \bigg|_{t = 0} (\sigma \otimes \rho) = - G_\rho (\sigma)
\end{equation*}
for any $\rho \in \St_1(\rA)$. Since $\sigma$ is invariant, $\map U_{t, \rho}(\sigma) = e^{G_\rho t} \sigma= \sigma$ for any $\rho \in \St_1(\rA)$ and for any $t \in \R$. It follows that $G_\rho (\sigma) =0$, and therefore $ G_\sigma (\rho) = 0$ for any $\rho \in \St_1(\rA)$, \textit{i.e.} $\sigma$ generates the trivial dynamics.

On the other hand, if the collision model is both stationary and universal and $(\tU_{t, \sigma})_{t\in\R} = \tI$  for every $t\in \R$, then $G_\sigma (\rho)= 0$ for any state $\rho$. Again, by Lemma~\ref{lemma:swap-of-generator-and-target}, $G_\rho (\sigma)= 0$. By the universality of the collision model, $\sigma$ is invariant under all reversible dynamics, hence $\sigma$ is an invariant state of system $\rA$. \qed

\medskip

\subsection{Proof of Theorem~2}

\begin{theorem}
	\label{thm:1to1-state-generator}
	In a theory satisfying DIN,  the group of reversible transformations $\grp{G}_\rA$ is a connected Lie group, with Lie algebra $\mathfrak{g}_\rA$.  The map $G  :  \St_1 (\rA) \to  \mathfrak{g}_\rA,  \,  \sigma \mapsto G_\sigma$ is injective if and only if system $\rA$ has a unique invariant state $\chi_\rA$.
\end{theorem}

\Proof By Proposition~\ref{prop:lie-group}, $\grp{G}_\rA$ is a Lie group. Furthermore, from the universality of the collision model, for each reversible transformation $\tT \in \grp{G}_\rA$, it exists a one-parameter subgroup $(\tU_{t,\sigma})_{t\in\R} \subset\grp{G}_\rA$ such that $\tT = \tU_{\hat{t},\sigma}$ for some $\hat{t}\in\R$, \textit{i.e.}, $\grp{G}_\rA$ is connected. By Theorem~\ref{thm:limit-cm}, for each system $\rA$, each collision model $(\rA,\tS_\tau)$ defines a map $G$ from the set $\St_1(\rA)$ to the generator algebra $\mathfrak{g}_\rA$ defined by 
\begin{equation*}
	G : \St_1(\rA) \rightarrow \mathfrak{g}_\rA :: \sigma \mapsto G_\sigma = (\tI \otimes u)  G_\text{TOT} (\tI \otimes \sigma )  \, ,
\end{equation*}
where $G_\text{TOT} := \frac{\d \tS_\tau}{\d \tau} \big|_{\tau = 0}$. By Lemma~\ref{lemma:invariant-state-generates-trivial-dynamics}, $G_\sigma = 0$ if and only if $\sigma = \chi$.		

If $G$ is injective, then the invariant state must be unique, otherwise multiple states would generate the same trivial dynamics.

On the other hand, let us assume UIS. Let $\sigma$ and $\rho \in \St_1(\rA)$ be such that $G(\sigma) =G (\rho)$. Since $\rho$ and $\sigma$ both generate the trivial dynamics if and only if $\rho = \sigma = \chi$ (see Lemma~\ref{lemma:invariant-state-generates-trivial-dynamics}), without loss of generality we can consider $\rho \ne \chi$. By Lemma~\ref{lemma:unique-invariant-implies-internal}, there exist a probability $p$ and a state $\rho_C \in \St_1(\rA)$ such that $\chi = p \rho + (1-p) \rho_C $. By linearity of $G$ we have
\begin{align*}
	G_\sigma - G_\rho &= 0 \\
	G_{\sigma - \rho} &= 0 \\
	p \, G_{\sigma-\rho} &= 0 \\
	G_{\chi} + p \, G_{\sigma - \rho} &= 0 \\
	G_{\chi - p\rho + p \sigma} &= 0 \\
	G_{(1- p)\rho_C + p \sigma} &= 0 \, .
\end{align*}
We notice that $(1- p)\rho_C + p \sigma$ is a convex combination between two valid states, thus it is a valid state. Since it generates the trivial dynamics, it must be the invariant state $\chi$, then
\begin{align*}
	\chi = (1- p)\rho_C + p \sigma \\
	\chi - (1- p)\rho_C =  p \sigma \\
	p \rho =  p \sigma \\
	\rho =  \sigma \, .
\end{align*}
\qed

\medskip

\subsection{Proof of Theorem~3}

Now we introduce an important property: self-duality \cite{koecher1958die, vinberg1960homogeneous}.

\begin{definition}[Strong self-duality]
	A theory is strongly self-dual if there exists an isomorphism $\Phi : \St_+(\rA) \to \Eff_+(A)$ that is symmetric and positive semi-definite, \textit{i.e.}, $\Phi(\rho) \circ \omega = \Phi(\omega) \circ \rho $ for all $\rho, \omega \in \St_+(\rA)$ and $\Phi(\psi)\circ \psi\ge0$ for all $\psi \in \St_+(\rA)$.
\end{definition}

In other words, strong self-duality states that the cones spanned by the sets $\St(\rA)$ and $\Eff(\rA)$ are isomorphic and that the pairing $e \circ \rho$ can be expressed as a Euclidean inner product $\< \omega_e, \rho\>$ over $\St_\R(\rA)$, for $\omega_e = \Phi(e)$. Since reversible transformations preserve normalization, for every $\map U \in \grp G_\rA$, $\< \map U \circ \omega, \map U \circ \rho\> = \< \omega, \rho\>$ for every $\omega, \rho \in \St(\rA)$.

A consequence of self-duality together with the Uniqueness of the Invariant State is the uniqueness of the deterministic effect.

\begin{lemma}
	\label{lemma:unique-det-eff-from-UIS}
	In a theory satisfying UIS and strong self-duality, every system has a unique deterministic effect.
\end{lemma}
\Proof Let $\chi \in \St_1(\rA)$ be the unique invariant state, $u \in \Eff_1(\rA)$ be a deterministic effect, and $\omega_u \in \St_1(\rA)$ the unique state corresponding to $u$ via self-duality, \textit{i.e.} $u = \Phi(\omega_u)$. Since $u(\rho) = 1$ for every state $\rho \in \St_1(\rA)$, for every $\psi \in \St_1(\rA)$ 
\begin{equation}
	u\circ \psi = u \circ \map U \circ \psi \, ,
\end{equation}
for every $\map U \in \grp G_\rA$. Therefore,
\begin{equation}
	\Phi(\psi) \circ \omega_u = \< \omega_u, \psi\> = u \circ \psi =  u \circ \map U \circ \psi  = \< \omega_u, \map U \circ \psi\> = \< \map U^{-1} \circ \omega_u,  \psi\> = \Phi(\psi) \circ \map U \circ \omega_u \,.
\end{equation}
Since the previous chain of equalities holds for any state $\psi \in \St_1(\rA)$ (namely, for every normalized effect in $\Eff_1(\rA)$), we conclude that $\omega_u = \map U^{-1} \circ \omega_u$ for all $\map U \in \grp G_\rA$ and therefore $ \omega_u \equiv \chi$. Since we chose $u$ arbitrarily and the self-duality correspondence is unique, UIS implies that $u$ is the unique deterministic effect of system $\rA$. \qed

\medskip

Theorem \ref{thm:1to1-state-generator} established a one-to-one correspondence between states and generators of the reversible dynamics.  The one-to-one correspondence between states and effects provided by self-duality turns the state-generator correspondence into an observable-generator (bijective) correspondence.

\begin{theorem}[Observable-generator duality]
	\label{thm:energy-obs}
	In a theory satisfying DIN, UIS, and strong self-duality, the elements of the Lie algebra $\mathfrak{g}_\rA$ are in one-to-one linear correspondence with the elements of the subspace  $\map O_{\rA}  :  =  \{ h  \in  \Eff_\R  (\rA) ~|~  h  \circ \chi_\rA =  0\}$, in such a way that every element $h  \in  \map O_\rA$ is invariant under the dynamics generated by the corresponding element $G_h \in  \mathfrak{g}_\rA$.   
\end{theorem}
\Proof By self-duality, $\tO_\rA$ is isomorphic to the affine subspace $\tS_\rA := \left\{ \omega \in \St_\R(\rA) ~|~ u(\omega) = 0  \right\}$, where $u$ is the deterministic effect of system $\rA$, \textit{viz.} Lemma~\ref{lemma:unique-det-eff-from-UIS}. In turn, $\tS_\rA$ can be more conveniently identified as
$$\tS_\rA := \left\{ \sum_i \lambda_i \sigma_i ~ \middle| ~\sum_i \lambda_i = 0 \; \text{ and } \; \sigma_i \in \St_1(\rA) \text{ for all }i  \right\} \, . $$
The fact that every state is proportional to a deterministic one is again a direct consequence of UIS and strong-duality, in particular, it is an equivalent condition to the uniqueness of the deterministic effect granted by Lemma~\ref{lemma:unique-det-eff-from-UIS}, see Ref.~\cite{dariano2017quantum}.

Now, let $G:\St_1(\rA) \to \mathfrak{g}_\rA$ be the linear function defined in Theorem~\ref{thm:1to1-state-generator}, injective under the UIS hypothesis. $G$ can be naturally extended to a linear function from the affine space $\tS_\rA$ to the algebra $\mathfrak{g}_\rA$ such that, for any $\omega = \sum_i \lambda_i \sigma_i \in \tS_\rA$, $G(\omega) = \sum_i \lambda_i G_{\sigma_i}$. The extended function $G$ is still injective. Indeed, let $\omega = \sum_i \lambda_i \sigma_i \in \tS_\rA$ such that $G(\omega)=0$ and let $N = \sum_i |\lambda_i|$ (note that, since $\sum_i \lambda_i = 0$, the sums of the positive and negative coefficients of the linear combination have the same absolute value, \textit{i.e.}  $\sum_{ i, \, \lambda_i >0 }\lambda_i = - \sum_{ i, \, \lambda_i <0 }\lambda_i$). Then
\begin{equation*}
	\begin{aligned}
		0 = G(\omega) = G \left( \frac{\omega}{N} \right) = \sum_{ i, \, \lambda_i >0 }\frac{\lambda_i}{N} G(\sigma_i) + \sum_{ i, \, \lambda_i <0 }\frac{\lambda_i}{N} G(\lambda_i) \, .
	\end{aligned}
\end{equation*}
Since both $\sum_{ i, \, \lambda_i >0 }\frac{\lambda_i}{N} \sigma_i$ and $-\sum_{ i, \, \lambda_i <0 }\frac{\lambda_i}{N} \sigma_i$ are proportional to convex combinations of normalized states with the same proportionality factor, say $\lambda\,  \rho$ and $\lambda\,\rho'$, respectively, for $\lambda>0$ and $\rho, \, \rho' \in \St_1(\rA)$, we can rewrite the above equation as
\begin{equation*}
	\begin{aligned}
		0 = G(\omega) = \lambda \left( G(\rho) - G(\rho') \right) \, .
	\end{aligned}
\end{equation*}
Given the injectivity of $G$ on $\St_1(\rA)$, it must be $\rho = \rho'$, proving the injectivity of $G$ on $\tS_\rA$.

Surjectivity is an immediate consequence of the universality assumption. Indeed, since for every $\tU \in \grp G_\rA$ there exist a time $t_*$ and a normalized state $\sigma \in \St_1(\rA)$ such that $\tU = e^{G_\sigma t_*}$, by linearity of the map $G$, the vector $t_* (\sigma - \chi) \in \tS_\rA$ is also a generator of $\tU$, so that $\mathfrak{g}_\rA \subseteq \{G(\omega) ~|~ \omega \in \tS_\rA \}$.

It is now left to check that $G_\omega(\omega) =0$ for every $\omega \in \tS_\rA$ ($h \in \tO_\rA$). Let $\omega = \sum_{ i } \lambda_i \sigma_i$ ($h = \sum_{ i } \lambda_i e_{\sigma_i}$) then,
\begin{equation}
	G_\omega(\omega) = \sum_{i,j} \lambda_i \lambda_j G_{\sigma_i}(\sigma_j) \, .
\end{equation}
By Lemma~\ref{lemma:trivial-dynamics-on-itself}, we can ignore all the terms in the sum where $i=j$:
\begin{equation}
	G_\omega(\omega) = \sum_{i\ne j} \lambda_i \lambda_j G_{\sigma_i}(\sigma_j) = \sum_{i < j} \lambda_i \lambda_j G_{\sigma_i}(\sigma_j) + \sum_{i > j} \lambda_i \lambda_j G_{\sigma_i}(\sigma_j) \,.
\end{equation}
Finally, by Lemma~\ref{lemma:swap-of-generator-and-target}, the two terms at the right end side of the previous chain of equations cancel each other. \qed

\medskip

\subsection{Observable-generator duality}

We define an observable on system $\rA$ as an element $X$ of the real vector space $\Eff_\R (\rA)$ \cite{chiribella2016entanglement}. For a linear combination of effects $\{e_j\}$ with coefficients $\{x_j\}$, the expectation value of the observable $X = \sum_j \, x_j \, e_j$  on the state $\rho$ is given by $ \< X \>_\rho : = X (\rho) = \sum_j \, x_j \, e_j (\rho)$, and can be estimated by performing suitable measurements containing the effects $\{e_j\}$ and by post-processing the outcomes.

Theorem~\ref{thm:energy-obs} states a one-to-one correspondence between the algebra of generators and the vector subspace $\tO_\rA := \{ h \in \Eff_\R(\rA)~|~h\circ \chi_\rA = 0\} \subset \Eff_\R(\rA)$, namely, $G_h \leftrightarrow h = \sum_i \lambda_i e_{\sigma_i}$ for $e_{\sigma_i} \in \Eff_1(\rA)$. Such correspondence naturally leads to a definition of energy observable.

\begin{definition}[Energy observable]
	\label{def:energy-obs}
	In a OPT satisfying DIN, UIS, and strong self-duality, for every $g \in \mathfrak{g}_\rA$, we define the \text{energy observable} associated to the  generator $g$ the functional $h = \sum_i \lambda_i \, e_{\sigma_i}$, such that $g = G_h$.  
\end{definition}

We will see in the following (Section~\ref{sec:energy-observable}), that under the assumption of CD and SS, the energy observable can be measured in a canonical way by performing an ideal measurement described by pure effects.

\section{Derivation of quantum theory}

We now show that finite-dimensional quantum theory can be uniquely characterized by DIN, plus four additional assumptions that appeared in the previous literature on quantum reconstructions. 

Some of our axioms will refer to the notion of \textit{perfect distinguishability}.
\begin{definition}[Perfectly distinguishable states]
	A set of pure states $\{ \psi_i \}_{ i=1 }^k$ is named {\em perfectly distinguishable} if there exists a measurement $\tM=\{e_j\}_{j=1}^k$  such that $e_j(\psi_i)=\delta_{ji}$ for every $i,j=1,\dots,k$.
\end{definition}
For every system $\rA$, we denote by $d_\rA$ the cardinality of the greatest set of perfectly distinguishable pure states. A set of perfectly distinguishable pure states with cardinality $d_\rA$ is said to be \textit{maximal}. In the following, without loss of generality we will consider system of cardinality $d$, hence dropping the label of the corresponding system. We name $d$ the \textit{capacity} of the system.

\subsection{The assumptions}
\label{sec:assumptions}

\begin{enumerate}
	\item \textbf{Dynamics from informational nonequilibrium (DIN).} For every system $\rA$, there exists at least one symmetric collision model that is both stationary and universal;
	
	\item \textbf{Causality (C)~\cite{chiribella2010probabilistic}.} The probability of preparations is independent of the choice of measurements and effects;
	\item {\bf Classical decomposability (CD)~\cite{barnum2014higher}.} Every state can be diagonalized. Namely, for every $\rho \in \St_1(\rA)$, there exist a maximal set of perfectly distinguishable pure states $\{ \psi_i\}_{i=1}^d \subseteq \Pur\St_1(\rA)$ and a probability distribution $\{ p_i \}_{i=1}^d$ such that $\rho = \sum_{i=1}^d p_i \psi_i $.
	The probability $p_i$ is called the eigenvalue of $\rho$ corresponding to the eigenvector $\psi_i$;
	\item \textbf{Purity preservation (PP)~\cite{chiribella2016entanglement}.} The sequential and/or parallel composition of pure transformations is a pure transformation;
	\item \textbf{Strong symmetry (SS)~\cite{barnum2014higher}.} Let $\{\psi_1,\dots,\psi_n\}$ and $\{\psi'_1,\dots,\psi'_n\}$ be two sets of perfectly distinguishable pure states of system $\rA$, then there exists a reversible transformation $\tT \in \grp{G}_\rA$ such that $\tT(\psi_i)=\psi'_i$ for every $i=1,\dots,n$.
	
\end{enumerate}

As we have already widely discussed about the DIN assumption, we discuss now about the informational meaning of the other four.

Causality can equivalently be stated as requiring uniqueness of the deterministic effect \cite{dariano2017quantum}. Not only that, but even the convex structure of an OPT can be derived from the causality assumption together with closeness of the set of transformation \cite{dariano2017quantum}.

Classical decomposability  \cite{barnum2014higher}, also known as {\em diagonalization} \cite{chiribella2015operational}, is the requirement that every state can be prepared as a convex mixture of perfectly distinguishable pure states. Diagonalization grants us the ability of comparing the information content of different states by decomposing them into probability distributions of perfectly distinguishable pure states. Furthermore, such decomposition lays the groundwork for sensible definitions of entropic notions, see for example Ref.~\cite{short2010entropy, barnum2012entropy, chiribella2016entanglement}. 

Purity preservation \cite{chiribella2016entanglement}, requires that  composing two pure transformations in parallel or in sequence gives rise to another pure transformation. This assumption comes naturally from our perception of reality and it indeed happens to be satisfied by all concrete examples of GPTs and OPTs known to date, with the notable exception of the ad hoc construction of Ref.~\cite{dariano2020classicality}.  

Finally, strong symmetry \cite{barnum2014higher} states that every two sets of perfectly distinguishable states can be reversibly converted one into the other if they have the same cardinality. Strong symmetry has a clear information-theoretic interpretation, as it guarantees that two sets that can perfectly encode the same amount of information are equivalent.

Classical decomposability and strong symmetry appear either as axioms or as key results in many information-theoretic derivations of quantum theory \cite{hardy2001quantum, chiribella2010probabilistic, chiribella2011informational, masanes2011derivation, dakic2011quantum, barnum2014higher} and often play a role in the study of information processing in other examples of GPTs  \cite{barrett2007information, janotta2011limits, dariano2014feynman, chiribella2016entanglement, chiribella2024bell}.

\subsection{Implications of strong symmetry and classical decomposability}

A special instance of SS is the transitivity property.
\begin{lemma}[Transitivity]
	\label{lemma:transitivity}
	In a theory satisfying SS, for every pair of pure states $\psi, \, \psi' \in \St(\rA)$, there exists a reversible transformation $\tT$ mapping one onto the other: $\tT(\psi) = \psi'$.
\end{lemma}
A well known corollary of transitivity is the uniqueness of the invariant state.
\begin{corollary}[Existence and uniqueness of the invariant state \cite{chiribella2010probabilistic}]
	\label{cor:existence-invariant-state}
	For every system, there is a unique invariant state.
\end{corollary}
\Proof
The proof is exactly the same of Corollaries~33 and~34 of Ref.~\cite{chiribella2010probabilistic}, where the thesis derives from finite dimensions and Proposition~\ref{lemma:transitivity}. \qed

\medskip

In Lemma~\ref{lemma:unique-invariant-implies-internal}, we proved that if the invariant state is unique, then it is also internal. Thanks to CD we can further prove that the invariant state is in fact the maximally mixed state.
\begin{lemma}[The invariant state is maximally mixed]
	\label{lemma:maximally-mixed}
	The invariant state $\chi$ is also maximally mixed. Namely, for any maximal set of perfectly distinguishable normalized pure states $\{ \psi_1 , \dots ,  \psi_d \}$, $$\chi = \frac{1}{d}\sum_{i=1}^d \psi_i \, .$$
\end{lemma}
\Proof Let us consider the invariant state $\chi \in \St_1(\rA)$. By the CD assumption, there exists a maximal set of perfectly distinguishable pure states $\{ \psi_i \}_{i=1}^d $ and a probability distribution $\{p_i \}_{i=1}^d$ such that $\chi = \sum_{i=1}^d p_i \psi_i$. Let $\{\psi'_i\}_{i=1}^d$ another set of states such that $\psi'_1 = \psi_2$, $\psi'_2 = \psi_1$ and $\psi'_i = \psi_i$ for all $i=1,\dots,d$. By SS, let $\tT \in \grp{G}_\rA$ the transformation such that $\tT(\psi_i)=\psi'_i$ for all $i=1,\dots,d$. Then, 
\begin{equation}
	\label{eq:max-mixed-first-decomposition}
	\chi = \tT(\chi) = p_1 \psi_2 + p_2 \psi_1 + \sum_{ i=3 }^d p_i \psi_i \, .
\end{equation}
Let $\{e_i\}_{i=1}^d \subset \Eff_1(\rA)$ be the measurement perfectly distinguishing the set $\{ \psi_i \}_{i=1}^d $, namely, $e_i(\psi_j)=\delta_{ij}$ for all $i,j=1,\dots,d$. By applying $e_1$ to both side of Eq.~\eqref{eq:max-mixed-first-decomposition} we get
\begin{equation}
	p_1 = e_1 \circ \chi = e_1 \circ \tT \circ \chi = p_2 \, .
\end{equation}
By repeating the same strategy for all possible pairs of indexes, it must be that $p_i = 1/d$ for all $i=1,\dots,d$. Finally, for any other maximal set of perfectly distinguishable pure states $\{\phi_i\}_{i=1}^n$, by SS it exists a transformation $\tS : \psi_i \mapsto \phi_i$ such that
\begin{equation}
	\chi = \tS(\chi) = \frac{1}{d} \sum_{ i=1 }^d \phi_i \, .
\end{equation} \qed

The combination of CP and SS not only implies UIS, but also strong self-duality. This result was firstly proved in Ref.~\cite{barnum2014higher}.

\begin{proposition}[Self-duality, Proposition 3 of \cite{barnum2014higher}]
	\label{prop:self-duality}
	CP and SS imply self-duality.
\end{proposition}

We are now in position to extend the previous results on the characterization of the invariant state to its corresponding element in the effects set: the deterministic effect $u$.

\begin{lemma}
	\label{lemma:invariant-measurement}
	For every maximal set of perfectly distinguishable pure states $\{\psi_i\}_{i=1}^d$, the set $\{e_{\psi_j}\}_{j=1}^d$ is a valid measurement (that perfectly discriminates between the $\psi_i$-s).
\end{lemma}
\Proof
By Lemma~\ref{lemma:maximally-mixed},
\begin{equation}
	\label{eq:invariant-diagonalized}
	\chi = \frac{1}{d} \sum_{i=1}^d \psi_i \, .
\end{equation}
Let $\{f_j\}_{j=1}^d$ be the measurement that perfectly discriminates the set of pure states $\{\psi_i\}_{i=1}^d$, and let $\{e_{\psi_j}\}_{j=1}^d$ be the set of effects dual to $\{\psi_i\}_{i=1}^d$. If $f_j$ is pure, then by uniqueness of the self-duality state-effect correspondence, $e_{\psi_j} \equiv f_j$. Let us suppose that $f_j = \alpha a_j + \beta b_j $, for some $\alpha, \, \beta \in \mathbb{R}$ and $a_j, \, b_j $ pure effects of system $\rA$. In order for $f_j$ to be a valid effect, $0 \le \alpha, \, \beta \le 1$. Also, by applying $f_j$ to $\chi$ diagonalized as in Eq.~\eqref{eq:invariant-diagonalized}
\begin{equation*}
	\begin{aligned}
		f_j (\chi) &= \frac{1}{d} \sum_{i=1}^d f_j \left(\psi_i \right) =  \frac{1}{d} \sum_{i=1}^d \delta_{ij} = \frac{1}{d}\\
		f_j (\chi) &= \alpha a_j (\chi) + \beta b_j (\chi) = \frac{1}{d}(\alpha +  \beta)
	\end{aligned}
\end{equation*}
where in the second line we used again Lemma~\ref{lemma:maximally-mixed} and the self-duality state-effect correspondence for the pure effects $a_j$ and $b_j$. Therefore, $f_j$, if coarse-grained, is a convex combination of pure effects. Since $f_j(\psi_j)=1$, then $a_j (\psi_j) = b_j (\psi_j) =1$, but from uniqueness of self-duality $a_j \equiv b_j \equiv e_{\psi_j}$. \qed

\medskip 

\begin{lemma}
	\label{lemma:extension-to-maximal-set}
	Every set of perfectly distinguishable pure states can be extended to a maximal set of perfectly distinguishable pure states.
\end{lemma}
\Proof 
Let $\{\psi_1, \dots, \psi_k\}$ be a set of perfectly distinguishable pure states. Let us also consider a maximal set of perfectly distinguishable pure states $\{\psi'_i\}_{i=1}^d$ and let $\{e'_j\}_{j=1}^d$ be the set of dual pure effects that compose the measurement that distinguishes them, see Lemma~\ref{lemma:invariant-measurement}.
By SS, it exists a reversible transformation $\tT$ such that
\begin{equation*}
	\begin{aligned}
		\tT(\psi'_i) &= \psi_i \, \text{ for every }i=1,\dots,k
	\end{aligned}
\end{equation*}
Therefore, $\{\tT(\psi'_i)\}_{i=1}^d =\{\psi_i\}_{i=1}^k \cup \{\tT(\psi'_i)\}_{i=k+1}^d $ is a set of $d$ pure states perfectly discriminated by $\{e'_j \circ \tT^{-1} \}_{j=1}^d$. To check that the latter is actually a measurement it is sufficient to notice that
\begin{equation*}
	\sum_{j=1}^d e'_j \circ \tT^{-1} = \left( \sum_{j=1}^d e'_j \right) \circ \tT^{-1} = u \circ \tT^{-1} = u \, .
\end{equation*}
\qed

\medskip

\begin{corollary}[\cite{guha2022untitled}]
	\label{cor:orthogonal-iff-perfectly-distinguishable}
	In every self-dual theory, a pair of pure states is perfectly distinguishable if and only if they are orthogonal respect to the inner product inherited by self-duality.
\end{corollary}
\Proof Suppose there exist two states $\psi_1,\psi_2\in\Pur(\rA)$ such that $\langle \psi_1 ,\psi_2 \rangle=0$. Now, due to the self-dual structure of the theory we can assign the extremal effects $e_i$ to $\psi_i$ for $i=1,2$ such that $e_i(\psi_j)=\langle\psi_i,\psi_j\rangle=\delta_{i,j}$ for $i,j =1,2$. We can always assume $e_i(\psi_i)=\langle\psi_i,\psi_i\rangle=1$ for $i=1,2$ without losing generality, if not, we can just define a new normalized inner product. Since $u(\psi)=1,\forall\psi\in\St(\rA)$, it is trivial that  $(u-e_i)(\psi_j)=1 - \delta_{ij}$. Hence by performing the measurement $\{e_i,u-e_i\}$ one can perfectly discriminate the pair of states. 

Conversely, if the set $\{\psi_1,\psi_2\}$ is perfectly distinguishable, then accordingly to Lemma~\ref{lemma:extension-to-maximal-set} it can be extended to a maximal set of perfectly distinguishable pure state $\{\psi_i\}_{i=1}^d $. By Lemma~\eqref{lemma:invariant-measurement}, the set of dual effects $\{e_{\psi_i}\}_{i=1}^d $ forms a valid measurement such that $e_{\psi_i}(\psi_j) = \< \psi_i, \psi_j \> = \delta_{ij}$ for every $i,j \in \{1, \dots, d \}$. \qed

\begin{proposition}[Corollary~21 of Ref.~\cite{chiribella2011informational}]
	\label{prop:diagonalization-vector-states}
	For every $\omega \in \St_\R(\rA)$ there exists a maximal set of perfectly distinguishable pure states $\{\psi_i\}_{i=1}^d$ and a set of real numbers $\{\lambda_i\}_{i=1}^d$ such that $\omega = \sum_{i=1}^{d} \lambda_i \, \psi_i$.
\end{proposition}
A direct consequence of strong self-duality and Proposition~\ref{prop:diagonalization-vector-states} is the following.
\begin{proposition}
	\label{prop:diagonalization-vector-effects}
	For every $X  \in \Eff_\R (\rA)$ there exists a maximal set of perfectly distinguishable pure states $\{\psi_i\}_{i=1}^d$ and a set of real numbers $\{\lambda_i\}_{i=1}^d$ such that $X = \sum_{i=1}^{d} \lambda_i \, e_{\psi_i}$.
\end{proposition}

\subsection{Implications of purity preservation, strong symmetry, and classical decomposability}
\label{sec:compression}

From the combination of strong symmetry and purity preservation, we show that there exist physical transformations that act as \textit{filters}. Intuitively, a filter is a projection from state space to one of its faces that satisfies some additional requirements. While we will explore in the following the precise mathematical definition of filter, we want to remark since the beginning what we mean here with physical. Indeed, these filters we will construct are not mere mathematical functions, but rather they are transformations of the theory, \textit{i.e.}, potentially physically implementable by an agent with adequate resources.

We start by providing the definition of face of the state space and some preliminaries properties.

\begin{definition}[Face]
	A face $F$ of $\St_+(\rA)$ ($\St_1(\rA)$) is a convex subset of $\St_+(\rA)$ ($\St_1(\rA)$) such that, for every $\rho \in F$, if $\rho = p \omega_1 + (1-p) \omega_2$, with $p \in (0,1)$, $\omega_1,\omega_2 \in \St_+(\rA)$ ($\St_1(\rA)$), then $\omega_1,\omega_2 \in F$. We define the capacity of the face $F$, $\#(F)$, as the cardinality of a maximal set of perfectly distinguishable pure states in $F$.
\end{definition}
It is worth noticing that the faces of $\St_+(\rA)$ and $\St_1(\rA)$ are in one-to-one correspondence. In particular, if $F$ is a face of $\St_1(\rA)$, then the corresponding face of $\St_+(\rA)$ is the set $\{ \lambda \, \rho ~|~ \rho \in F, \lambda \ge 0 \}$. Vice versa, for $F$ face of $\St_+(\rA)$, the corresponding face in $\St_1(\rA)$ is simply $F \cap \St_1(\rA)$.

Given $S \subseteq \St_+(\rA)$ ($\St_1(\rA)$), we say that the face $F$ is \textit{generated} by $S$ if $F$ is the smallest face of $\St_+(\rA)$ ($\St_1(\rA)$) such that $S \subseteq F$.
\begin{proposition}[Proposition 2 of Ref.~\cite{barnum2014higher}]
	CD and SS imply that every face of $\St_+(\rA)$ is generated by a set of perfectly distinguishable pure states. Moreover, any set of perfectly distinguishable pure states $\{ \psi_1, \dots, \psi_k \} \subseteq F$ generates $F$ if and only if $k = \#(F)$.
\end{proposition}

Given a face $F \subseteq \St_\R(\rA)$, we denote by $F^\perp$ the subspace of $\St_\R(\rA)$ corresponding to the orthogonal complement of $\Span (F)$. Also, given a face $F$, we define $F'\coloneqq F^\perp \cap \St_+(\rA)$.
\begin{proposition}[Proposition 7 of Ref.~\cite{barnum2014higher}]
	\label{prop:face-maximal-set}
	CD and SS imply that for every face $F$ of $\St_+(\rA)$, the set $F'$ is a face of $\St_+(\rA)$ of capacity $d-k$, where $d$ and $k$ are the capacities of $\St(\rA)$ and of $F$, respectively. Furthermore, if $\{ \psi_i \}_{i=1}^{k'}$ with $k' < k$ is a set of perfectly distinguishable pure states that is contained in $F$, then it can be extended to a maximal set of perfectly distinguishable pure states of $F$.
\end{proposition}

The following lemma is a rearrangement of Lemma 11 of Ref.~\cite{barnum2014higher}.
\begin{lemma}
	\label{lemma:unique-det-eff-face}
	CD and SS imply that, for every face $F$ of $\St_+ (\rA)$, there is a unique effect $u_F$ such that $u_F(\rho) = 1$ for every $\rho \in F \cap \St_1(\rA)$ and $u_F(\rho) = 0$ for every $\rho \in F' \cap \St_1(\rA)$. In particular, $u_F$ is the coarse-graining of the pure effects dual to any set of perfectly distinguishable pure states generating $F$.
\end{lemma}
\Proof Let $A$ be a system of cardinality $d$ and $F \subset \St_+ (\rA) $ a face of cardinality $k$. Also, let $\{\psi_i\}_{i=1}^k$ be a set of perfectly distinguishable normalized pure states generating $F$. We define $u_F := \sum_{i=1}^k e_{\psi_i}$, where $e_{\psi_i}$ is the pure effect dual to $\psi_i$. For any state $\rho \in F \cap \St_1(\rA)$, there exist a probability distribution $(p_j)$ and a set of perfectly distinguishable normalized pure states $\{\phi_j\}_{j=1}^k$ in $F \cap \St_1(\rA)$ such that $\rho = \sum_{j=1}^k p_j \phi_j$. By lemma~\ref{lemma:extension-to-maximal-set}, the set $\{\psi_i\}_{i=1}^k$ can be extended to a maximal set of perfectly distinguishable pure state of system $\rA$, say $\{\psi_i\}_{i=1}^k \cup \{\psi_i\}_{i=k+1}^d$. Since every element of the set $\{\psi_i\}_{i=k+1}^d$ is in the orthogonal complement of $F$, every element of $\{\psi_i\}_{i=k+1}^d$ is also orthogonal to $\rho \in F$. Then
\begin{equation*}
	1 = u(\rho) = \sum_{i=1}^d e_{\psi_i}(\rho) = \sum_{i=1}^k  e_{\psi_i}(\rho) = u_F(\rho) \, .
\end{equation*}
Now, let $\{\phi_j\}_{j=1}^k$ be another set of perfectly distinguishable normalized pure states generating $F$ and let $\{e_{\phi_j}\}_{j=1}^k$ be the set of dual pure effects. We denote by $u'_F := \sum_{j=1}^k e_{\phi_j}$, by $u_{F'} = u - u_F = \sum_{i=k+1}^d e_{\psi_i}$, and by $u'_{F'} = u - u'_F = \sum_{j=k+1}^d e_{\phi_j}$, where $\{\phi_j\}_{j=k+1}^d$ is the extension from proposition~\ref{prop:face-maximal-set} to $\{\phi_j\}_{j=1}^k$. By strong symmetry, let $\tT$ be the reversible transformation mapping $\phi_j$ to $\psi_i$ for $i=1, \dots, d$, we have
\begin{equation*}
	u_F + u_{F'} = u = \tT u = \tT u_F + \tT u_{F'} = u'_F +  u'_{F'} \, .
\end{equation*}
Since $u_F$ and $u'_{F'}$ are orthogonal to each other, and similarly for $u_{F'}$ and $u'_F$, necessarily $u_F = u'_F$. \qed

\medskip

For any face $F$ of $\St_+(\rA)$, we call $u_F$ the deterministic effect of the face $F$. We now introduce another useful result.

\begin{lemma}
	\label{lemma:iff-condition-for-a-face}
	Let $\rA$ be a system of cardinality $d$ and $F\subset \St_+(\rA)$ a face of cardinality $k < d$. Let $u_F$ be the deterministic effect of $F$, then, for every $\rho \in \St_1(\rA)$, 
	\begin{equation*}
		\rho \in F \cap \St_1(\rA) \, \, \Leftrightarrow \, \, u_F(\rho) = 1 \, .
	\end{equation*}
\end{lemma}

\Proof
\begin{itemize}
	\item[$\Rightarrow)$] We already proved this direction in lemma~\ref{lemma:unique-det-eff-face}.
	\item[$\Leftarrow)$] Let $\rho \in \St_1(\rA)$, $u_F( \rho ) =1$ and by contradiction let us assume that $\rho \notin F$. Then it exists at least one pure state $\phi \in F^\perp \cap \St_1(\rA)$ such that $\<\phi, \rho \> > 0$. Since $\phi \in F^\perp$, for every $\{ \psi_i \}_{i=1}^k$ maximal set of perfectly distinguishable normalized pure states of $F$, $\< \phi, \psi_i\> =0$ for every $i=1,\dots,k$, therefore $\{\psi_1,\dots,\psi_k,\phi\}$ is a set of perfectly distinguishable normalized pure states that can be extended to a maximal one, namely $\{ \psi_1, \dots, \psi_k, \phi \} \cup \{ \psi_{k+2}, \dots, \psi_d\} $. According to Lemma~\ref{lemma:invariant-measurement}, the corresponding dual effects form a measurement $\{e_{\psi_1}, \dots, e_{\psi_k}, e_\phi , e_{\psi_{k+2}}, \dots, e_{\psi_d}  \}$. Finally,
	\begin{equation*}
		u(\rho) = u_F(\rho) + e_\phi(\rho) + \sum_{i=k+2}^d e_{\psi_i}  (\rho) = 1+ \< \phi, \rho \> + \sum_{i=k+2}^d e_{\psi_i}( \rho ) > 1 \, .
	\end{equation*}
\end{itemize}
\qed

\medskip

Now, we devote our attention to the characterization of faces of multi-partite systems.
\begin{lemma}
	\label{lemma:face-multipartite-system}
	Let $\rA$ and $\rA'$ be systems of cardinality $d$ and $d'$, respectively, and let $\{\psi_i\}_{i=1}^{d}$ and $\{\psi'_i\}_{i=1}^{d'}$ be two maximal sets of perfectly distinguishable pure states of $\St(\rA)$ and $\St(\rA')$, respectively. Also, let $F$ be the face of $\St_+(\rA)$ generated by $\{\psi_i\}_{i=1}^{k}$. Then the set $\{\psi_i \otimes \psi_0'\}_{i=1}^{k}$ generates the face $F \otimes \{\psi'_0\}$ in $\St_+(\rA\rA')$.
\end{lemma}
\Proof Let us name $\tilde{F}$ the face generated by $\{\psi_i \otimes \psi_0'\}_{i=1}^{k}$. Obviously, $F \otimes \{\psi'_0\} \subseteq \tilde{F}$ and also $\psi_i \otimes \psi_0' \in F \otimes \{\psi'_0\}$ for every $i = 1, \dots, k$. Since $\tilde{F}$ is the minimal face including the elements  $\{\psi_i \otimes \psi_0'\}_{i=1}^{k}$, to prove the equivalence between $F \otimes \{\psi'_0\} $ and $ \tilde{F}$ it is only left to show that $F \otimes \{\psi'_0\} $ is a face.
Let $\rho \otimes \psi'_0$ an arbitrary normalized state in $ F \otimes \{\psi'_0\}$ such that $\rho \otimes \psi'_0 = p \sigma + (1-p) \sigma_C$, for $p \in (0,1)$ and $\sigma, \sigma_C \in \St_1(\rA\rA')$ and, without loss of generality, let $\sigma$ be pure. Then $\rho = (\tI \otimes e_{\psi'_0}) \rho \otimes \psi'_0 = p \sigma^* + (1-p) \sigma^*_C$, where we defined $\sigma^* := (\tI \otimes e_{\psi'_0}) \sigma$ and $ \sigma^*_C := (\tI \otimes e_{\psi'_0}) \sigma_C$. By purity preservation, $\sigma^*$ is pure. Also, $1 = u(\rho) = p \, u(\sigma^*) + (1-p) \, u(\sigma^*_C)$ implies that $u(\sigma^*) = 1$. Therefore, by self-duality, it exist a pure effect $e_{\sigma^*}$ such that $(e_{\sigma^*}  \otimes e_{\psi'_0} ) \sigma = 1$. By uniqueness of self duality, $\sigma = \sigma^* \otimes \psi'_0$ and therefore $\sigma \in F \otimes \{\psi'_0\} $. \qed

\medskip

\begin{definition}[Projection, Definition~13 of Ref.~\cite{barnum2014higher}]
	Let $\rA$ be a system with cone $\St_+(\rA)$. Projections are linear operators $\Pi: \St_\R(\rA) \to \St_\R(\rA)$ with $\Pi^2 = \Pi$. A projection is positive if for every $\rho \in \St_+(\rA)$, $\Pi(\rho) \in \St_+(\rA)$. Positive projections $\Pi$ and $\Pi_C$ are called complementary if $\Im \Pi \cap \St_+(\rA) = \ker \Pi_C \cap \St_+(\rA)$ and vice versa. A positive projection $\Pi$ is complemented if there exists a positive projection $\Pi_C$ such that $\Pi$ and $\Pi_C$ are complementary.
\end{definition}

\begin{definition}[Filters and projectivity, Definition~14 of Ref.~\cite{barnum2014higher}]
	A filter is a positive projection $\Pi: \St_\R(\rA) \to \St_\R(\rA)$ which is
	\begin{enumerate}[i)]
		\item complemented,
		\item has a complemented dual $\Pi^*$, and
		\item is normalized, \textit{i.e.}, it satisfies $u_\rA \circ \Pi (\rho) \le u_\rA \circ \rho$ for all $\rho \in \St_+(\rA)$.
	\end{enumerate}
	The space $\St_\R(\rA)$ is called projective if every face of $\St_+(\rA)$ is the positive part, $\Im \Pi \cap \St_+(\rA)$, of the image of a filter $\Pi$.
\end{definition}

We now show that the combination of C, CD, PP, and SS implies that the state space is projective. Namely, for every face $F$ of the positive cone of states $\St_+(\rA)$, it exists a filter whose positive image is indeed $F$. Remarkably, such filter is an element of $\Transf(\rA)$, therefore a physical transformation of the theory. 

\begin{proposition}[Existence of filters and projectivity]
	\label{prop:projectivity}
	C, CD, PP, and SS imply that the for any system $\rA$ the space $\St_\R$ is projective. Namely, for every face $F$ of $\St_+(\rA)$, it exists a filter $\Pi_F$ (that is also a physical transformation, \textit{i.e.}, $\Pi_F \in \Transf(\rA)$) for which $F = \Im\Pi_F \cap \St_+(\rA)$.
\end{proposition}
\Proof The proof works constructively. We explicitly build a physical transformation and we check that \textit{i)} it is a positive projection and that \textit{ii)} it is also a filter.

The statement is trivial for $F$ being the whole cone of positive states, then $\Pi_F\equiv I$. Let then $F$ be a non-trivial face of $\St_+(\rA)$, $\{\psi_i\}_{i=1}^k$ a set of perfectly distinguishable normalized pure states generating $F$ and let $\{\psi_i\}_{i=1}^d$ be its extension to a maximal set of perfectly distinguishable normalized pure states of $\St_1(\rA)$, with $k < d$.

Let also $\rA'$ be another system of capacity greater or equal to 2, and let $\{\psi'_0, \psi'_1\}$ be a maximal set of perfectly distinguishable normalized pure states of $\St_1(\rA')$. We consider a reversible transformation $\tS_F \in\Transf(\rA\otimes \rA')$ (whose existence is granted by strong symmetry) that maps the set of perfectly distinguishable normalized pure states $\{ \psi_i \otimes \psi_j'\}_{(i,j) \in \{1,\dots,d\} \times \{0,1\}}$ to itself as follows
\begin{equation}
	\label{eq:physical-equiv-map-for-compression}
	\tS_F:
	\begin{cases}
		\psi_i \otimes \psi'_0 \mapsto \psi_i \otimes \psi_0'  &\quad  \text{ for every } i = 1 ,\dots, k  \\
		\psi_i \otimes \psi'_0 \mapsto \psi_i \otimes \psi_1'  &\quad  \text{ for every } i = k+1, \dots, d  \\
		\psi_i \otimes \psi'_1 \mapsto \psi_i \otimes \psi_1'  &\quad  \text{ for every } i = 1, \dots, k  \\
		\psi_i \otimes \psi'_1 \mapsto \psi_i \otimes \psi_0'  &\quad  \text{ for every } i = k+1, \dots, d  \\
	\end{cases} \, .
\end{equation}
As usual, given a pure state $\phi$, we denote by $e_\phi$ the corresponding unique pure effect from self duality. We then define the map $\Pi_F:\St(\rA) \to \St(\rA)$ as follows
\begin{equation}
	\label{eq:filter}
	\Pi_F :=
	\begin{aligned}
		\Qcircuit @C=1.2em @R=.8em @!R {   &    
			&
			\qw\poloFantasmaCn{\rA}&
			\multigate{1}{\tS_F}&
			\qw&
			\qw\poloFantasmaCn{\rA}&
			\qw&
			\qw& 
			\multigate{1}{\tS_F^{-1}}&
			\qw\poloFantasmaCn{\rA}&
			\qw \\
			&
			\prepareC{\psi'_0}&
			\qw\poloFantasmaCn{\rA'}&
			\ghost{\tS_F}&
			\qw\poloFantasmaCn{\rA'}&
			\measureD{e_{\psi'_0}}&
			\prepareC{\psi'_0}&
			\qw\poloFantasmaCn{\rA'}&
			\ghost{\tS_F^{-1}}&
			\qw\poloFantasmaCn{\rA'}&
			\measureD{e_{\psi'_0}}\\
		} 
	\end{aligned}\, .
\end{equation}
Clearly, $\Pi_F \in \Transf(\rA)$ by construction. In particular, the fact the $\Pi_F$ is a physical transformation trivially implies that \textit{i)} $\Pi_F$ is linear on $\St_\R(\rA)$ and can therefore be naturally extended to a map $\Pi_F : \St_\R(\rA) \to \St_\R(\rA)$, and \textit{ii)} $\Pi_F(\rho) \in \St_+(\rA)$ for any $\rho \in \St_+(\rA)$.

To conclude that $\Pi_F$ is a positive projection it is left to show that $\Pi_F^2 = \Pi_F$. Se start by proving that for any $\rho \in \St_+(\rA)$, $\Pi_F(\rho) \in F$. The first block composing $\Pi_F$ in Eq.~\eqref{eq:filter}, namely the transformation
\begin{equation*}
	\begin{aligned}
		\Qcircuit @C=1.2em @R=.8em @!R {   &    
			&
			\qw\poloFantasmaCn{\rA}&
			\multigate{1}{\tS_F}&
			\qw\poloFantasmaCn{\rA}&
			\qw& \\
			&
			\prepareC{\psi'_0}&
			\qw\poloFantasmaCn{\rA'}&
			\ghost{\tS_F}&
			\qw\poloFantasmaCn{\rA'}&
			\measureD{e_{\psi'_0}}&
		}
	\end{aligned} \, ,
\end{equation*}
maps elements of $\St_+(\rA)$ to $F$. This fact is a direct consequence of Lemma~\ref{lemma:iff-condition-for-a-face} and it becomes evident by writing the deterministic effect in terms of the maximal set $\{\psi_i\}_{i=1}^d$, $u = \sum_{i=1}^d e_{\psi_i}$, and by decomposing the deterministic effect of the face $F$ as $u_F = \sum_{i=1}^k e_{\psi_i}$ (cf. Lemma~\ref{lemma:unique-det-eff-face}), then by the action of $\tS_F$ as defined in Eq.~\eqref{eq:physical-equiv-map-for-compression}
\begin{equation}
	\label{eq:deterministic-effect}
	\begin{aligned}
		\Qcircuit @C=1.2em @R=.8em @!R {   &    
			&
			\qw\poloFantasmaCn{\rA}&
			\multigate{1}{\tS_F}&
			\qw\poloFantasmaCn{\rA}&
			\measureD{u}& \\
			&
			\prepareC{\psi'_0}&
			\qw\poloFantasmaCn{\rA'}&
			\ghost{\tS_F}&
			\qw\poloFantasmaCn{\rA'}&
			\measureD{e_{\psi'_0}}&
		}
	\end{aligned}
	= 
	\begin{aligned}
		\Qcircuit @C=1.2em @R=.8em @!R {   &    
			&
			\qw\poloFantasmaCn{\rA}&
			\multigate{1}{\tS_F}&
			\qw\poloFantasmaCn{\rA}&
			\measureD{u_F}& \\
			&
			\prepareC{\psi'_0}&
			\qw\poloFantasmaCn{\rA'}&
			\ghost{\tS_F}&
			\qw\poloFantasmaCn{\rA'}&
			\measureD{e_{\psi'_0}}&
		} 
	\end{aligned} \, .
\end{equation}
Analogously, the transformation
\begin{equation*}
	\begin{aligned}
		\Qcircuit @C=1.2em @R=.8em @!R {   &    
			&
			\qw\poloFantasmaCn{\rA}&
			\multigate{1}{\tS_F^{-1}}&
			\qw\poloFantasmaCn{\rA}&
			\qw& \\
			&
			\prepareC{\psi'_0}&
			\qw\poloFantasmaCn{\rA'}&
			\ghost{\tS_F^{-1}}&
			\qw\poloFantasmaCn{\rA'}&
			\measureD{e_{\psi'_0}}&
		}
	\end{aligned}
\end{equation*}
must send arbitrary element of $\St_+(\rA)$ to $F$.

We now get back to prove $\Pi_F^2 = \Pi_F$. Eq.~\eqref{eq:deterministic-effect} can be further expanded as follows
\begin{equation}
	\begin{aligned}
		\Qcircuit @C=1.2em @R=.8em @!R {   &    
			&
			\qw\poloFantasmaCn{\rA}&
			\multigate{1}{\tS_F}&
			\qw\poloFantasmaCn{\rA}&
			\measureD{u}& \\
			&
			\prepareC{\psi'_0}&
			\qw\poloFantasmaCn{\rA'}&
			\ghost{\tS_F}&
			\qw\poloFantasmaCn{\rA'}&
			\measureD{e_{\psi'_0}}&
		}
	\end{aligned}
	= 
	\begin{aligned}
		\Qcircuit @C=1.2em @R=.8em @!R {   &    
			&
			\qw\poloFantasmaCn{\rA}&
			\multigate{1}{\tS_F}&
			\qw\poloFantasmaCn{\rA}&
			\measureD{u_F}& \\
			&
			\prepareC{\psi'_0}&
			\qw\poloFantasmaCn{\rA'}&
			\ghost{\tS_F}&
			\qw\poloFantasmaCn{\rA'}&
			\measureD{e_{\psi'_0}}&
		} 
	\end{aligned}
	= 
	\begin{aligned}
		\Qcircuit @C=1.2em @R=.8em @!R {   &    
			&
			\qw\poloFantasmaCn{\rA}&
			\measureD{u_F}& \\
			&
			\prepareC{\psi'_0}&
			\qw\poloFantasmaCn{\rA'}&
			\measureD{e_{\psi'_0}}&
		} 
	\end{aligned}  \, ,
\end{equation}
meaning that, if $\phi \in F$, then $\tS_F (\phi \otimes \psi'_0) \in F \otimes \{ \psi'_0\}$, the face of which $u_F\otimes e_{\psi'_0}$ is the deterministic effect. By Lemma~\ref{lemma:face-multipartite-system},
\begin{equation}
	\begin{aligned}
		\Qcircuit @C=1.2em @R=.8em @!R {   &    
			\prepareC{\phi}&
			\qw\poloFantasmaCn{\rA}&
			\multigate{1}{\tS_F}&
			\qw\poloFantasmaCn{\rA}& \\
			&
			\prepareC{\psi'_0}&
			\qw\poloFantasmaCn{\rA'}&
			\ghost{\tS_F}&
			\qw\poloFantasmaCn{\rA'}&
		}
	\end{aligned}
	= 
	\begin{aligned}
		\Qcircuit @C=1.2em @R=.8em @!R {   &    
			\prepareC{\phi'}&
			\qw\poloFantasmaCn{\rA}& \\
			&
			\prepareC{\psi'_0}&
			\qw\poloFantasmaCn{\rA'}&
		} 
	\end{aligned} \, ,
\end{equation}
for some $\phi' \in F$. At this point, the following part of the circuit of Eq.~\eqref{eq:filter}, $\tI_\rA \otimes (\psi'_0 \circ e_{\psi'_0} )$, acts as the identity on $\phi' \otimes \psi'_0$ and the last part $(\tI_\rA \otimes e_{\psi'_0})\tS^{-1}$ brings the state back to the original input state $\phi$. In conclusion, we just proved that if $\phi \in F$, then $\Pi_F(\phi)=\phi$. This property, in combination with the fact proved above that $\Pi_F(\rho)\in F$ for any $\rho \in \St_+(\rA)$ not only implies that $\Pi_F^2 = \Pi_F$, but also that $\Im \Pi_F \cap \St_+(\rA) = F$.

To show that the positive projection $\Pi_F$ is also a filter we start by showing that it is complemented. Let $F_C$ be the face of system $\rA$ generated by $\{\psi_{k+1}, \dots, \psi_d\}$. By proposition~\ref{prop:face-maximal-set}, $F_C = F^\perp \cap \St_+(\rA)$, where $F^\perp$ is the orthogonal complement of $F$. Analogously to $\Pi_F$ for $F$, for $F_C$ we can consider a reversible transformation $\tS_C \in \grp{\rA\rA'}$ such that
\begin{equation}
	\tS_C:
	\begin{cases}
		\psi_i \otimes \psi'_0 \mapsto \psi_i \otimes \psi_1'  &\quad  \text{ for every } i = 1 ,\dots, k  \\
		\psi_i \otimes \psi'_0 \mapsto \psi_i \otimes \psi_0'  &\quad  \text{ for every } i = k+1, \dots, d  \\
		\psi_i \otimes \psi'_1 \mapsto \psi_i \otimes \psi_0'  &\quad  \text{ for every } i = 1, \dots, k  \\
		\psi_i \otimes \psi'_1 \mapsto \psi_i \otimes \psi_1'  &\quad  \text{ for every } i = k+1, \dots, d  \\
	\end{cases} \, .
\end{equation}
and its associated positive projection
\begin{equation*}
	\Pi_C :=
	\begin{aligned}
		\Qcircuit @C=1.2em @R=.8em @!R { &    
			&
			\qw\poloFantasmaCn{\rA}&
			\multigate{1}{\tS_C}&
			\qw&
			\qw\poloFantasmaCn{\rA}&
			\qw&
			\qw& 
			\multigate{1}{\tS_C^{-1}}&
			\qw\poloFantasmaCn{\rA}&
			\qw \\
			&
			\prepareC{\psi'_1}&
			\qw\poloFantasmaCn{\rA'}&
			\ghost{\tS_C}&
			\qw\poloFantasmaCn{\rA'}&
			\measureD{e_{\psi'_1}}&
			\prepareC{\psi'_1}&
			\qw\poloFantasmaCn{\rA'}&
			\ghost{\tS_C^{-1}}&
			\qw\poloFantasmaCn{\rA'}&
			\measureD{e_{\psi'_1}}\\
		}
	\end{aligned} \, .
\end{equation*}
Then it is straightforward to check that $\Im(\Pi_C) \cap \St_+(\rA) = F_C =  \ker(\Pi_F)\cap \St_+(\rA)$ and that $\Im(\Pi_F) \cap \St_+(\rA) = F = \ker(\Pi_C)\cap \St_+(\rA)$. Furthermore, by strong self-duality, the property of being complemented immediately translates to the dual projectors $\Pi_F^*$ and $\Pi_C^*$, since $\Pi_F^*=\Pi_F$ and $\Pi_C^*=\Pi_C$. Property \textit{iii)} in the definition of a filter is automatically checked since $\Pi_F \in \Transf(\rA)$.

Since we already proved that $\Pi_F$ is surjective on $F$, given the arbitrariness of $F$, we conclude that $\St_\R(\rA)$ is projective. \qed

We are finally in position to state the last result of this section, it will be a useful corollary in the derivation of the speed bounds.
\begin{corollary}
	\label{cor:compression}
	For every couple of pure states $\phi, \psi$ of system $\rA$, there exists a face $F\subset\St(\rA)$ of capacity 2 such that $\phi,\psi \in F$.
\end{corollary}
\Proof
If $\< \phi , \psi \> = 0$, the thesis trivially follows by considering the face of $\rA$ generated by $\{\phi, \psi \}$. If not, let $\{\psi, \psi_1, \dots, \psi_{d-1}\}$ be a maximal set of perfectly distinguishable pure state. Let $\Pi$ the compression map on the face generated by $\{\psi_1, \dots, \psi_{d-1}\}$. Then let us consider the state 
\begin{equation}
	\psi^\perp \coloneqq \frac{\Pi \phi}{u (\Pi \phi) } \, .
\end{equation}
In particular,
\begin{equation*}
	u(\Pi(\phi)) = \sum_{i=1}^{d-1} \< \psi_i , \Pi \phi \> + \< \psi ,  \Pi \phi \> = \sum_{i=1}^{d-1} \< \Pi \psi_i ,  \phi \> + \< \Pi \psi ,  \phi \> = \sum_{i=1}^{d-1} \< \psi_i ,  \phi \> = 1- \< \psi , \phi \> \eqqcolon 1-p \, ,
\end{equation*}
where in the last step we defined $p \coloneqq \< \psi , \phi \> $. Since $\psi^\perp$ is a pure state, also
\begin{equation*}
	1 = \< \psi^\perp , \psi^\perp \> = \frac{1}{(1-p)^2}  \< \Pi \phi , \Pi \phi \> = \frac{1}{(1-p)^2} \< \phi , \Pi \phi \> \, ,
\end{equation*}
from which we get the relation
\begin{equation}
	\label{eq:phi-phi-perp}
	\< \phi , \Pi \phi \> = (1-p)^2 \, .
\end{equation}
By construction, $\psi^\perp$ is a pure state orthogonal to $\psi$. Furthermore,
\begin{equation}
	\< \psi , \phi \> + \< \psi^\perp , \phi \> = p +  \frac{1}{1-p} \< \Pi \phi , \phi \> = p + (1-p) = 1 \, ,
\end{equation}
where in the one before the last step we inserted Eq.~\eqref{eq:phi-phi-perp}. In conclusion, by Lemma~\ref{lemma:iff-condition-for-a-face}, $\phi$ lies in the 2-capacity face generated by $\{\psi, \psi^\perp\}$. \qed

\subsection{Reconstructing quantum theory}

Here, we derive the structure of quantum theory on finite-dimensional Hilbert space from our five axioms of section~\ref{sec:assumptions}. To start, we show that the combination of C, CD,  PP, and SS  implies that the states of the theory form a irreducible Euclidean Jordan algebra \cite{jordan1934algebraic, barnum2020composites}.

\begin{proposition}
	\label{prop:eja}
	C, CD,  PP, and SS  implies that the states of the theory form a irreducible Euclidean Jordan algebra.
\end{proposition}
\Proof The proof is an adaptation of the proofs of Theorem~16 of Ref.~\cite{barnum2014higher} and/or of Theorem~8 of Ref.~\cite{barnum2017ruling}. In both cases, the respective authors refer to results of Alfsen and Shultz. In particular, Theorem~9.33 in Ref.~\cite{alfsen2012geometry} implies that finite-dimensional systems that satisfy strong self-duality, that are projective, and whose filters preserve purity, have state spaces affinely isomorphic to the state spaces of Euclidean Jordan algebras. By Proposition~\ref{prop:self-duality}, every system is self-dual. Also, by Proposition~\ref{prop:projectivity}, every system space is projective and the filters associated to every face that provide the projectivity condition also preserve purity, since the filters we have constructed are pure and the composition of pure transformations is pure (PP assumption). \qed 

\medskip

This result  implies in particular that every theory satisfying  C, CD, PP, and SS also satisfies strong self-duality and UIS.  We can combine this fact with Theorem \ref{thm:energy-obs}, where we showed that DIN, UIS, and strong self-duality imply the observable-generator duality, to leverage a result of Barnum,   M\"uller, and Ududec  (cf. Theorem~31 of Ref.~\cite{barnum2014higher}), who showed that irreducible Jordan algebras plus observable-generator duality  imply quantum theory on complex Hilbert spaces. In conclusion, the combination of Proposition~\ref{prop:eja} and the result of Ref.~\cite{barnum2014higher} implies that in a theory satisfying C, CD,  PP, SS, and DIN, every system is a simple complex matrix algebra.

\section{Speed bound}

\subsection{Energy observable}
\label{sec:energy-observable}
In principle, the well-posedeness of  Definition~\ref{def:energy-obs} only relies on two notions: a collision model and the self-duality property. As already discussed, collision models can be defined in any possible physical theory with a continuous dynamics without any other additional assumption. A collision model is sufficient to build a state-generator correspondence, even if not injective in general. For what concern self-duality, its role is simply to extend this correspondence to observables.
In theories satisfying DIN, CD, and SS, not only the energy observables are in a one-to-one correspondence with the generators of the collisional dynamics, but every energy observable $h$ can also be measured in a canonical way, by performing an ideal measurement described by pure effects (cf. Proposition~\ref{prop:diagonalization-vector-effects}). Explicitly, the energy observable of Definition~\ref{def:energy-obs} can be diagonalized as
\begin{equation}
	\label{eq:h}
	h =  \sum_{i=1}^d  \, x_i\, e_{\psi_i} \, ,
\end{equation}
for a set of real numbers $\{x_i\}_{i=1}^d$ and $\{\psi_i\}_{i=1}^d$ maximal set of perfectly distinguishable normalized pure states. 

However, a sound notion of energy observable should capture the two key aspects of the dynamics: (1) the trajectories for all possible initial states, and (2) the rate at which such trajectories are traveled. Referring to Eq.~\eqref{eq:h}, the state space trajectories are uniquely determined by the ratios between the different values of the coefficients $x_i$-s, while the rate depends (but is not only specified) by their absolute value. Indeed, the rate of the collisional dynamics $U_{t,h} = e^{G_h t}$ is not described by $h$ alone, but it also depends on the particular collision model chosen, \textit{i.e.} on the particular subgroup $\map S_t$ governing the collisions.

Therefore, to take into account the magnitude of both the functional $h$ of the collisional dynamics $g= G_h$ and of the generator of the collision model $G_\text{TOT}= \frac{\d \map S_t}{\d t} \big|_{t=0}$, the rate of the collisional dynamics can be properly expressed in terms of the maximum singular value of the generator $G_h = \frac{\d U_{t,h}}{\d t} \big|_{t=0}$, which corresponds to the maximum rate for the change of vectors under the linear map $U_{t,h} = e^{G_h t}$.

In conclusion, we update the notion of energy observable by defining the \textit{canonical energy observable} associated to a generator $G_h \in \mathfrak{g}_\rA$ to be the linear functional
\begin{equation}
	\label{new-H}
	H := \lambda_{\max}(G_h) \sum_{i=1}^d  \, x_i\, e_{\psi_i} \, ,
\end{equation}
where $\lambda_{\max}(G_h)$ is the maximum singular values of $G_h$ and $h = \sum_{i=1}^d  \, x_i\, e_{\psi_i} \in \map O_\rA$ is the unique observable corresponding to $G_h$ via Theorem~\ref{thm:energy-obs}. As mentioned above, the pure effects $\{e_{\psi_i}\}_{i=1}^d$ form  a measurement, viz. Lemma~\ref{lemma:invariant-measurement}, which we interpret as the ideal energy measurement, with outcomes $\{1,\dots,  d\}$ associated to the possible energy values  $E_i :=\lambda_{\max}(G_h) \cdot x_i$, for $i=1,\dots,d$, respectively. Eq.~(\ref{new-H}) gives a canonical way to estimate the expectation value of the energy for every possible input state.

\subsection{Representation of the group of reversible transformations}

In Proposition~\ref{prop:lie-group} we proved that, for every system $\rA$, the group of reversible transformation is a Lie group. Therefore, reversible transformations can be represented by orthogonal matrices from $\St_\R(\rA)$ to itself. Furthermore, under the universality assumption of the collision model, every reversible transformation is connected to the identity by some one-parameter subgroup of $\grp{G}_\rA$. This implies that the representation of the reversible transformations is in fact the group, or a subgroup thereof, of the special orthogonal matrix group, the connected part of the orthogonal group. In turn, the Lie algebra $\mathfrak{g}_\rA$ of the generator of the dynamics will the be the algebra, or a subalgebra thereof, of the real skew-symmetric matrices.

\subsection{Skew-symmetric matrix decomposition}
\label{sec:maximum-singular-value}

We start by proving a technical lemma that will also be directly useful for the derivation of the speed bound.
\begin{lemma}
	\label{lemma:skew-sum}
	For any real skew-symmetric matrix $S$, there exist two real orthogonal matrices $Q_1$ and $Q_2$ and a positive number $c \ge \lambda_{\max}(S)/2$, such that $S = c(Q_1 + Q_2)$, where $\lambda_{\max}(S)$ is the maximum singular value of $S$. 
\end{lemma}
\Proof Let $S$ be a $N\times N$ real skew-symmetric matrix, namely $S^\top = -S$. We first consider the case of $N$ being even, the odd case being a simple generalization. Any real skew-symmetric matrix $S$ can be written as $S = O \Sigma O^\top$, for $O\in O(N)$, the group of $N\times N$ real orthogonal matrices, and $\Sigma$ a block diagonal matrix of the following form:
\begin{equation}
	\label{eq:skew-symmetric}
	\Sigma = 
	\begin{pmatrix}
		\Lambda_1 & 0  & \dots & 0 \\
		0 & \Lambda_2 &   & \vdots \\
		\vdots &  & \ddots & 0 \\
		0 & \dots & 0 & \Lambda_{N/2}
	\end{pmatrix}
	\quad \text{ where } \Lambda_i = \begin{pmatrix}
		0 & \lambda_i \\
		-\lambda_i & 0 \end{pmatrix}
\end{equation}
for $\lambda_i \ge 0$ for every $i=1,\dots, N/2$. Now, let us consider orthogonal matrices of the following block diagonal form
\begin{equation*}
	Q^{(s)} = 
	\begin{pmatrix}
		R_1^{(s)} & 0  & \dots & 0 \\
		0 & R_2^{(s)} &   & \vdots \\
		\vdots &  & \ddots & 0 \\
		0 & \dots & 0 & R_{N/2}^{(s)}
	\end{pmatrix}
	\quad \text{ where } R_i^{(s)} =  \begin{pmatrix}
		s \cos(\theta_i) &  \sin(\theta_i) \\
		- \sin(\theta_i) & s \cos(\theta_i) \end{pmatrix}
\end{equation*}
for $s=\pm 1$ and $\theta_i \in \mathbb{R}$ for $i=1,\dots,N/2$. The sum of $Q^{(+1)} + Q^{(-1)}$ will then return a block diagonal matrix where the $i$-th $2\times 2$ diagonal block is 
\begin{equation*}
	2 \begin{pmatrix}
		0 &  \sin(\theta_i) \\
		- \sin(\theta_i) & 0 \end{pmatrix} \, .
\end{equation*}
By multiplying the above matrix sum for a positive scalar $c$, we look for a set of coefficients $\{c,\theta_1,\dots,\theta_{N/2} \}$ such that the following system of equations is satisfied
\begin{equation*}
	\begin{cases}
		2c \sin(\theta_1) = \lambda_1 \\
		\vdots \\
		2c \sin(\theta_{N/2}) = \lambda_{N/2} 
	\end{cases} \, .
\end{equation*}
Rewriting the system as $\{ \theta_i = \arcsin{\lambda_i/2c} \}_{i=1}^{N/2}$, a solution can be found by setting $c = \max \{ \lambda_i/2 \}$, such that $0 \le \lambda_i/2c \le 1$ for all $i=1,\dots, N/2$ (note that $\max_{i} \{ \lambda_i \}$ is the maximum singular value of $S$). Finally
\begin{equation*}
	S = O \Sigma O^\top = O c(Q^{(+1)} + Q^{(-1)}) O^\top = c  ( O Q^{(+1)} O^\top+ O Q^{(-1)} O^\top ) \, ,
\end{equation*}
where both $O Q^{(+1)} O^\top$ and $O Q^{(-1)} O^\top$ are again orthogonal matrices.

If $N$ is odd, the matrix $\Sigma$ in Eq.~\eqref{eq:skew-symmetric} has always at least one row and column of zeros, without losing generality we can assume that those are the $N$-th row and the $N$-th column, respectively. Then, to achieve the same result we just need to require that the $N$-th row and column of $Q^{(s)}$ are all zero but the element $Q^{(s)}_{N,N} = s $, for $s = \pm 1$. \qed

\medskip 

We want to remark that such decomposition is not unique. It is particularly clear from the proof of the previous lemma that $S$ can be decomposed as $c (Q_1 + Q_2)$, for any  $c \ge \lambda_{\max}(S)/2$. The value $\lambda_{\max}(S)$ has however a particular significance. Taken the matrix $S$, a suitable basis can be chosen such that $S$ assumes the block-diagonal form of Eq.~\eqref{eq:skew-symmetric}. Then, $\lambda_{\max}(S)$ coincides with the greatest of the coefficients $\lambda_i$ that appear in the block-diagonal form of $S$ (cf. Eq.~\eqref{eq:skew-symmetric}), namely it represents the maximum rate of the evolution $e^{St}$ in the real vector space spanned by the states of the system.

\subsection{Energy invariance under time evolution}

Before proving the main result of this section, we start with two preliminary lemmas.
\begin{lemma}
	\label{lemma:unique-greatest-eigenvalue}
	Given a state $\rho$, let $p^*$ be the greatest eigenvalue of $\rho$ in some diagonalization, then $p^*$ is the greatest eigenvalue in any diagonalization. 
\end{lemma}
\Proof Let $\rho \in \St(\rA)$ and $\rho = \sum_{i=1}^{d} p_i \psi_i$ a diagonalization of $\rho$ by some set of perfectly distinguishable pure states $\{\psi_i\}_{i=1}^d$. Without loss of generality, we can assume that $p_i \ge p_{i+1}$ for $i=1,\dots,d-1$. Now, let $\rho = \sum_{j=1}^{d}p'_j \psi'_j$ another diagonalization of $\rho$ given by the set of perfectly distinguishable pure states $\{\psi'_j\}_{j=1}^d$, again such that $p'_j \ge p'_{j+1}$ for all $j=1,\dots,d-1$. Now, let $p_1 \ge p'_1$ (if not, we can just switch the two diagonalization) and let $e_1$ the unique pure effect corresponding to $\psi_1$. The, we have
\begin{equation}
	\begin{aligned}
		e_1(\rho) &= \sum_{j=1}^{d}p'_j \, e_1(\psi'_j) \\
		p_1 &= \sum_{j=1}^{d}p'_j \, e_1(\psi'_j) \le p'_1 \sum_{j=1}^{d}  e_1(\psi'_j) = p'_1 \, d\, e_1(\chi) = p'_1 \, ,
	\end{aligned}
\end{equation}
where in the last equality on the right hand side we used Lemma~\ref{lemma:maximally-mixed} to write $d \, \chi = \sum_{j=1}^{d} \psi'j$ and  to find that $e_1(\chi)=1/d$. Since by hypothesis it was that $p_1 \ge p'_1$, it must be that $p_1 = p'_1$.	\qed

\medskip

\begin{proposition}
	\label{lemma:unique-diagonalization}
	Given a system $\rA$ of capacity $d$, for any $p \in \left(1/2,1\right]$ and any $\psi_1,\psi_2 \in \Pur\St_1(\rA)$ perfectly distinguishable, the state $\rho = p \psi_1 + (1-p) \psi_2$ does not have other diagonalizations.
\end{proposition}
\Proof  By contradiction, let $\rho = \sum_{j=1}^{d}p'_j \psi'_j$ be another diagonalization of $\rho$ and, without loss of generality, let us assume that $p_j \ge p_{j+1}$ for all $j=1,\dots,d-1$. By Lemma~\ref{lemma:unique-greatest-eigenvalue}, $p'_1 = p$ and since $p > 1/2$, $p'_1>p'_j$ for any $j=2,\dots,d$. Now, let $e_1$ be the unique pure effect corresponding to $\psi_1$. Note that, by Lemma~\ref{lemma:maximally-mixed}, $\sum_{j=1}^d e_1(\psi'_j) = d\, e_1(\chi) = 1$, therefore $\{ e_1(\psi'_j)\}_{j=1}^d$ is a probability distribution. Then
\begin{equation}
	p = e_1(\rho) = p \, e_1(\psi'_1) +\sum_{j=2}^{d} p'_j \, e_1(\psi'_1) \,.
\end{equation}
Since $\sum_{j=1}^{d}  e_1(\psi'_j) =1$ and $p > p'_j$ for all $j=2\dots,d$, it must be that $e_1(\psi'_j) = \delta_{1j}$. By the uniqueness of the state-effect correspondence, $\psi_1 = \psi'_1$. Finally, it also holds that
\begin{equation}
	\psi_2 = \frac{\rho - p\psi_1}{(1-p)} =  \sum_{j=2}^{d}\frac{p'_j}{(1-p)}\psi'_j \,.
\end{equation}
Since $\psi_2$ is pure, it must be that $p'_2 = (1-p)$ and that $\psi'_2 = \psi_2$. \qed

\medskip

\begin{proposition}[Stationarity of the eigenstates]
	\label{prop:stationarity-eigen}
	Given a vector $\sigma = \sum_i  x_i \psi_i \in \St_\R(\rA)$,  its eigenstates $\{\psi_i\}_i$ are invariant under the collisional dynamics $ \tU_{t} = e^{G_\sigma t}$, where $G_\sigma$ is defined as in Eq.~\eqref{eq:state-generator-correspondence}.
\end{proposition}
\Proof  Let $(\tU_{t})_{t \in \R}$ be the collisional dynamics generated by $\sigma \in \St_\R(\rA)$. As a trivial consequence of the stationarity at equilibrium assumption, $ G_\rho \rho =  0$.
Now, for any set of perfectly distinguishable pure states, say $\{ \psi_i\}_{i=1}^d$, it follows that 
\begin{equation}
	\label{eq:invariance-state-on-itself}
	G_{\psi_i} \psi_i = 0 \quad \textnormal{ for every } i = 1, \dots, d,
\end{equation}
or equivalently $ \tU_{\psi_i, t} \psi_i = \psi_i$ for every $t \in \R$, $i = 1, \dots, d$. Let now $\rho = p\,  \psi_1 + (1-p) \, \psi_2$, for $p \in \left(1/2,1\right]$. Again, $ \tU_{\rho, t} \rho = \rho$ and by linearity, $ \rho = p\, \tU_{\rho, t} \psi_1 +  (1-p) \, \tU_{\rho, t} \psi_2$. Since $\tU_{ \rho, t} \psi_1$ and $\tU_{\rho, t} \psi_2$ are pure and orthogonal, by uniqueness of the diagonalization of $\rho$, viz. Lemma~\ref{lemma:unique-diagonalization}, $\tU_{\rho, t} \psi_i = \psi_i$ for $i=1,2$. Equivalently, $G_\rho \psi_i = 0 $ for $i=1,2$. By linearity of the state-generator correspondence function, cf. Eq.~\eqref{eq:state-generator-correspondence}, $G_\rho \psi_i = p\, G_{\psi_1} \psi_i + (1-p)\, G_{\psi_2} \psi_i = 0$ for $i =1,2$, that, combined with Eq.~\eqref{eq:invariance-state-on-itself}, gives us
\begin{equation*}
	\label{eq:invariance-state-on-next-one}
	\begin{aligned}
		G_{\psi_1} \psi_2 = 0 \, , \\
		G_{\psi_2} \psi_1 = 0 \, .
	\end{aligned}
\end{equation*}
By repeating the same argument for every couple of states in $\{ \psi_i\}_{i=1}^d$, we eventually obtain
\begin{equation}
	\label{eq:invariance-orthogonal-states}
	G_{\psi_i} \psi_j = 0 \quad \textnormal{ for every } i,j = 1, \dots, d.
\end{equation}
Since the previous argument works for any set of perfectly distinguishable states, without loss of generality let $\sigma$ be diagonalized as $\sum_{i=1}^n  p_i\, \psi_i$, with $n \le d$. Then, $G_\sigma = \sum_{i=1}^n  p_i \, G_{\psi_i}$. Finally, by Eq.~\ref{eq:invariance-orthogonal-states}, we get that
\begin{equation}
	G_\sigma \psi_j = \sum_{i=1}^n  p_i \, G_{\psi_i} \psi_j = 0 \quad  \textnormal{ for every } j = 1,\dots,d \, ,
\end{equation}
that is equivalent to $\tU_{t, \sigma} \psi_j = \psi_j$ for every $j=1,\dots,d$. \qed

\medskip

Let us consider the generator of a collisional dynamics, $G_\sigma$, and let $H = \sum_i  E_i   e_{\psi_i} = \sum_i  E_i   \< \psi_i , \cdot \>$ be the canonical energy observable associated to the generator $G_\sigma$. Then the expectation value of the observable $H$ for any state $\rho$ is invariant under the time evolution generated by $G_\sigma$:
\begin{equation}
	\label{eq:energy-conservation}
	\begin{aligned}
		\< H \>_{\rho(t)} &= \sum_i  E_i  \< \psi_i , \rho(t) \> = \sum_i E_i  \< \psi_i , e^{G_\sigma t} \rho \> = \\
		&= \sum_i E_i  \<  e^{- G_\sigma t} \psi_i , \rho \> \stackrel{\star}{=} \sum_i E_i  \< \psi_i , \rho \> = \< H \>_\rho \, .
	\end{aligned}
\end{equation}
Above, the equality marked as $\stackrel{\star}{=}$ is a direct consequence of Proposition~\ref{prop:stationarity-eigen}. Not only that, the entire probability distribution of the measurement $\{ e_{\psi_i} \}_{i=1}^d$ is invariant under time evolution. Indeed, for any $i \in \{1, \dots, d\}$
\begin{equation}
	\begin{aligned}
		p(i | \rho(t)) &=  \< \psi_i , \rho(t) \> = \< \psi_i , e^{G_\sigma t}  \rho \> = \\
		&=   \<  e^{- G_\sigma t} \psi_i ,  \rho \> \stackrel{\star}{=} \< \psi_i , \rho \>  = p(i | \rho) \, ,
	\end{aligned}
\end{equation}
where again the equality marked as $\stackrel{\star}{=}$ is a direct consequence of Proposition~\ref{prop:stationarity-eigen}.

\subsection{Evolution speed, notation and preliminary results}

For convenience of the reader, we report here the definition from the main text about the notion of speed in operational probabilistic theories: 
\begin{definition}[Speed of state change]
	Let $(\tU_t)_{t\in\R}$ be a dynamics, let $\{\rho_t = \tU_t  \rho~|~  t\in  \R\}$ be the trajectory of an initial state $\rho$,  and let $t_0$ and $t_1 \ge t_0$ be two moments of time.   The average speed from time $t_0$ to time $t_1$ is  defined as 
	\begin{equation}
		v_\rho  (t_0, t_1)\coloneqq \frac{\| \rho_{t_1}  - \rho_{t_0} \|}{t_1-t_0} \, , 
	\end{equation}
	where $\norm{\cdot}$ is an arbitrary  norm on $\St_\R(\rA)$. Similarly, the instantaneous speed at time $t$  is defined as   $v_\rho  (t)  = \lim_{  \delta t \to 0}    v_\rho  ( t ,  t+  \delta t )$. 		
\end{definition}
In the following we will take the norm $\norm{\cdot}$ to be the the Euclidean norm induced by self-duality, namely $  \|  \sum_{j} \, c_j  \,  \rho_j\|  : =  \sqrt{  \sum_{j,k}   \,  c_j,  c_k   e_{\rho_j} (\rho_k)}$ for every linear combination of  states $(\rho_j)$ with real coefficients $(c_j)$. With this choice of norm,  the instantaneous speed is constant along the trajectory  for every reversible dynamics with time-independent generator  $ \tU_{t}  = e^{G t}$. Indeed, for any $t \in \R$ and any $\rho \in \St(\rA)$ one has
\begin{align*}
	v_\rho  (t)  &= \lim_{  \delta t \to 0}    \frac{\| \rho_{t + \delta t}  - \rho_{t} \|}{\delta t} =  \lim_{  \delta t \to 0}    \frac{\| \tU_{\delta t} \rho_t  - \rho_t \|}{\delta t} = \lim_{  \delta t \to 0}    \frac{\| \rho_t + G \rho_t \delta t + O(\delta t^2) - \rho_t \|}{\delta t} = \\
	&=   \lim_{  \delta t \to 0} \norm{G \rho_t + O(\delta t) } = \norm{G \, \rho_t} = \norm{G \, \tU_{ t} \, \rho} = \norm{\tU_{ t} \, G \, \rho} = \norm{ G \rho} \, ,
\end{align*}
where we used the facts that $\tU_{t}$ is an orthogonal matrix and that it commutes with its generator $G$.

\begin{proposition}[Evolution speed]
	\label{prop:evolution-speed}
	For every reversible dynamics with time-independent generator one has the bound  
	\begin{equation}
		\label{eq:instant-speed}
		v_\rho   (  t_0,  t_1) \le \| G\, \rho \|  \equiv v_\rho (t)  \qquad \forall t_0,t_1,t  \in \R\, .
	\end{equation}
\end{proposition}

\Proof Given the dynamics $\tU_t=e^{G t}$ and a state $\rho \in \St(\rA)$, we have
\begin{align*}
	\|\rho_{t_1} - \rho_{t_0} \|^2 = \norm{ \int_{t_0}^t\frac{\d}{\d s} \tU_s \rho \, \d s }^2 = \norm{ \int_{t_0}^t G \tU_s \rho   \, \d s }^2 = \norm{ \int_{t_0}^t  \tU_s G \rho  \, \d s }^2 \le \\
	\le \left(  \int_{t_0}^t \norm{ \tU_s G \rho  }\, \d s \right)^2 = \left(  \int_{t_0}^t \norm{ G \rho  }\, \d s \right)^2 = (t-t_0)^2 \norm{ G \rho  }^2 \, .
\end{align*}
By rearranging the terms we have the thesis. \qed

\medskip

In the case of collisional dynamics we can refine the result of Proposition~\ref{prop:evolution-speed} even further.

\begin{lemma}
	\label{lemma:speed-bound-eigenstates}
	Given a collisional dynamics $(\tU_{t, h})_{t \in \R}$, let $\{ \psi_1, \dots, \psi_d \}$ ($\{ e_1, \dots, e_d  \}$) be the set of perfectly distinguishable pure states (effects) diagonalizing $h$. Then, the instantaneous speed of any state $\rho $ for any time $t \in \R$ is bounded according to the following relation
	\begin{equation}
		\label{eq:speed-bound}
		v_\rho(t)^2 \equiv \| G_h \rho \|^2 \le \lambda_{\max}^2   \left[ \< \rho, \rho \> - \sum_{k=1}^d \< \rho, \psi_i \>^2 \right]  \, ,
	\end{equation}
	where $\lambda_{\max}$ is the maximum singular value of $G_h$.
\end{lemma}
\Proof Let $G_h = c (Q_1+Q_2)$ with $c = \lambda_{max}/2$, cf. Lemma~\ref{lemma:skew-sum}. We can rewrite it as
\begin{equation*}
	G_\sigma = c W (I + Z),
\end{equation*}
where $I$ is the identity matrix, and the matrices $W=Q_1$ and $Z = W^\top Q_2$ are orthogonal.

Let $h$ be the vector generating $G_h$ via the collision model. Then, according to Proposition~\ref{prop:stationarity-eigen} the eigenstates $\{ \psi_1,\dots,\psi_d \}$ of $h$ are also eigenstates with eigenvalue 0 of the generator $G_h$. That set of $d$ orthogonal vectors can be extended to a basis of the whole vector space $\rA \simeq \mathbb{R}^N$. Let $\{ \psi_1,\dots,\psi_d \} \cup \{ \epsilon_1,\dots,\epsilon_{N-d}\}$ be such basis.

We now prove that $Z$ (and consequently $I+Z$) has a 2-block diagonal form, one acting on the subspace $\Span \{ \psi_1,\dots,\psi_d \}$ and the other on the subspace $\Span \{ \epsilon_1,\dots,\epsilon_{N-d}\}$.

Being $W$ orthogonal, $\ker\{W\}=\{0\}$. On the other hand, $G_h \psi_i = 0$ for all $i=1,\dots,d$. Therefore
\begin{align*}
	(I+Z)\psi_i=& \,0 \quad \forall \, i \\
	Z \psi_i =& -\psi_i \quad \forall \, i \,.
\end{align*}
Namely, the $\psi_i$-s are eigenvectors with eigenvalue $-1$ of the orthogonal matrix $Z$. Denoting with $\{e_j\}_{j=1}^d$ the set of normalized pure effects dual to the eigenvectors $\{ \psi_i\}_{i=1}^d$, we have
\begin{equation*}
	Z = -\sum_{i=1}^d \psi_i \circ e_i \oplus E \, ,
\end{equation*}
with $E \in \text{O}(N-d)$ an orthogonal matrix acting on $\Span \{ \epsilon_1,\dots,\epsilon_{N-d}\}$. For a matter of convenience, we can rewrite $Z$ as follows
\begin{equation*}
	Z =  I_{d\times d} \oplus E - 2 \sum_{i=1}^d \psi_i \circ e_i \, ,
\end{equation*}
where $I_{d\times d}$ is the $d \times d$ identity matrix acting on $\Span \{ \psi_1,\dots,\psi_d \}$, making $I_{d\times d} \oplus E \in \text{O}(N)$.

Now, given an arbitrary state $\rho$, let us consider $\|G_h \rho\|^2$. Since $\norm{G_h \rho}^2 = \<G_h \rho, G_h \rho \> = \< \rho, G_h^\top G_h \rho \> $, we focus on $G_h^\top G_h $:
\begin{align*}
	G_h^\top G_h &= c^2(I+Z^\top) W^\top W (I+Z) \\
	&= c^2 (I+Z^\top) (I+Z) = c^2 (2I+Z+Z^\top) \\
	&= c^2 \left[ 2I_{N \times N} + I_{d\times d} \oplus E + I_{d\times d} \oplus E^\top - 4 \sum_{i=1}^d \psi_i \circ e_i \right]
	\, .
\end{align*}
Since both $I_{d \times d} \oplus E$ and $I_{d \times d} \oplus E^\top$ are orthogonal matrices, they cannot increase the inner product. Indeed, let $O$ be an orthogonal matrix acting on $\mathbb{R}^K$ and $\vec{v} \in \mathbb{R}^K$, then by the Cauchy-Schwarz inequality
\begin{equation*}
	\< \vec{v}, O\vec{v}\> \le \|\vec{v}\| \cdot \| O\vec{v} \| \le \|\vec{v}\| \cdot \| O\| \|\vec{v} \| \le \|\vec{v}\|^2 = \< \vec{v}, \vec{v}\>
\end{equation*}
Therefore, for every state $\rho \in \St(\rA)$, we have 
\begin{equation}
	\label{eq:generator-bound}
	\begin{aligned}
		\< \rho, G_h^\top G_h \rho \> &= c^2\left[ 2\< \rho, \rho \> + \< \rho,  (I_{d\times d} \oplus E) \rho \> +  \< \rho, (I_{d\times d} \oplus E^\top) \rho \> - 4 \< \rho, \sum_{i=1}^d \psi_i \circ e_i (\rho) \> \right] \le \\
		& \le 4c^2 \left[ \< \rho, \rho \>  -  \< \rho, \sum_{i=1}^d \psi_i \circ e_i (\rho) \> \right] = 4c^2 \left[ \< \rho, \rho \>  -   \sum_{i=1}^d \< \rho, \psi_i \>^2 \right] \, .
	\end{aligned}
\end{equation}
Finally, by substituting $c = \lambda_{\max}/2$ in the previous equation we obtain the thesis. \qed

\medskip

\begin{lemma}
	\label{lemma:speed-bound-pure-generator}
	Let $(\tU_{t, \psi})_{t \in \R}$ be the collisional dynamics generated by $\psi \in \Pur\St (\rA)$. The instantaneous speed of any state $\rho $ for any time $t \in \R$ is bounded according to the following relation
	\begin{equation}
		\label{eq:speed-bound-pure-generator}
		v_\rho(t)^2 \equiv \| G_\psi \rho \|^2 \le \lambda_{\max}^2   \left[ \< \rho, \rho \>- \< \rho, D \rho \> \right]
	\end{equation}
	where $D$ is the identity matrix on $\Span \{ \rho \in \St(\rA) | \< \rho , \psi \> = 0 \}$ and $0$ elsewhere.
\end{lemma}
\Proof When the collisional dynamics is generated by a pure state $\psi$, every state that is orthogonal to $\psi$ is invariant under the evolution $(\tU_{t, \psi})_{t \in \R}$. Indeed, let consider $\rho \in \St(\rA)$ such that $\< \rho, \psi \> = 0$. Since $\psi$ is not included in any decomposition of $\rho$, $\rho$ cannot be internal and its diagonalization has at least one null eigenvalues: $\rho = \sum_{i=1}^k p_i \phi_i \in \St(\rA)$, with $k < d$ and $p_i > 0$ for $\{ \phi_1, \dots \phi_k\}$ set of perfectly distinguishable pure states.  Then $\{ \psi, \phi_1 , \dots, \phi_k \}$ is a set of perfectly distinguishable pure states that can be extended to a maximal one. By Proposition~\ref{prop:stationarity-eigen}, $G_\psi \phi_i = 0$ for every $i=1, \dots, k$, thus $G_\psi \rho = 0$. In conclusion $G_\psi \rho = 0$, for every $\rho \in \Span \{ \rho \in \St(\rA) | \< \rho , \psi \> = 0 \}$. In particular, Eq.~\eqref{eq:speed-bound} reads as $\| G_\psi \rho \|^2 \le \lambda_{\max}^2   \left[  \< \rho, \rho \> - \< \rho, D \rho \> \right]$ where $D$ is the identity matrix on $\Span \{ \rho \in \St(\rA) | \< \rho , \psi \> = 0 \}$ and $0$ elsewhere. \qed

\subsection{Proof of Theorem~4}

\begin{theorem}[Speed bound]
	\label{theorem:speed-bound}
	DIN, C, CD, PP, and SS imply that the time $\Delta t$ taken by a system to transition from an initial state $\rho_{t_0}$ to a final state $\rho_{t_1}$ through a reversible dynamics is lower bounded as
	\begin{equation}
		\label{eq:speed-bound-final}
		\Delta t \ge \frac{D(\rho_{t_0}, \rho_{t_1})}{\Delta H} \, ,
	\end{equation}
	where $D(\rho, \rho') := \| \rho - \rho' \|/\sqrt{2}$ is the normalized Euclidean distance between the states $\rho$ and $\rho'$, while $\Delta H : = \sqrt{ \<H^2\>_{\rho} - \< H\>^2_{\rho}}$ is the standard deviation of the canonical energy observable $H$ associated to the dynamics, and $H^2$ is the observable defined as $H^2:=\sum_i E_i^2 e_{\psi_i}$.
\end{theorem}

We prove the above theorem in two parts. Firstly, in Section~\ref{sec:proof-thm3-part1} we prove Theorem~\ref{theorem:speed-bound} for collisional dynamics targeting pure states. Secondly, in Section~\ref{sec:proof-thm3-part2} we generalize the proof to arbitrary target states finally deriving Eq.~\eqref{eq:speed-bound-final}.

\subsubsection{Proof of theorem~4: evolution of pure states}
\label{sec:proof-thm3-part1}

Let $\sigma \in \St_\R(\rA)$ diagonalized as $\sigma = \sum_{i=1}^d x_i\psi_i$ and let $G_\sigma$ be the generator of the collisional dynamics generated by $\sigma$. Also, let $\phi_1 \in \Pur(\rA)$ be the target state undergoing the evolution generated by $G_\sigma$. Lemma~\ref{lemma:speed-bound-pure-generator} gives
\begin{equation*}
	\| G_\sigma \phi_1 \|^2 = \| G_{\phi_1} \sigma \|^2 = \< G_{\phi_1} \sigma , G_{\phi_1} \sigma \> = \< \sigma , G_{\phi_1}^\top G_{\phi_1} \sigma \> \le \lambda_{\max}^2   \left[  \< \sigma, \sigma \> - \< \sigma, D \sigma \> \right]
\end{equation*}
where $D$ is the identity matrix on $\Span \{ \rho \in \St(\rA) | \< \rho , \phi_1 \> = 0 \}$ and $0$ elsewhere.
In particular, for any set of perfectly distinguishable pure states extending $\{\phi_1 \}$, namely $\{\phi_1 \} \cup \{ \phi_2, \dots, \phi_d \}$ we have
\begin{equation}
	\label{eq:speed-bound-steps}
	\begin{aligned}
		\| G_\sigma \phi_1 \|^2 & \le \lambda_{\max}^2   \left[  \< \sigma, \sigma \> - \< \sigma, D \sigma \> \right] \le  \lambda_{\max}^2   \left[ \sum_{i,j =1}^d x_i x_j \< \psi_i , \psi_j \> - \sum_{i,j =1}^d  \sum_{k=1}^d x_i x_j \< \psi_i, \phi_k \> \< \phi_k , \psi_j \> \right] \\			
		&= \lambda_{\max}^2 \left[ \sum_{i=1}^d x_i^2 \left( 1 - \sum_{k=1}^{d} \< \psi_i , \phi_k \>^2 \right) - \sum_{i \ne j} x_i x_j  \left(  \sum_{k=1}^d \< \psi_i , \phi_k \> \< \phi_k ,  \psi_j \> \right) \right] 
	\end{aligned}
\end{equation}
By Corollary~\ref{cor:compression}, for every index $i$ in the sum, we can extend $\{\phi_1\}$ to a maximal set of perfectly distinguishable pure states $\{\phi_1\} \cup \{ \phi_k^{(i)} \}_{k=2}^{d}$ such that $\<\psi_i, \phi_1\> + \< \psi_i, \phi^{(i)}_2\> = 1$ and $\sum_{k=3}^{d} \< \psi_i, \phi^{(i)}_k\> = 0$. Let us also define by $F^{(i)}$ the face generated by $\{ \phi_1 , \phi^{(i)}_2 \}$ (notice that $\psi_i \in F^{(i)}$ by construction), by $\Pi_{F^{(i)}}$ the compression map on $F^{(i)}$, and by $\zeta_j^{(i)} \coloneqq N \, \Pi_{F^{(i)}} \psi_j$, where $N$ is a normalization factor, namely $N = 1 / u ( \Pi_{F^{(i)}} \psi_j) $. Note that other than $\{ \phi_1, \phi^{(i)}_2 \}$, another maximal set of perfectly distinguishable pure states of $F^{(i)}$ is represented by $\{\psi_i, \zeta_j^{(i)}\}$ for any $j \ne i$, indeed $\zeta_j^{(i)}$ is pure and $\< \psi_i , \zeta_j^{(i)}\> = N \< \psi_i , \Pi_{F^{(i)}} \psi_j \> = N \< \Pi_{F^{(i)}}\psi_i ,  \psi_j \> = N \< \psi_i ,  \psi_j \> = 0$. In this way Eq.~\eqref{eq:speed-bound-steps} simplifies to 
\begin{align}
	\| G_\sigma \phi_1 \|^2 &\le \lambda_{\max}^2  \left[  \sum_{i}  x_i^2  \left(  1 - \< \psi_i, \phi_1 \>^2 - (1 - \< \psi_i, \phi_1 \>)^2 \right) - \sum_{i \ne j}  x_i x_j  \left(   \< \psi_i , \phi_1 \> \< \phi_1, \psi_j \>  + \< \psi_i , \phi_2^{(i)} \> \< \phi_2^{(i)}, \psi_j \> \right)  \right] = \notag \\
	&= \lambda_{\max}^2  \left[ 2 \sum_{i}  x_i^2  \< \psi_i, \phi_1 \> \left(  1 - \< \psi_i, \phi_1 \> \right) - \sum_{i \ne j}  x_i x_j     \< \psi_i , \phi_1 \> \< \phi_1 , \psi_j \>  - \sum_{i \ne j}  x_i x_j  \, \< \psi_i , \phi_2^{(i)} \>  \< \phi_2^{(i)}, \zeta_j^{(i)} \>  /N  \right] = \notag \\
	&= \lambda_{\max}^2  \left[ 2 \sum_{i}  x_i^2  \< \psi_i, \phi_1 \> - \sum_i x_i^2  \< \psi_i, \phi_1 \>^2 -  \< \sigma , \phi_1 \>^2 - \sum_{i \ne j}  x_i x_j  \, \left( 1 - \< \psi_i, \phi_1 \> \right) \left( 1- \< \phi_1 , \zeta_j^{(i)} \>  \right) /N \right] = \notag \\
	&= \lambda_{\max}^2  \left[ 2 \sum_{i}  x_i^2  \< \psi_i, \phi_1 \> - \sum_i x_i^2  \< \psi_i, \phi_1 \>^2 -  \< \sigma , \phi_1 \>^2  - \sum_{i \ne j}  x_i x_j  \, \left( 1 - \< \psi_i, \phi_1 \> - \< \phi_1 , \zeta_j^{(i)} \> + \< \psi_i, \phi_1 \>  \< \phi_1 , \zeta_j^{(i)} \> \right) /N \right] = \notag \\
	&= \lambda_{\max}^2  \left[ 2 \sum_{i}  x_i^2  \< \psi_i, \phi_1 \> - \sum_i x_i^2  \< \psi_i, \phi_1 \>^2 -  \< \sigma , \phi_1 \>^2  - \sum_{i \ne j}  x_i x_j  \,  \< \psi_i , \phi_1 \> \< \phi_1, \zeta_j^{(i)} \> /N \right] = \notag \\
	&= \lambda_{\max}^2  \left[ 2 \sum_{i}  x_i^2  \< \psi_i, \phi_1 \> - \sum_i x_i^2  \< \psi_i, \phi_1 \>^2 -  \< \sigma , \phi_1 \>^2  - \sum_{i \ne j}  x_i x_j  \,  \< \psi_i , \phi_1 \> \< \phi_1, \psi_j \> \right] = \notag \\
	&= 2\lambda_{\max}^2  \left[  \sum_{i}  x_i^2  \< \psi_i, \phi_1 \> -  \< \sigma , \phi_1 \>^2 \right] = \notag \\
	&= 2\lambda_{\max}^2  \left[  \sum_{i}  x_i^2  \< \psi_i, \phi_1 \>  -  \left( \sum_{i}  x_i      \< \psi_i , \phi_1 \> \right)^2  \right]\, , \label{eq:preliminary-bound}
\end{align}
where from line 4 to line 5 of the previous equation we used the fact that $1 = \< \psi_i, \phi_1 \> + \< \zeta_j^{(i)} , \phi_1 \>$, because $\{ \psi_i, \zeta_j^{(i)}\}$ is a maximal set of perfectly distinguishable pure states of $F^{(i)}$.

Now, let $H = \lambda_{\max} \sum_{i=1}^d x_i e_{\psi_i} $ be the canonical energy observable associated to $G_\sigma$, \textit{viz.} Eq.~\eqref{new-H}. By calculating the variance of $H$ for the pure state $\phi_1$, we have $\langle H^2\rangle_{\phi_1} =  \lambda_{\max}^2 \sum_{i=1}^d x_i^2 e_{\psi_i}(\phi_1) = \, \lambda_{\max}^2 \sum_{i=1}^d x_i^2 \langle \psi_i, \phi_1 \rangle $ and $\left\langle H\right\rangle^2_{\phi_1} = \lambda_{\max}^2 \left( \sum_{i} x_i \langle \psi_i, \phi_1 \rangle\right)^2$. Namely,
\begin{equation}
	\label{eq:variance-pure-generator}
	\Delta H^2 = \langle H^2\rangle - \left\langle H\right\rangle^2 = \lambda_{\max}^2 \sum_{i=1}^d x_i^2 \langle \psi_i, \phi_1 \rangle - \lambda_{\max}^2 \left( \sum_{i=1}^d x_i \langle \psi_i, \phi_1 \rangle\right)^2  \, . 
\end{equation}

Finally, comparing Eqs.\eqref{eq:preliminary-bound} and~\eqref{eq:variance-pure-generator} together, we get
\begin{equation*}
	\begin{aligned}
		\norm{G_\sigma \phi_1}^2 \le  2  \Delta H^2  \, .
	\end{aligned}
\end{equation*} \qed

\subsubsection{Proof of theorem~4: general case}
\label{sec:proof-thm3-part2}

As before, let $G_\sigma$ be the generator of the collisional dynamics generated by $\sigma = \sum_{i=1}^d p_i \psi_i$ and let $\rho = \sum_{j=1}^d q_j \psi'_j $ be another state of system $\rA$. By the convexity of the norm squared we get
\begin{equation*}
	\begin{aligned}
		\norm{G_\sigma \rho}^2 &=  \norm{\sum_{j=1}^d q_j G_\sigma  \psi'_j}^2 \le  \sum_{j=1}^d q_j \norm{ G_\sigma  \psi'_j}^2 \,.
	\end{aligned}
\end{equation*}
Now, for every $j$ in the sum we can use the result of the previous section, namely
\begin{equation}
	\label{eq:bound-variance}
	\norm{G_\sigma \rho}^2 \le  \sum_{j=1}^d  q_j \norm{G_\sigma  \psi'_j}^2 \le 2 \sum_{j=1}^d q_j \Delta H_{\psi'_j}^2  \le  2 \Delta H_{\rho}^2
	\,,
\end{equation}
where we used the notation $\Delta H_{\psi'_j}^2$ and $\Delta H_{\rho}^2$ to point out that the variance of $H$ was calculate respect to $ \psi'_j$ and $\rho$, respectively, and in the last step we used the concavity of the variance function.

Finally, combining the bound on the average speed in Proposition.~\eqref{prop:evolution-speed} with Eq.~\eqref{eq:bound-variance}, we have that for any two instants of time $t_0$ and $t_1 \in \R$ and for any state $\rho \in \St(\rA)$
\begin{equation*}
	\begin{aligned}
		v_\rho (t_0, t_1) &\le \sqrt 2 \Delta H \\
		\frac{\norm{\rho_{t_1} - \rho_{t_0} } } {t_1 - t_0 } &\le \sqrt 2 \Delta H \\
		\Delta t := t_1 - t_0 &\ge \frac{\norm{\rho_{t_1} - \rho_{t_0} } } {\sqrt 2 \Delta H}  \\
		\Delta t &\ge \frac{ D(\rho_0,  \rho_1) } { \Delta H} \, ,
	\end{aligned}
\end{equation*}
where we introduced the normalized Euclidean distance between the states $\rho_0$ and $\rho_1$ as $D(\rho_0,  \rho_1)  :  =    \norm{ \rho_1  -  \rho_0 } /\sqrt 2 $. \qed
	
\end{document}